\documentclass{aa}  

\usepackage{graphicx}
\usepackage{txfonts}
\newcommand{\iram}{IRAM-30\,m}

\newcommand{\lsim}{\raisebox{-.4ex}{$\stackrel{<}{\scriptstyle \sim}$}}
\newcommand{\gsim}{\raisebox{-.4ex}{$\stackrel{>}{\scriptstyle \sim}$}}
\newcommand{\farc}{\mbox{$.\!\!^{\prime\prime}$}}

\newcommand{\mloss}{\mbox{$\dot{M}$}}

\newcommand{\my}{\mbox{$M_{\odot}$~yr$^{-1}$}}
\newcommand{\ls}{\mbox{$L_{\odot}$}}
\newcommand{\msun}{\mbox{$M_{\odot}$}}
\newcommand{\md}{\mbox{$M_{\rm d}$}}
\newcommand{\mhii}{\mbox{$M_\ion{H}{ii}$}}
\newcommand{\mtot}{\mbox{$M_{\rm H_2}$}}
\newcommand{\mco}{\mbox{$M_{\rm H_2}^{\rm hot}$}}
\newcommand{\rsun}{\mbox{$R_{\odot}$}}
\newcommand{\rs}{\mbox{$R_{\star}$}}

\newcommand{\lstar}{\mbox{$L_{\star}$}}
\newcommand{\mstar}{\mbox{$M_{\star}$}}
\newcommand{\rin}{\mbox{$R_{\rm in}$}}

\newcommand{\nlyc}{\mbox{$N_{\rm LyC}$}}
\newcommand{\kms}{\mbox{km\,s$^{-1}$}}

\newcommand{\vexp}{\mbox{$V_{\rm exp}$}}
\newcommand{\vexslow}{\mbox{$V_{\rm slow}$}}
\newcommand{\vexfast}{\mbox{$V_{\rm fast}$}}

\newcommand{\vrot}{\mbox{$V_{\rm rot}$}}
 
\newcommand{\vsys}{\mbox{$V_{\rm sys}$}} 
\newcommand{\vlsr}{\mbox{$V_{\rm LSR}$}}

\newcommand{\h}{$^{\rm h}$}
\newcommand{\m}{$^{\rm m}$}

\newcommand{\te}{\mbox{$T_{\rm e}$}}

\newcommand{\td}{\mbox{$T_{\rm d}$}}
\newcommand{\teff}{\mbox{$T_{\rm eff}$}}

\newcommand{\dense}{\mbox{$n_{\rm e}$}}

\newcommand{\jb}{\mbox{Jy\,beam$^{-1}$}}

\newcommand{\tha}{\mbox{$\theta_{\rm a}$}}
\newcommand{\thd}{\mbox{$\theta_{\rm d}$}}
\newcommand{\thw}{\mbox{$\theta_{\rm w}$}}

\newcommand{\htal}{\mbox{H30$\alpha$}}
\newcommand{\htnal}{\mbox{H39$\alpha$}}
\newcommand{\hcg}{\mbox{H55$\gamma$}}
\newcommand{\hce}{\mbox{H51$\epsilon$}}
\newcommand{\hsd}{\mbox{H63$\delta$}}

\newcommand{\bn}{\mbox{$b_n$}}

\newcommand{\docem}{$^{12}$CO}
\newcommand{\trecem}{$^{13}$CO}

\def\snu#1{\ifmmode {S_\nu\,\propto\,\nu^{#1}}
          \else \hbox{$S_\nu$\,$\propto$\,$\nu^{#1}$}\fi}
\def\cm#1{\ifmmode {\,{\rm cm^{-#1}}}                  
        \else \hbox{$\,${\rm cm$^{\rm -#1}$}}\fi}
\def\raw {\ifmmode\rightarrow\else$\rightarrow$\fi}
\def\ex#1{\ifmmode {\times 10^{#1}}         
        \else \hbox{{$\times 10^{\rm #1}$}}\fi}

\begin{document} 
\title{A rotating fast bipolar wind and disk system 
  around the B[e]-type star MWC\,922}



   \author{ C.~S\'anchez Contreras\inst{1}
          \and
          A.~B\'aez-Rubio\inst{2}
          \and 
	  J.~Alcolea\inst{3} 
          \and
          A. Castro-Carrizo\inst{4}
          \and 
          V.~Bujarrabal\inst{5}
	  \and
          J. Mart\'in-Pintado\inst{2}
          \and
          D.\,Tafoya\inst{6}}

  \institute{Centro de Astrobiolog{\'i}a (CSIC-INTA), ESAC, Camino Bajo del Castillo s/n, Urb. Villafranca del Castillo,
  E-28691 Villanueva de la Ca\~nada, Madrid, Spain\\ \email{csanchez@cab.inta-csic.es}
  \and
  Centro de Astrobiolog{\'i}a (CSIC-INTA), Ctra de Torrejón a Ajalvir, km 4, 28850 Torrej\'on de Ardoz, Madrid, Spain
  \and
  Observatorio Astron\'omico Nacional (IGN), Alfonso XII
  No 3, 28014 Madrid, Spain
  \and
  Institut de Radioastronomie Millimetrique, 300 rue de la Piscine, 38406 Saint Martin d’Heres, France
  \and Observatorio Astron\'omico Nacional
  (IGN), Ap 112, 28803 Alcal\'a de Henares, Madrid, Spain
  \and  Department of Space, Earth and Environment, Chalmers University of Technology, Onsala Space Observatory, 439 92 Onsala, Sweden
  }
   \date{Received; accepted}

   \date{}

 
   \abstract{We present interferometric observations with the Atacama
     Large Millimeter Array (ALMA) of the free-free continuum and
     recombination line emission at 1 and 3\,mm of the "Red Square
     Nebula" surrounding the B[e]-type star MWC922. The unknown
     distance to the source is usually taken to be $d$=1.7-3\,kpc.
     The unprecedented angular resolution (up to $\sim$0\farc02) and
     exquisite sensitivity of these data unveil, for the first time,
     the structure and kinematics of the emerging, compact ionized
     region at its center. We imaged the line emission of \htal\ and
     \htnal, previously detected with single-dish observations, as
     well as of H51$\epsilon$, H55$\gamma$, and H63$\delta$, detected
     for the first time in this work. The line emission is seen over a
     full velocity range of $\sim$180\,\kms\ arising in a region of
     diameter $<$0\farc14 (less than a few hundred au) in the maser
     line H30$\alpha$, which is the most intense transition reported
     here. We resolve the spatio-kinematic structure of a nearly
     edge-on disk rotating around a central mass of
     $\sim$10\,\msun\ ($d$=1.7\,kpc) or $\sim$18\,\msun\ ($d$=3\,kpc),
     assuming Keplerian rotation.  Our data also unveil a fast
     ($\sim$100\,\kms) bipolar ejection (a jet?)  orthogonal to the
     disk. In addition, a slow ($<$15\,\kms) wind may be lifting off
     the disk. Both, the slow and the fast winds are found to be
     rotating in a similar manner to the ionized layers of the
     disk. {\sl This represents the first empirical proof of rotation
       in a bipolar wind expanding at high velocity
       ($\sim$100\,\kms)}.  The launching radius of the fast wind is
     found to be $<$30-51\,au i.e., smaller than the inner rim of the
     ionized disk probed by our observations. We believe that the fast
     wind is actively being launched, probably by a disk-mediated
     mechanism in a (accretion?) disk around a possible compact
     companion. We have modelled our observations using the radiative
     transfer code MORELI. This has enabled us to describe with
     unparalleled detail the physical conditions and kinematics in the
     inner layers of MWC\,922, which has revealed itself as an ideal
     laboratory for studying the interplay of disk rotation and
     jet-launching. Although the nature of MWC\,922 remains unclear,
     we believe it could be a $\sim$15\,\msun\ post-main sequence star
     in a mass-exchanging binary system. If this is the case, a more
     realistic value of the distance may be $d$$\sim$3\,kpc.}


   \titlerunning{ALMA continuum and mm-RRL emission maps of MWC\,922}
   \authorrunning{S\'anchez Contreras et al.}
   \maketitle

%

\section{Introduction}
\label{intro}

Jets are one of the most intriguing phenomena in Astrophysics, yet
their formation and interaction with their surroundings remains a
challenging topic mainly due to the lack observational data of the
relevant regions.  Our recent pilot study of millimeter radio
recombination lines (mm-RRLs) in a sample of young Planetary Nebula
(yPN) candidates with the \iram\ radiotelescope \citep[][hereafter
  CSC+17]{san17}, showed that mm-RRLs are optimal tracers to probe the
deepest regions at the heart ($\lsim$150au) of these objects, from
where collimated fast winds, presumably responsible for the onset of
asphericity and polar acceleration in these late evolutionary stages,
are launched \citep[see e.g.][for a review]{bal02}.

Studying the jet-launching sites is difficult due to their small
(sub-arcsec) angular sizes and because they are usually heavily
obscured by optically thick circumstellar dust. For this reason, the
wind collimation mechanism in yPNe and related objects (supposedly
hosting mass-exchanging binaries and disks) remains largely
unknown. Progress in this field requires sensitive, high-angular
resolution observations at mm/submm wavelengths of {\em key} objects,
i.e.\,objects that harbour on-going fast bipolar ejections and other
structures with an essential role in jet-formation, such as rotating
disks. The results presented in this work demonstrate that the IR
excess B[e] star MWC\,922 is of one of the very few such objects.

MWC\,922 was dubbed The Red Square Nebula after the discovery of a
large-scale ($\sim$5\arcsec-sized) nebulosity in the near-infrared
with a remarkable X-shaped morphology surrounding the central star
\citep{tut07}. This nebular structure has its symmetry axis oriented
nearly along the NW-SE direction and it is crossed by a dark band at its
center (oriented along the NE-SW direction) that represents an equatorial
dusty disk. An evolved nature has been suggested for MWC\,922 based on
similarities of its nebular morphology and spectral energy
distribution (SED) with the Red Rectangle, a well known post-AGB nebula
\citep[e.g.][]{coh75,jur95,wat98,buj16}. For a comprehensive
description of MWC\,922's characteristics, see, for example, CSC+17
and references therein, or the most recent studies by \cite{weh17} and
\cite{bal19}. These works report on 300\,nm-2.5$\mu$m spectroscopy of
the rich emission line spectrum in this target, including the
detection of PAH and CO first and second overtone emission
\citep{weh17}, and deep narrow-band optical imaging showing the
presence of a $\sim$[0.8-10]\ex{4} years old, pc-scale jet
perpendicular to the disk and expanding at
$\sim$500\,\kms\ \citep{bal19}.

MWC\,922 readily revealed itself as an outstanding source in our pilot
mm-RRLs study (CSC+17). The drastic transition from single-peaked
mm-RRL profiles at 3\,mm (\htnal\ and H41$\alpha$) to double-peaked
profiles at 1\,mm (H31$\alpha$ and \htal) indicated maser
amplification of the highest frequency lines. To our knowledge, only a
few other sources with mm-RRLs masers are known in the literature,
namely, MWC 349A, $\eta$\,Carinae, Cepheus A HW2, MonR2-IRS2, the
PN Mz3 \citep{mar89a,cox95,jim11,jim13,abr14,ale18} and tentatively
towards the starburst galaxy NGC\,253 \citep{bae18}.  The mm-RRL
profiles in MWC\,922 suggested rotation of the ionized gas, probably
arranged in a disk+wind system around a $\sim$8\,\msun\ central mass
(CSC+17). The ionized core of MWC\,922 was modelled using the non-LTE
radiative transfer core MORELI \citep[MOdel for REcombination
  LInes,][]{bae13}, adopting a double-cone geometry for the ionized
disk and wind system. The size of the mm-continuum and mm-RRLs
emitting region was inferred to be $\lsim$150\,au, containing a total
mass of ionized gas of \mhii$\sim$4\ex{-5}\,\msun\ (adopting $d$=1.7\,pc)

As it is the case of many gas/dust enshrouded B-type emission line
stars, the nature of MWC\,922 is uncertain. This is
partially due to the unknown distance to the source. The latter is
usually taken to be in the range $d$=1.7-3\,pc, where the low and high
ends are respectively the distance to the nearby Ser OB1
association\footnote{The Ser OB1 association belongs to the open
  galactic cluster Messier 16, for which a distance of
1.74$^{+0.13}_{-0.12}$\,pc
  has been derived from
  Gaia DR2 \citep{kuh19}.}  (with which MWC\,922 may or may not be
related) and the near kinematic distance deduced from the Galactic
rotation curve and the systemic radial velocity of the source,
\vlsr=32-33\,\kms\ (CSC+17).
For this range of distances, the total luminosity of MWC\,922, after
extinction correction, is \lstar$\sim$[1.8-5.9]\ex{4}\ls\ (CSC+17). The
effective temperature of MWC\,922 is also not well established, although it
probably lies within the range \teff$\sim$20000-30000\,K \citep{rud92}.

Throughout this paper, we use $d$=1.7\,kpc by default because this is
the most commonly used value in the literature and the one adopted in
our original model of MWC\,922 (CSC+17), which facilitates
comparison. However, as a result from our analysis, we believe that a
value of $d$$\sim$3\,kpc may be more realistic (\S\,\ref{dis-hrd}).

\section{Observations and data reduction}
\label{sec-obs}

\begin{table}[ht!]
\centering
\caption{Central frequency, bandwidth, velocity resolution, spectral lines and continuum fluxes in the different
  spectral windows (SPWs).} \label{tab:spws}
\begin{tabular}{r c c l c}
\hline
\hline
Center       &  Bandwidth     &   $\Delta$v  & Line & Continuum \\ 
(GHz)        &     (MHz)      &   (km/s)            &        & Flux (mJy) \\    
\hline
\multicolumn{5}{c}{\sl Band 6. Obs. DATE: 2017-10-28 to 2017-10-28} \\  
232.90  & 1875    & 1.257 & & 194 \\ 
231.90  & 469 & 0.631 & H30$\alpha$ & 191\\ 
230.54  & 469 & 0.635 & $^{12}$CO\,2-1 & 184\\ 
217.50  & 1875    & 1.346 & & 189\\ 
215.40  & 1875    & 1.359 & H51$\epsilon$ & 185 \\ 
\multicolumn{5}{c}{\sl Band 3. Obs. DATE: 2017-09-28 to 2017-09-29} \\ 
110.20  & 234 & 0.664 & $^{13}$CO\,1-0 & 122 \\
109.54  & 234 & 0.668 & H55$\gamma$ & 118 \\ 
109.00  & 469 & 2.686 & & 118 \\ 
107.10  & 234 & 2.734 & & 116 \\
106.74  & 234 & 0.343 & H39$\alpha$ & 117 \\ 
 97.00   & 1875 & 3.019 & & 107 \\  
 95.35   & 1875 & 3.071 & H63$\delta$ & 106\\  
\hline                  
\end{tabular}
\tablefoot{Absolute flux calibration uncertainties are $\sim$5\%.}
\end{table}

The observations were performed with the ALMA 12-m array as part of
our projects 2016.1.00161.S and 2017.1.00376.S. A total of twelve
different spectral windows (SPWs) within Band 3 (3\,mm) and \mbox{Band
  6} (1\,mm) were observed to map the emission of several mm-RRLs as
well as the continuum towards the B[e] star MWC\,922
(Table\,\ref{tab:spws}).  The Band 3 and 6 observations were done
separately in two different $\la$2\,hr-long execution blocks in
September and October 2017, respectively. The data were obtained with
44-48 antennas, with baselines ranging from 41.4\,m to 14.9\,km for
Band 3 and from 113.0\,m to 13.9\,km for \mbox{Band 6}. The maximum
recoverable scale is $\sim$0\farc8 and $\sim$0\farc35 at 3 and 1\,mm,
respectively.

The total time spent on the science target, MWC\,922, was
about 49\,min in Band 3 and 33\,min in \mbox{Band 6}. A number of sources
(J\,1733-1304, J\,1825-1718, J\,1828-2123, J\,1830-1606, J\,1835-1513,
and J1924-2914) were also observed as bandpass, complex gain, and flux
calibrators. The flux density adopted for J\,1733-1304 is 2.958\,Jy at
108.983\,GHz and for J\,1924-2919 is 3.717 Jy at 231.868\,GHz.


An initial calibration of the data was performed using the automated
ALMA pipeline of the Common Astronomy Software Applications
(CASA\footnote{\tt https://casa.nrao.edu/}, versions 4.7.2 and 5.1.1).
Calibrated data were used to identify the line-free channels and to
obtain initial images of the lines and continuum in the different
SPWs.  When strong source emission was detected, the noise in those
channel maps was dominated by secondary lobes triggered by residual
calibration errors. Therefore, given the high signal-to-noise ratio
(S/N$>$100) achieved in the continuum images and the absence of
significant/complex large-scale structure in MWC\,922 in our
observations (see \S\,\ref{results}), we self-calibrated our data
using the initial model of the source derived from the standard
calibration to improve the sensitivity and fidelity of the images.
Self-calibration as well as image restoration and deconvolution was
done using the GILDAS\footnote{{\tt
    http://www.iram.fr/IRAMFR/GILDAS}.} software MAPPING.

We created continuum images for each of the 12 SPWs using line-free
channels. Line emission cubes were produced after subtracting the
corresponding continuum (i.e.\,the continuum of the SPW containing the
line). We used a spectral resolution of $\Delta$v=1.5-4.5\,\kms,
depending on the S/N of the maps; typically $\Delta$v=1.5\,\kms\ for
the \htal\ and \htnal\ transitions and $\Delta$v=4.5\,\kms\ for the
non-$\alpha$ lines.  By default, we used the Hogbom deconvolution
method with a robust weighting scheme,
which results in angular resolutions down to $\sim$22$\times$32\,mas at
1\,mm and $\sim$50$\times$65\,mas at 3\,mm.
The typical rms noise level in the line-free
$\Delta$v=1.5\,\kms\ ($\Delta$v=4.5\,\kms) channels of our spectral
cubes is $\sigma$$\sim$1.1\,m\jb\ ($\sigma$$\sim$0.4\,m\jb). The rms
noise level range in the continuum maps is $\sim$0.10-0.15 and
0.04-0.09\,m\jb\ for the individual SPWs within Band 3 and Band 6,
respectively.

Finally, to better characterize some of the small-scale structural
details of the compact source, we analysed the distribution of clean
components by producing a yet higher angular resolution version of the
images (referred to as super-resolution images) by imposing a circular
restoring beam of 15 and 30\,mas at 1 and 3\,mm, respectively. These
figures, shown in the Appendix, are used to discern between similar
models with slightly different input parameters.

\section{Observational results}
\label{results}

\subsection{Continuum}
\begin{figure}[hb!]
   \centering 
   \includegraphics[width=0.85\hsize]{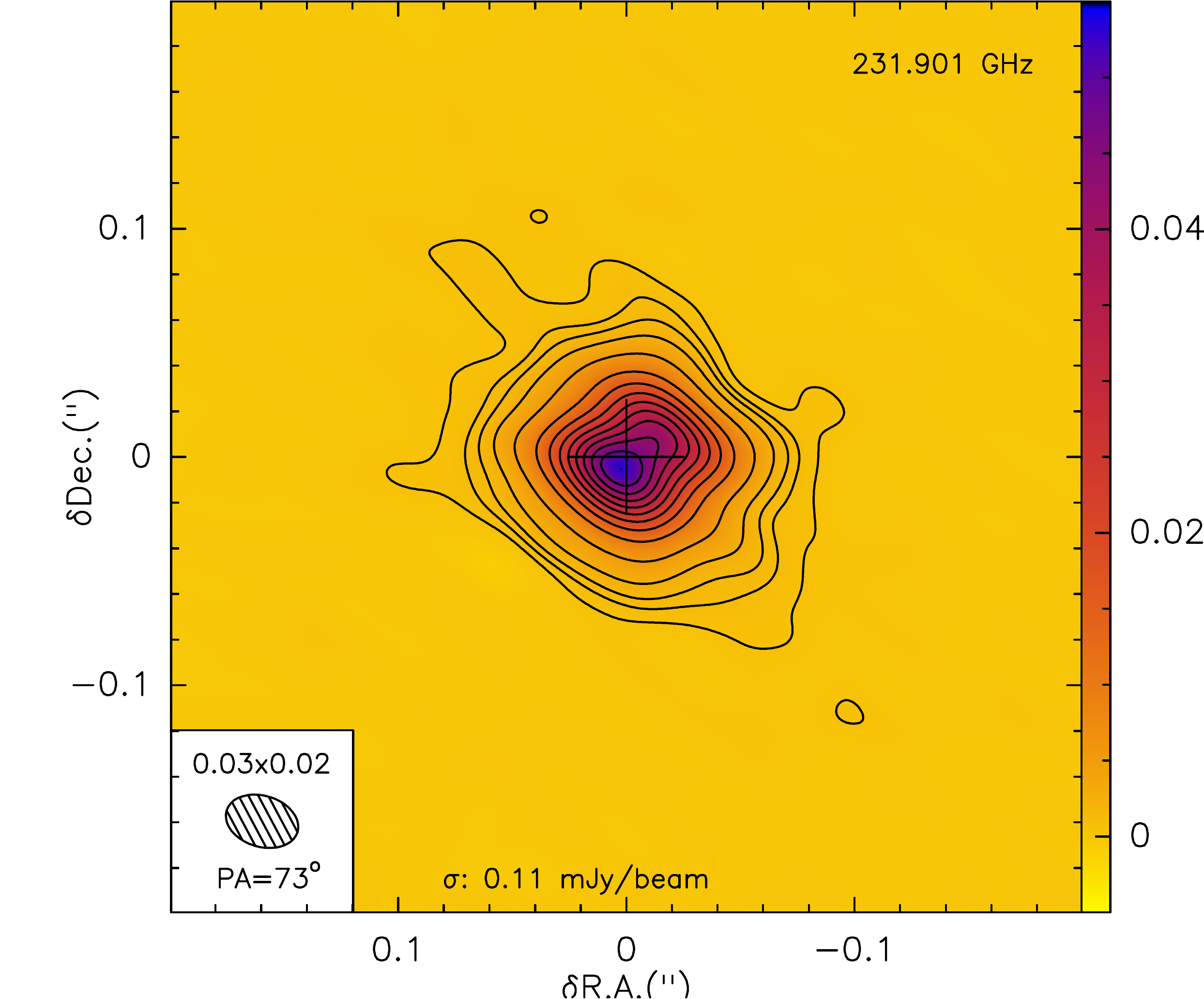}
   \includegraphics[width=0.85\hsize]{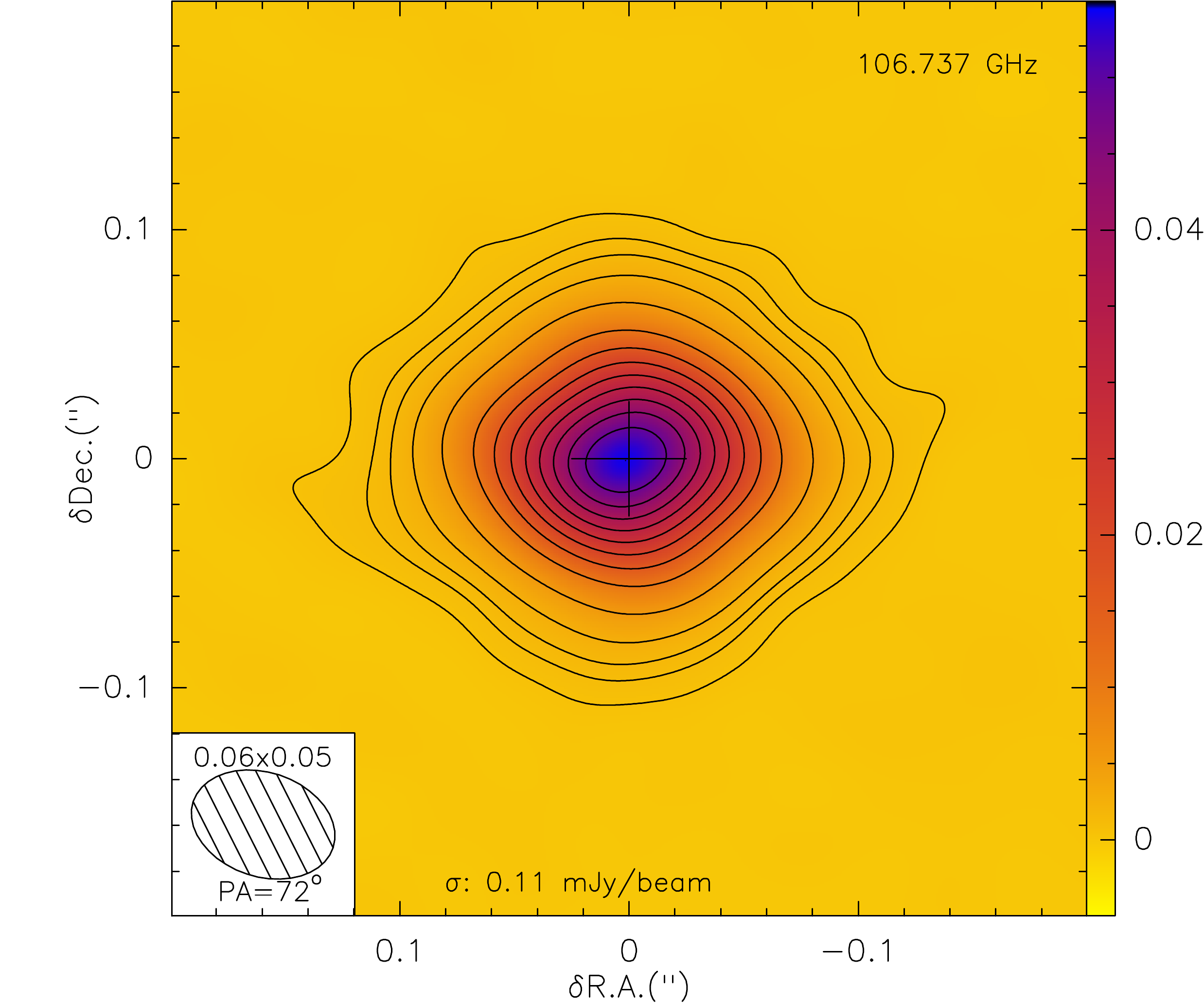}
    \caption{Maps of the continuum emission adjacent to the
      H30$\alpha$ (231.9\,GHz) and H39$\alpha$ (106.7\,GHz)
      lines. Contour levels are at 1\%, 2\%, 3\%, 5\%, and from 10\%
      to 100\% by 10\%\ of the peak (53.0 and 54.5\,m\jb\ at 1 and
      3\,mm, respectively). The central cross marks the
        3\,mm-continuum surface brightness peak.
   \label{f-cont-acc}}      
   \end{figure}

Continuum emission maps at frequencies near the 
H30$\alpha$ (231.9\,GHz) and H39$\alpha$ (106.7\,GHz) lines 
are shown in Fig.\,\ref{f-cont-acc}.  No significant differences in
the surface brightness distribution of the continuum
at other SPWs within the same band (not shown) were
found. The 3\,mm continuum surface brightness peaks at coordinates
RA=18\h21\m16\fs0570 and Dec=$-$13\degr01\arcmin25\farc72 (J2000).
Throughout this paper, these coordinates have been adopted in defining
the (0\arcsec, 0\arcsec) positional offsets in all figures
illustrating image data.

The continuum emission, which is spatially resolved, appears as a
square- or rectangular-like nebulosity of full ($\sim$3$\sigma$-level)
dimensions $\sim$0\farc12$\times$0\farc16 (convolved with the beam) at
1\,mm. Its long axis is oriented along PA$\sim$45\degr, i.e.\,
similarly to the dark equatorial band crossing the center of the NIR
nebula (\S\,\ref{intro}). The central brightest parts of the
1\,mm-continuum nebulosity are elongated orthogonally to the equator
where it exhibits a subtle waist. 
The ALMA continuum maps with super-resolution (15 and 30\,mas at 1\,mm and
3\,mm, respectively) are shown in Fig.\,\ref{f-cont-HiRes}. These maps
reveal an X-shape morphology of the continuum emitting region
reminiscent of the
NIR and mid-IR nebulae observed at much larger angular scales
\citep[from $\approx$1\arcsec\ to
  $\approx$100\arcsec,][]{tut07,lag11}.  The X-shape of the
mm-continuum maps is consistent with the free-free emission arising in a
biconical shell inscribed in a larger, and predominantly neutral,
rotating disk that is illuminated and photoionized by the central
source, as we hypothesized in CSC+17.

   \begin{figure*}[htbp!]
   \centering 
   \includegraphics[width=0.560\hsize]{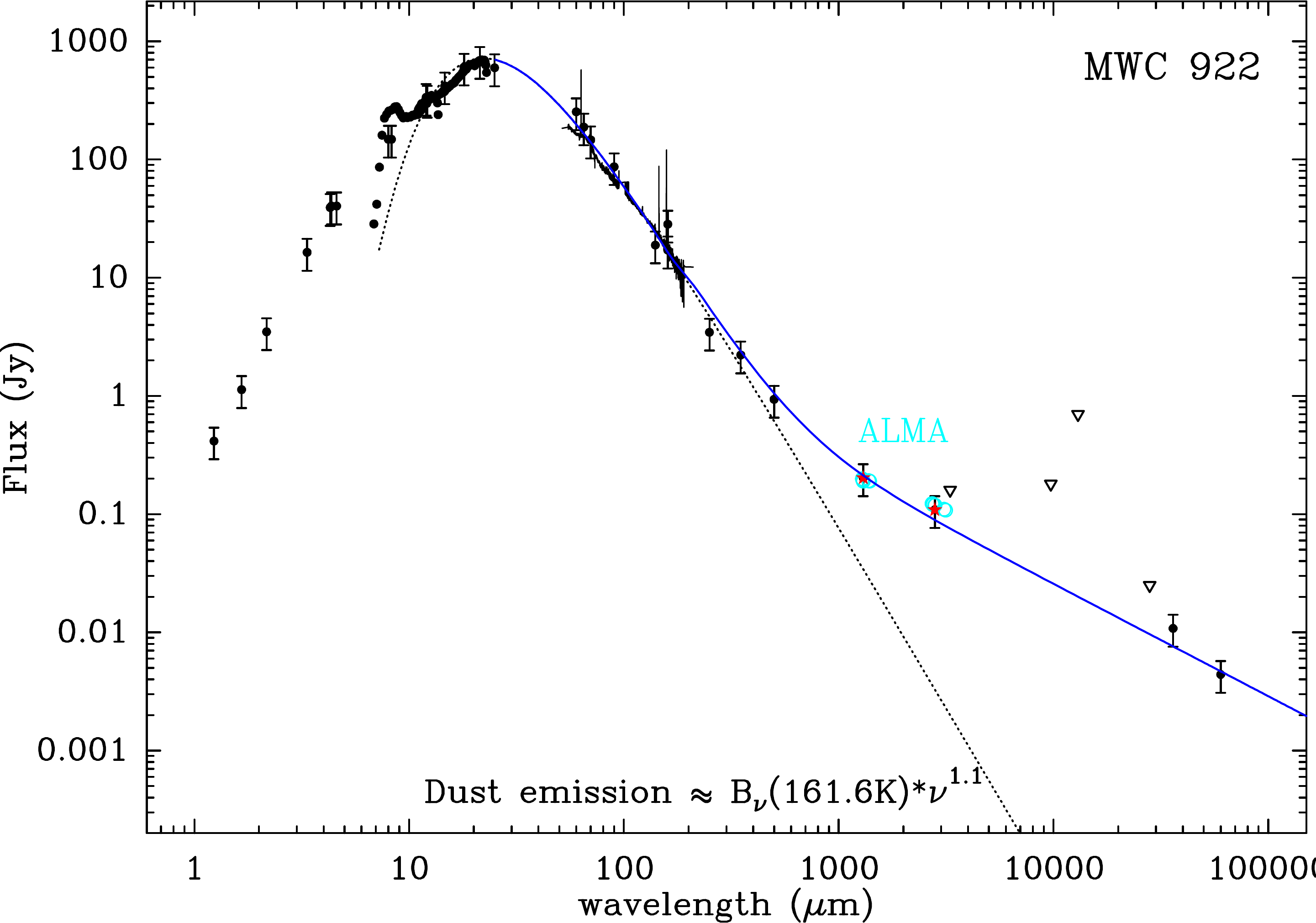}
   \includegraphics[width=0.410\hsize]{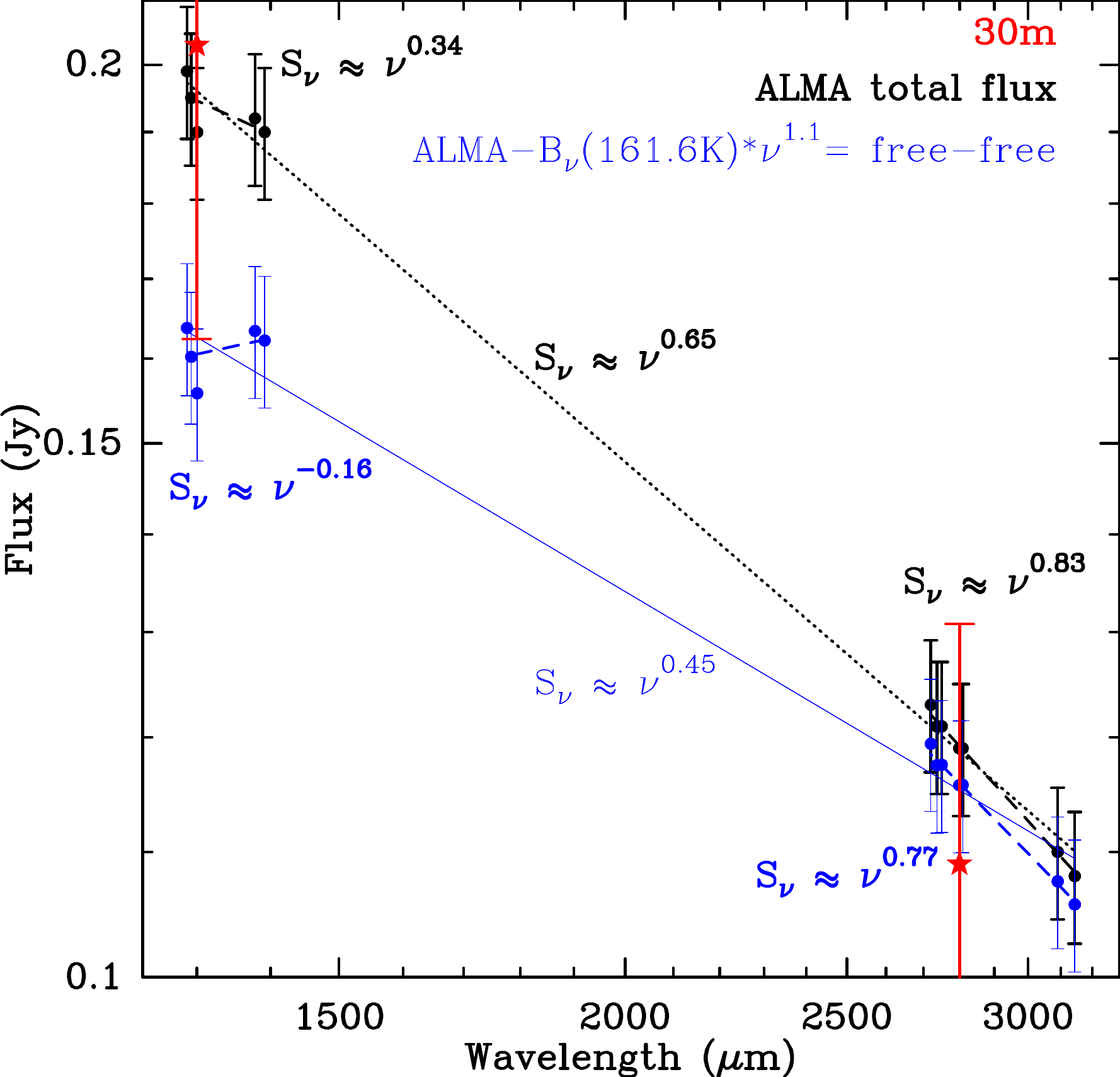}   
       \caption{ {\bf Left)} SED of MWC\,922 as in CSC+17, after
         adding PACS and SPIRE data \citep{mol16,jr18}, and including
         our ALMA continuum flux measurements (cyan, Table\,\ref{tab:spws}). {\bf Right)}
         ALMA mm-continuum fluxes before
         (black) and after (blue) subtraction of the $\sim$162\,K
         dust emission contribution (dotted line in the left
         panel). Red star-like symbols represent single-dish
         \iram\ data (CSC+17). 
          \label{f-sed}
        }
   \end{figure*}

The total continuum fluxes integrated over the whole source at
twelve different frequencies are given in Table\,\ref{tab:spws}. These
fluxes are shown in Fig.\,\ref{f-sed} together with the SED of MWC\,922 from the near-IR to the radio
domain. The comparison of the ALMA fluxes with those obtained at
similar frequencies with the \iram\ single-dish telescope
(CSC+17) indicates that there are no significant interferometric flux
losses in our data. The spectral index deduced from our ALMA
1\,mm-to-3\,mm data is $\alpha$$\sim$0.65 (Fig.\,\ref{f-sed}-right),
which is consistent with predominantly free-free
continuum from an ionized wind, as already known (CSC+17). The spectral slope of the
continuum is rather different if we compute it using separately Band 3 (3\,mm)
and \mbox{Band 6} (1\,mm) continuum fluxes (\snu{0.83} and \snu{0.34}, respectively).

As can be seen in the left panel of Fig.\,\ref{f-sed}, there is a
significant contribution ($\sim$15-20\%) of dust emission to the total flux at
1\,mm. The dust emission in MWC\,922 in the mid-to-far IR (from
$\sim$12 to $\sim$300\,$\mu$m) is well represented by a modified
black-body at $\sim$162\,K with a dust opacity frequency dependence of
$\sim$$\nu$$^{1.1}$ (dotted line in Fig.\,\ref{f-sed}-left).  After
subtraction of the dust emission contribution to the
total continuum flux measured with ALMA (blue symbols in
Fig.\,\ref{f-sed}-right), the spectral index of the remaining
free-free emission at 3\,mm is smaller than, but similar to, the
initial value ($\alpha$=0.77$\pm$0.04). However, at 1\,mm the
free-free continuum flattens notably and even appears to reach a
negative spectral index ($\alpha$=$-$0.16$\pm$0.3). Although the
spectral index at 1\,mm is poorly determined, mainly due to the
uncertainties in the fit of the underlying dust emission and the
scatter of the ALMA data-points, this behaviour suggests that the
free-free emission is optically thin around 1\,mm. The positive
spectral index at 3\,mm, however, is consistent with partially
optically thick free-free emission. One can thus expect the free-free
turnover frequency to be somewhere in between these two wavelengths.

\subsection{Radio recombination lines}

We have mapped with ALMA a total of five radio recombination lines in
the mm-wavelength range: H30$\alpha$ (231.9\,GHz) and H39$\alpha$
(106.7\,GHz), previously detected and reported by CSC+17, as well as
H51$\epsilon$ (215.7\,GHz), H55$\gamma$ (109.5\,GHz), and H63$\delta$
(96.0\,GHz), which are first time detections.

\begin{figure*}[htbp!]
   \centering 
     \includegraphics[width=0.28\hsize]{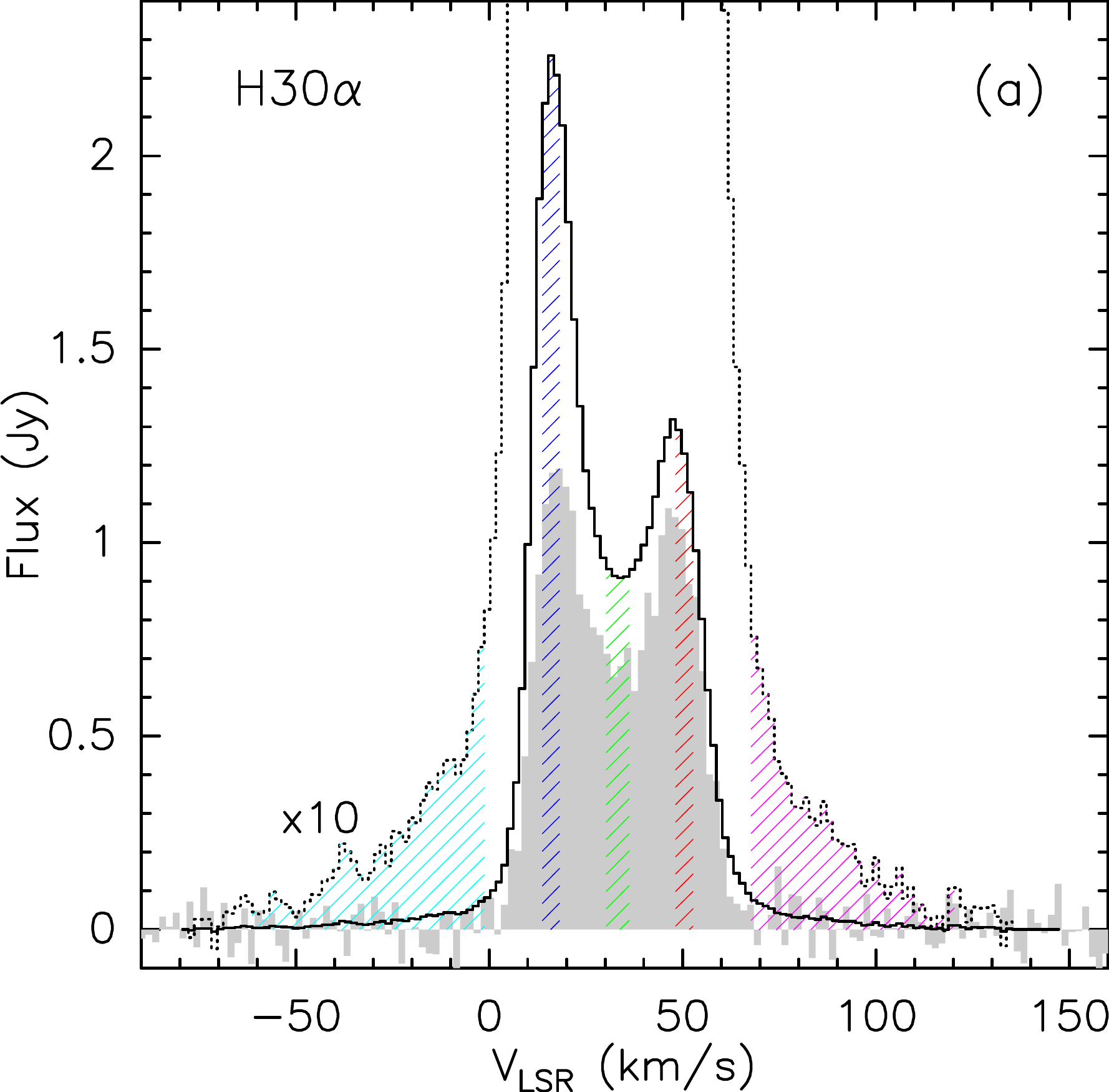}
   \includegraphics*[bb= 0 0 644 594,width=0.255\hsize]{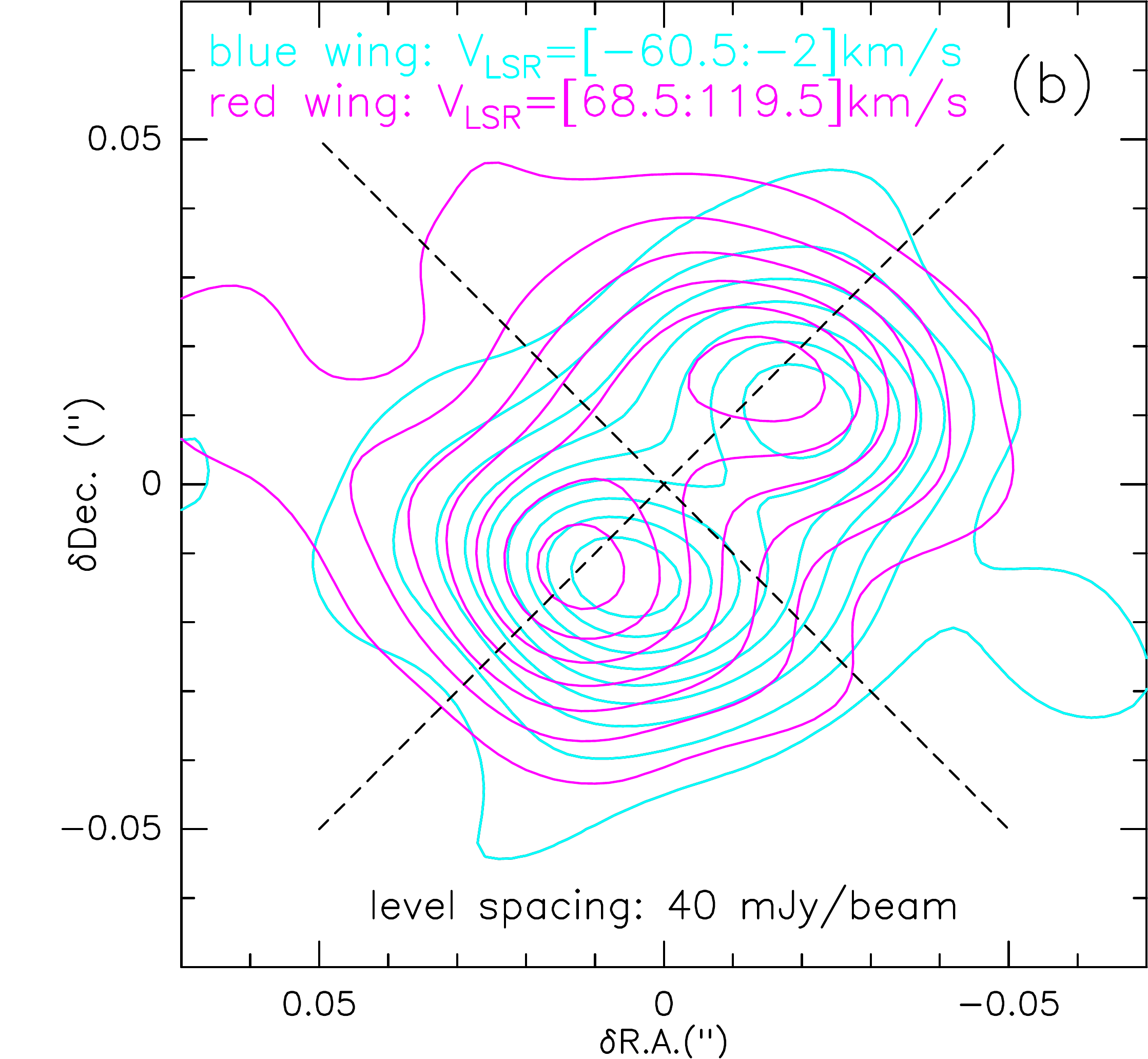}
   \includegraphics*[bb= 95 0 644 594,width=0.222\hsize]{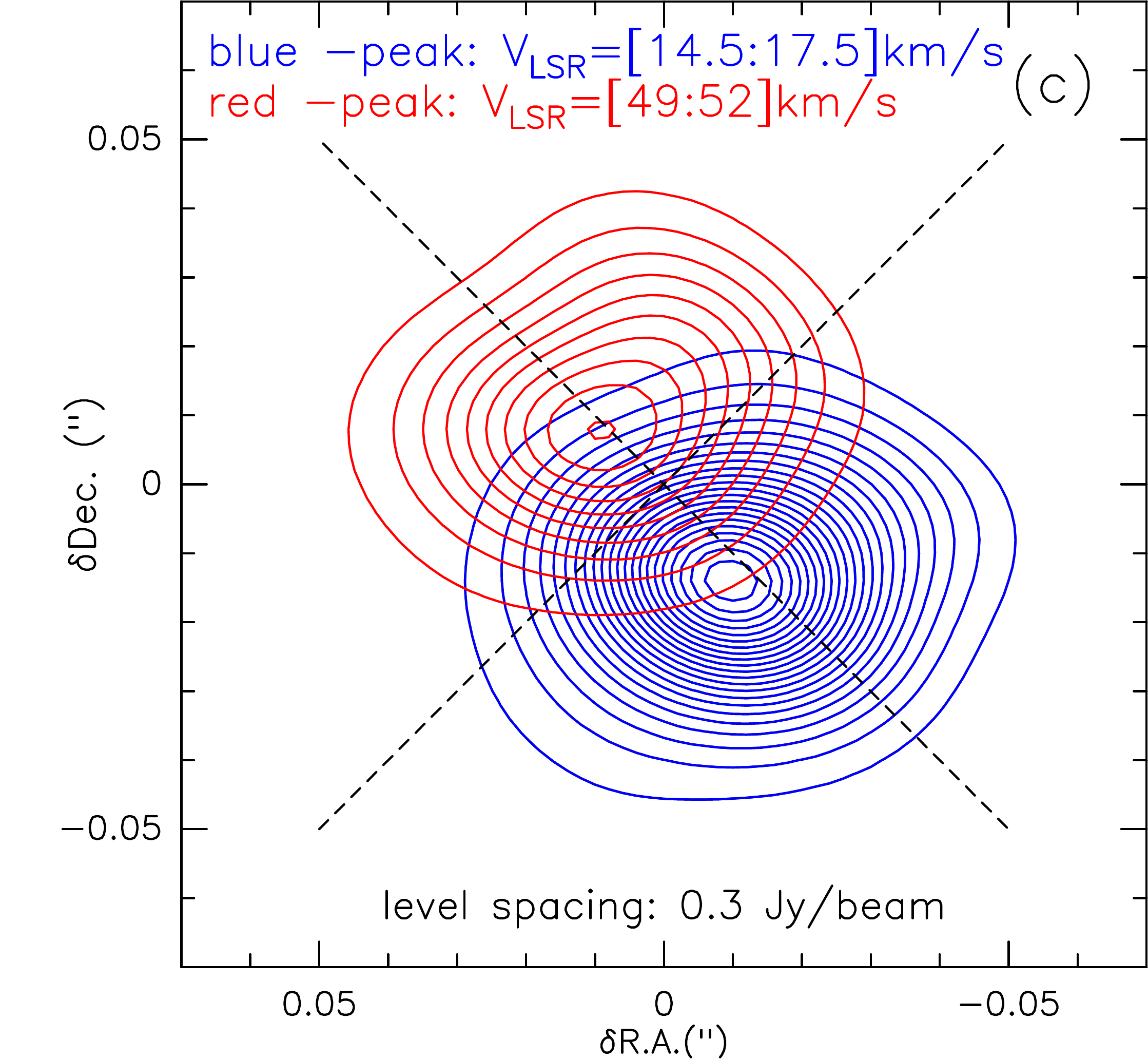}
   \includegraphics*[bb= 95 0 644 594,width=0.222\hsize]{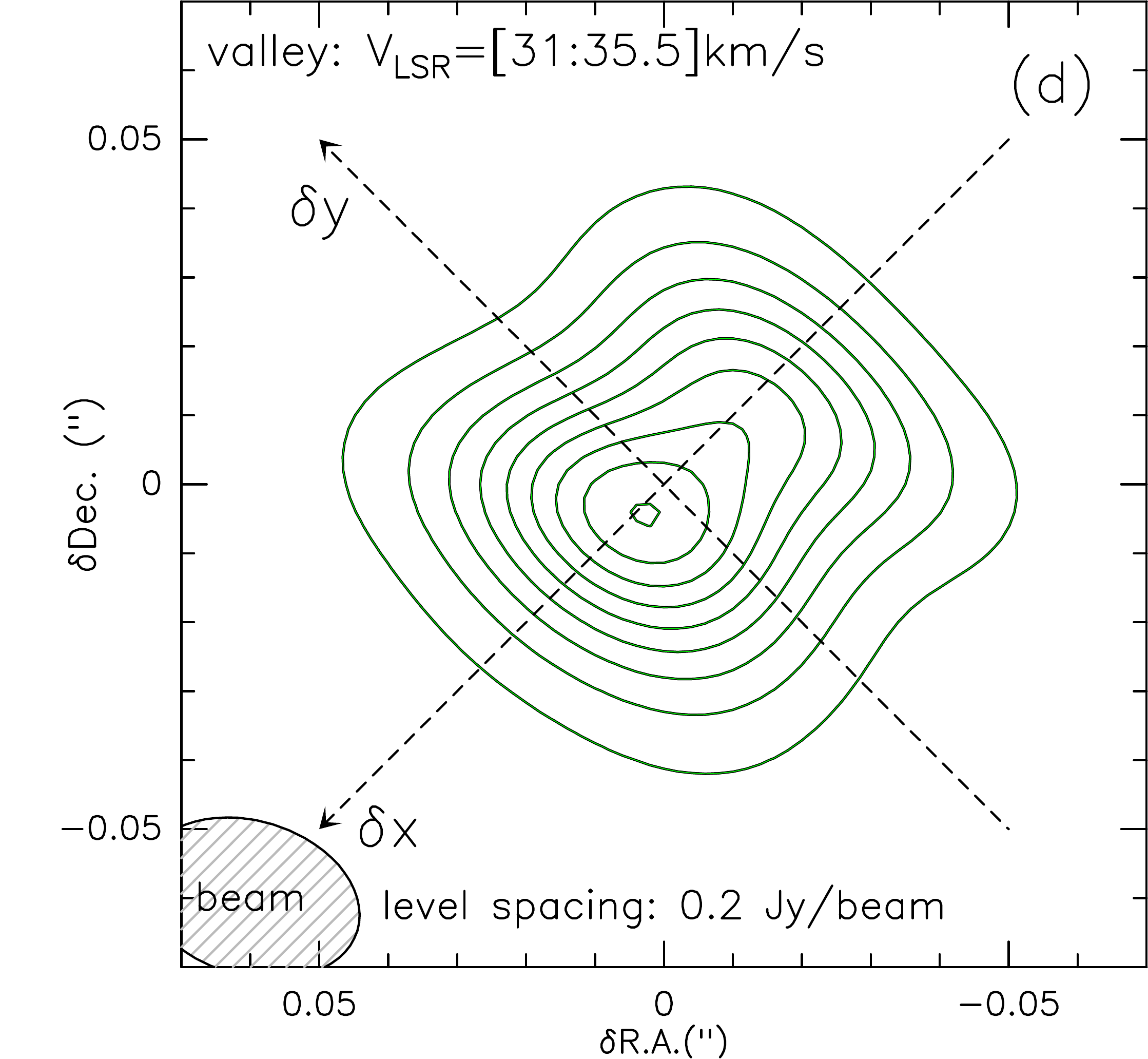} \\
   \vspace{0.5cm}
     \includegraphics[width=0.45\hsize]{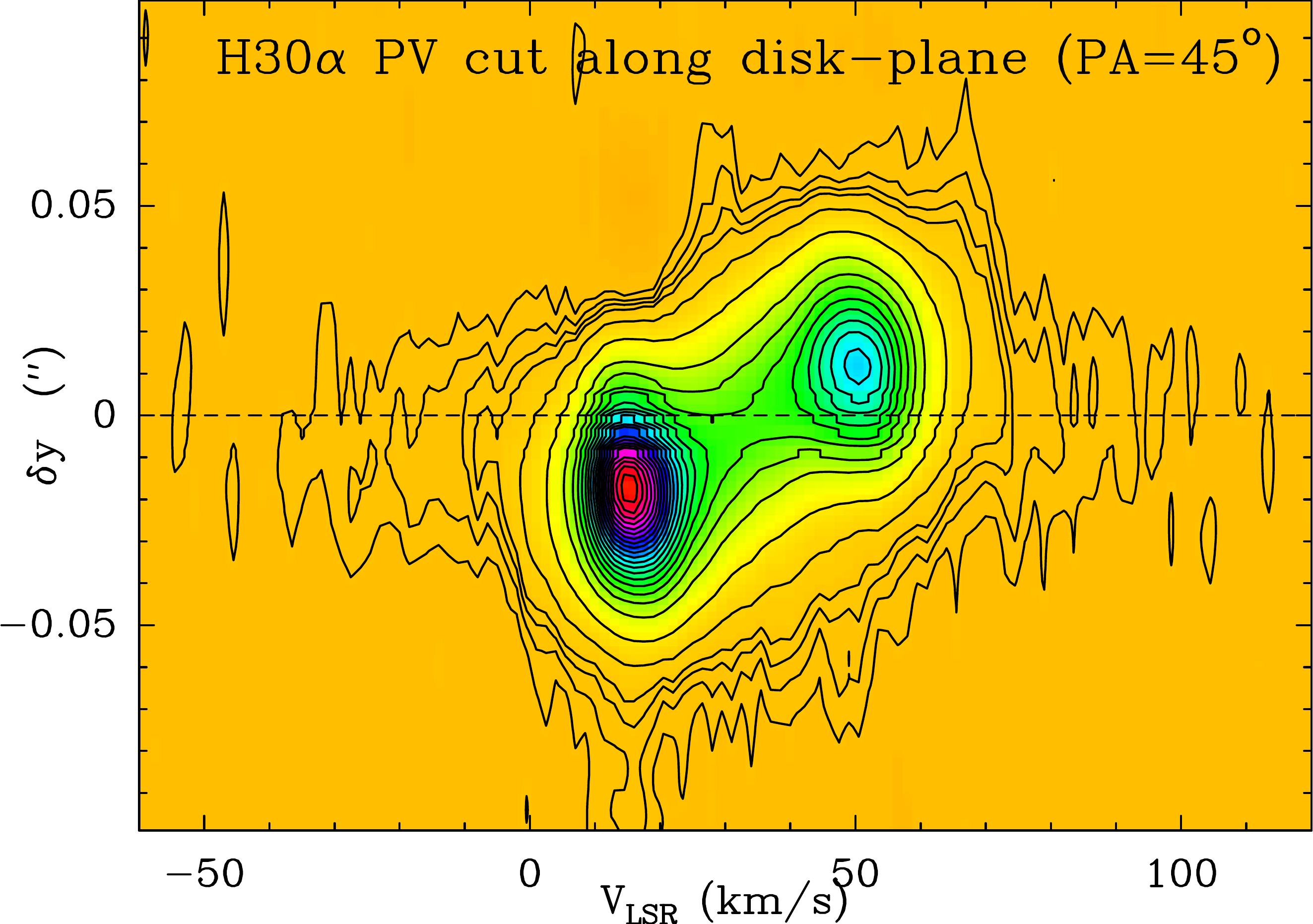} 
     \includegraphics[width=0.45\hsize]{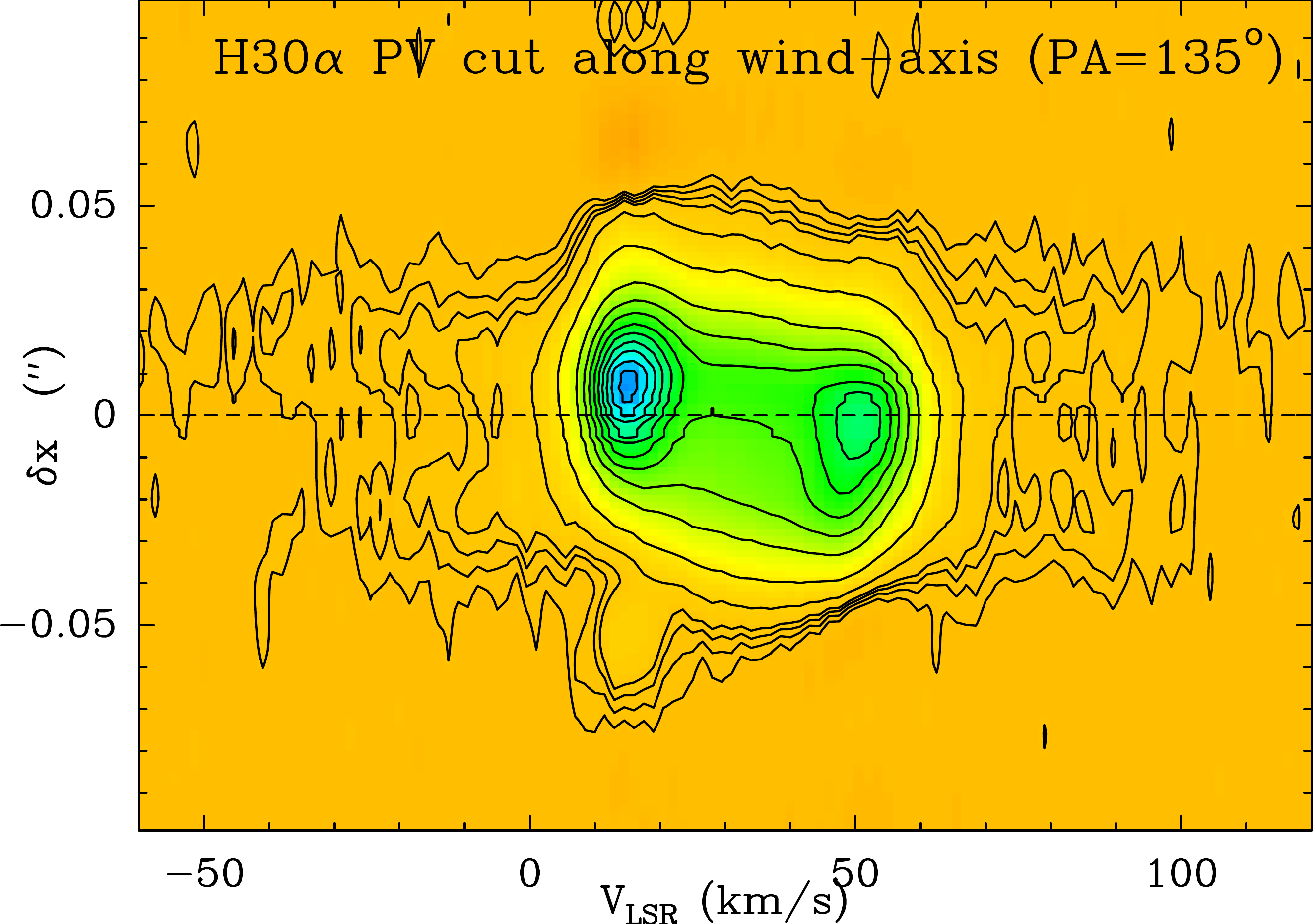} 

   \caption{Summary of ALMA data of the \htal\ line (see also
     Fig.\,\ref{f-cubeH30a}). {\bf Top) (a)} Integrated spectrum of
     the H30$\alpha$ line obtained with ALMA (black lines) and with
     the \iram\ antenna (grey histogram, CSC+17). The hatched-coloured
     regions indicate the different velocity intervals used to
     generate the maps in panels b,c, and c. The ALMA profile
     multiplied by a factor 10 (dotted line) is overplotted for a
     clearer view of the broad, line wings. In panels ({\bf b}), ({\bf
       c}), and ({\bf d}) we show the H30$\alpha$ emission maps
     integrated over the line wings, the peaks, and the valley,
     respectively. {\bf Bottom)} Position velocity cuts of the
     H30$\alpha$ line through the center of MWC\,922 along the disk
     plane (left, PA=45\degr) and the perpendicular direction
     (right).
     Levels are 2.5, 5, 7.5, 10, 20, and from 50 to 1700 by
     75 m\jb.
     \label{f-h30a-all}}      
   \end{figure*}

The observational results from our ALMA data of the \htal\ and
\htnal\ lines are summarized in Figs.\,\ref{f-h30a-all} to
\ref{f-h39a-all}. The complete velocity-channel maps of these
transitions are reported in the Appendix, in
Figs.\,\ref{f-cubeH30a}-\ref{f-cubeH39a}.

The H30$\alpha$ maser shows a double-horn line profile, with two
prominent peaks of emission at \vlsr$\sim$15.3 and 50.3\,\kms\ (Fig.\,\ref{f-h30a-all}). 
The high sensitivity of ALMA unveils weak
$\sim$180\,\kms-wide wings in the \htal\ profile, within the velocity range 
\vlsr$\sim$[$-$60,120]\,\kms.

As expected, the H30$\alpha$ line emission arises from a region that is comparable,
in shape and dimensions, to the continuum-emitting area. 
The emission from the red and blue peaks are displaced to the NE
(defined by us as the $y$-axis) and SW side of the nebula,
respectively, i.e., along the equatorial waist of the large-scale NIR
nebula (Fig. \ref{f-h30a-all}c). This is consistent with an equatorial
structure in rotation, which is also apparent in the position velocity
cut shown in the bottom-left panel of Fig.\,\ref{f-h30a-all}. The
relative separation between the red- and blue-peak emitting areas
along the equator is about $\delta y$=0\farc03.  For an edge-on
rotating disk, the velocity gradient observed implies a rotation
velocity of $\sim$17\,\kms\ at a radial distance of about
25\,au$\times$$\frac{d}{1.7\,kpc}$. Assuming Keplerian rotation, this
implies a value for the central mass of
$\sim$8\,\msun$\times$$\frac{d}{1.7\,kpc}$, in agreement with the
value inferred from our pilot study of single-dish mm-RRLs (CSC+17).

The spatial distribution of the red and blue emission peaks  
is extended along the direction perpendicular to the disk-plane 
(PA$\sim$135\degr, defined by us as the $x$-axis), with an angular size at half-intensity (at 3$\sigma$
level) of $\sim$0\farc04 ($\sim$0\farc09). 
This indicates that the H30$\alpha$ line emission, as the continuum,
is produced over a range of latitudes in a biconical shell and not in
a flat, equatorially thin disk, in agreement with our initial model of
the source presented in CSC+17. This is also directly visible in the
maps of the {\it valley} of the H30$\alpha$ transition, i.e.\,the
spectral region between the two emission peaks
(Fig.\,\ref{f-h30a-all}d). The pinched-waist rectangular
shape of these maps
stands for an underlying X-shaped morphology,
which is expected from a biconical emitting surface 
and that indeed becomes evident in our super-resolution maps
(Fig.\,\ref{f-h30a-all-moreli-hires}). 
Rotation in mid-to-high latitude regions is also visible in the PV cuts along the direction of the disk-plane at
different $\delta x$ offsets shown in Fig.\,\ref{f-pvoffsets}.

The H30$\alpha$ wing emission arises from a 
$\sim$0\farc095$\times$0\farc065 peanut-shaped region elongated in
the direction perpendicular to the disk-plane
(Fig. \ref{f-h30a-all}b). This is consistent with the presence of a fast bipolar 
wind orthogonal to the rotating disk. 
The distribution of the wing emission, concentrated towards the disk
rotation axis, indicates that the opening angle of the fast wind is
smaller than that of the biconical surface of the disk.
That is, the fast wind runs through the interior of
(occupies higher latitudes than) the disk, where the red- and
blue-emission peaks are predominantly formed. The position-velocity cut
along the polar axis of the wind (PA=135\degr,
Fig.\,\ref{f-h30a-all}, bottom-right) does not put on view a significant
velocity gradient in this direction, which is consistent with the
polar axis of the fast flow lying in the plane of the sky or very
close to it. The full width of the wings ($\sim$180\,\kms) indicates a lower limit to the
expansion velocity of the fast polar wind of $\sim$90\,\kms.

As we can see in
Fig.\,\ref{f-h30a-all}b, the two relative maxima of the \htal\ red-
and blue-wing emitting regions are $\sim$0\farc036 apart from each
other ($\sim$61\,au at 1.7\,kpc). This is compatible with the presence
of a central cavity of radius $\la$30\,au at the center of the ionized
region that surrounds MWC\,922 (see \S\,\ref{moreli}).
The peanut-like emitting regions of the red- and blue-wings do not lie
exactly on the disk's revolution axis (PA=135\degr). In fact, the red-
(blue-) wing emitting regions are slightly offset (by $\sim$0\farc01)
towards the NE (SW) direction. The overall offset of these two
features denotes a velocity gradient perpendicular to the outflow axis
that is an explicit indication for rotation in the fast wind. This is
also visible in the PV cuts along the disk-plane
(Fig.\,\ref{f-h30a-all}, bottom-left, and Fig.\,\ref{f-pvoffsets}) and
it is corroborated by our model (as we will show in \S\,\ref{moreli}).

   \begin{figure}[htbp!]
     \centering
      \includegraphics[width=0.75\hsize]{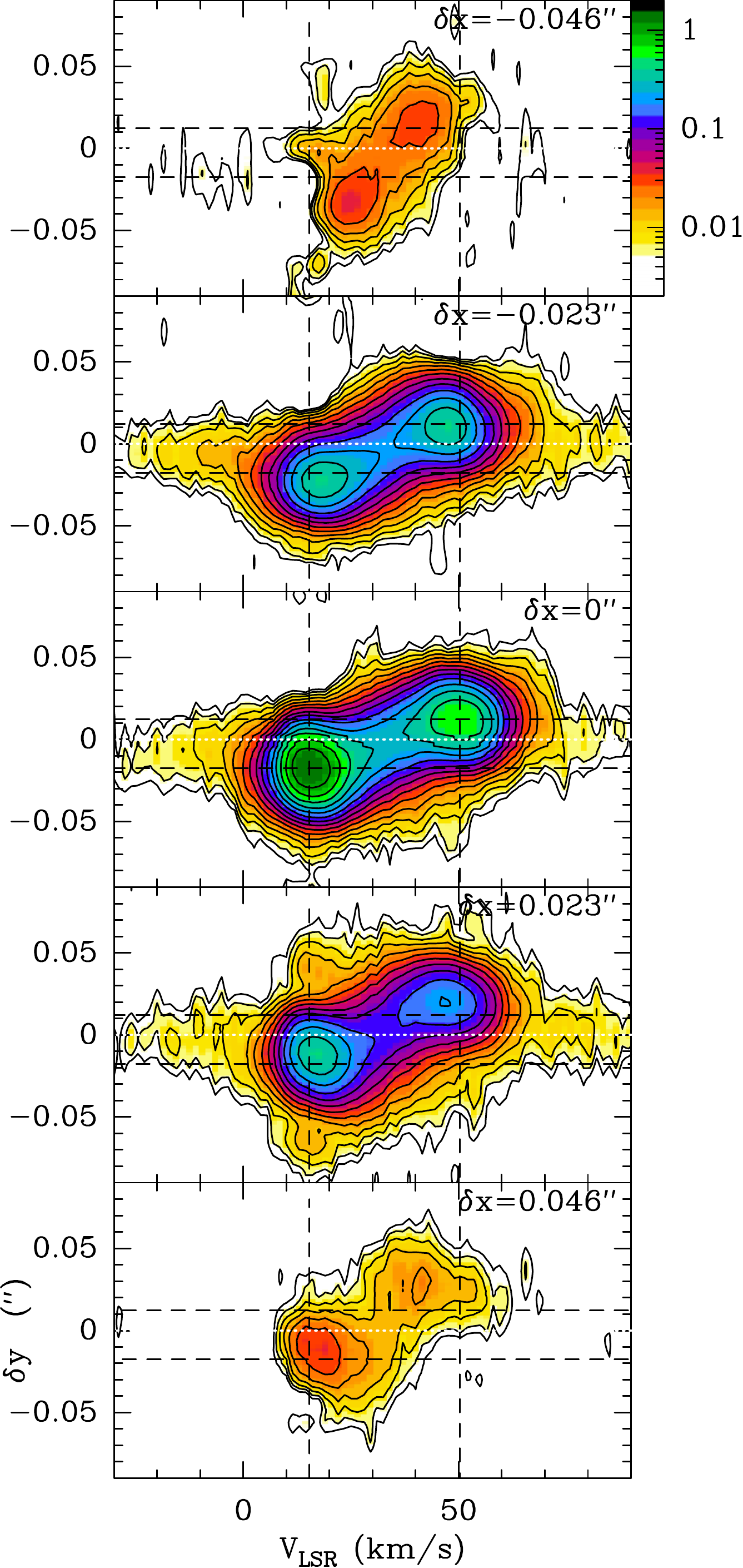}
   \caption{Position-velocity (PV) cuts of the \htal\ line along the
     direction of the equator (PA=45\degr) at different $\delta x$
     offsets ($\delta x$=$-$0\farc046, $\delta x$=$-$0\farc023,
     $\delta x$=0\arcsec, $\delta x$=+0\farc023, and $\delta
     x$=+0\farc046). The dashed lines mark the position and velocity
     of the red- and blue-emission peaks at offset $\delta
     x$=0\arcsec. Offset $\delta y$=0 is indicated by the white dotted
     line. The signature of rotation is very clear even at the largest
     offsets. Levels are 3.5$\times$1.5$^i$\,($i$=1,2,3,...)\,m\jb; the units of the wedge are m\jb. 
               \label{f-pvoffsets}
        }
   \end{figure}

   \begin{figure*}[htbp!]
   \centering 
     \includegraphics[width=0.28\hsize]{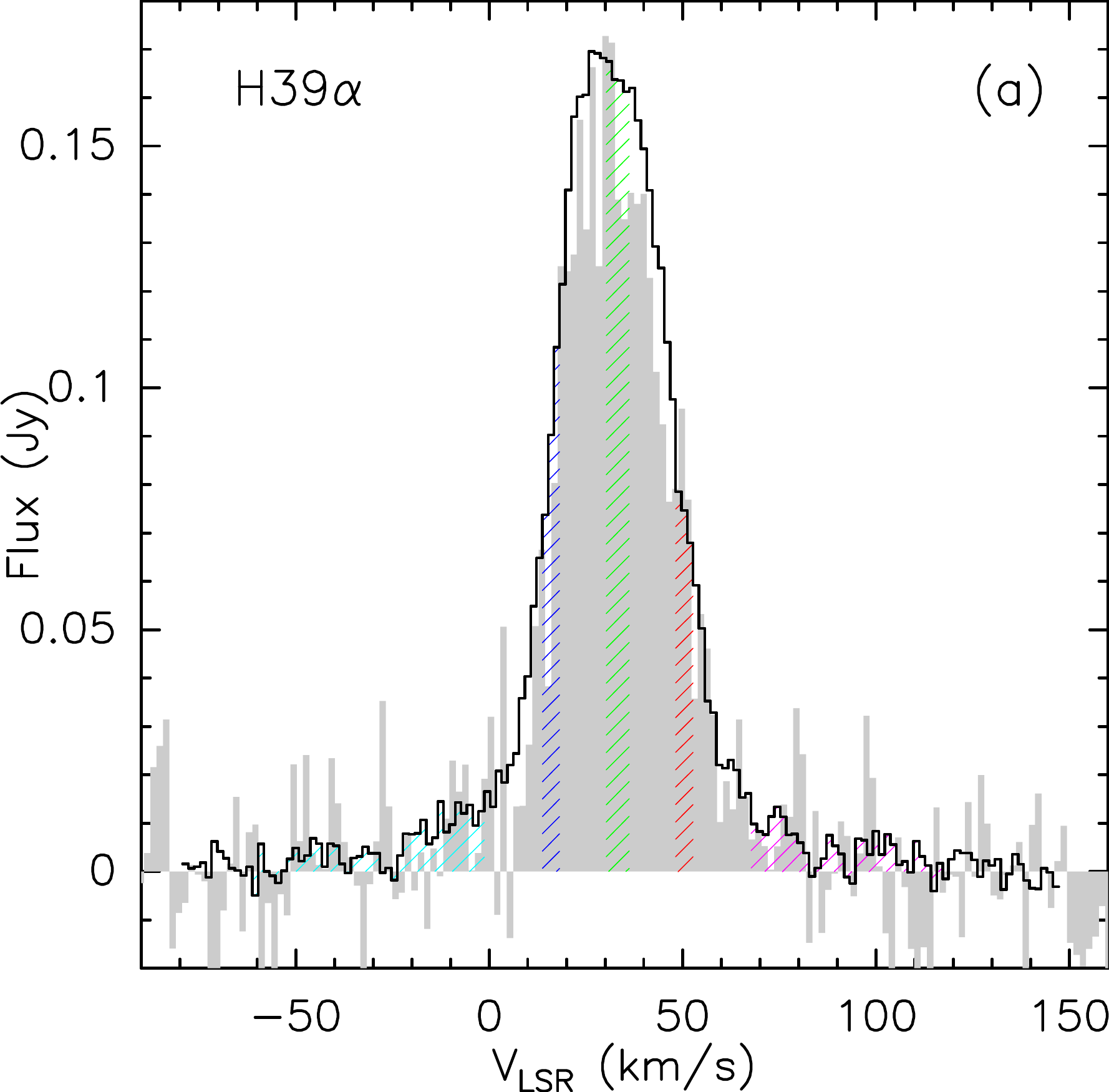}
   \includegraphics*[bb= 0 0 644 594,width=0.255\hsize]{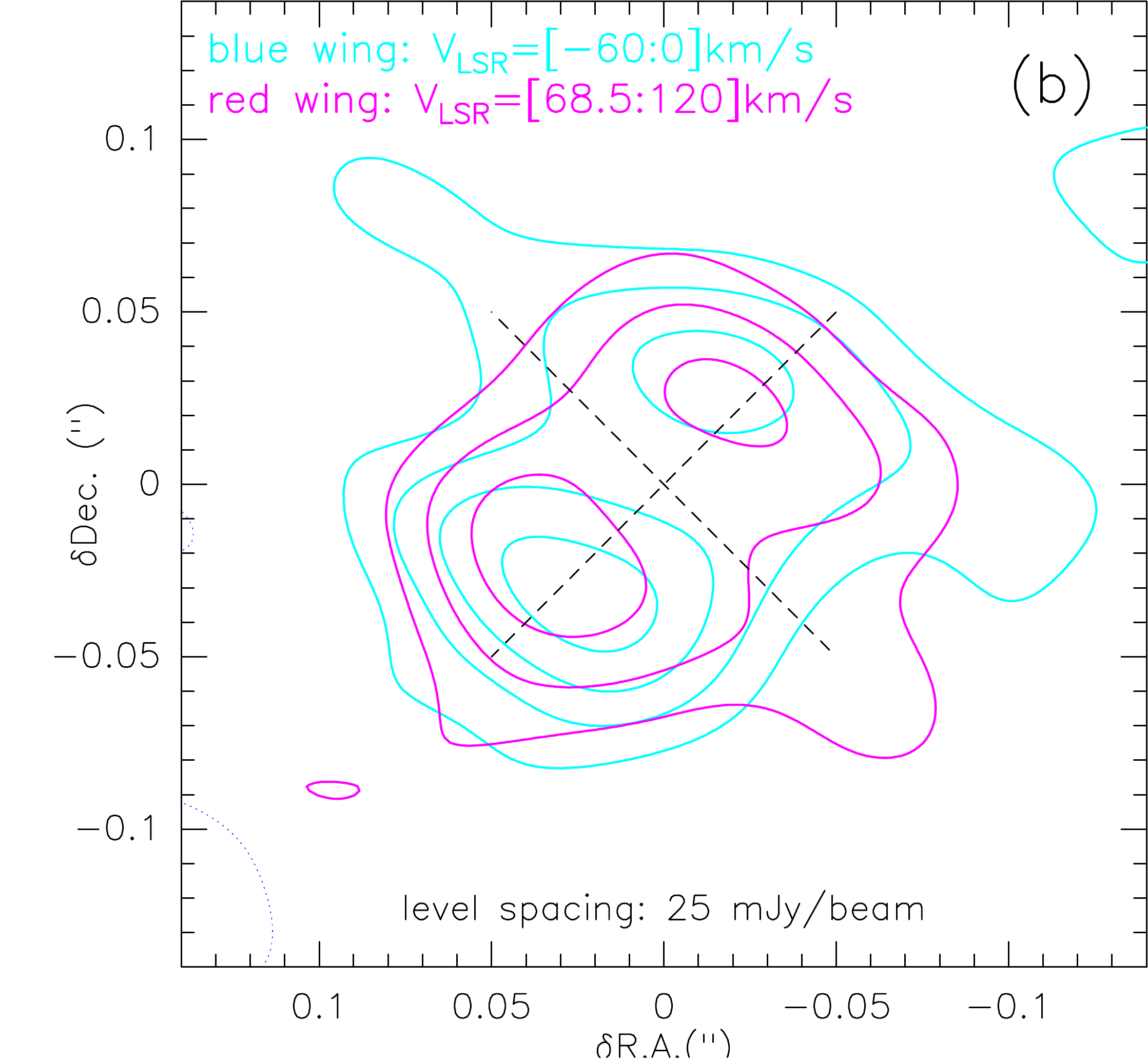}
   \includegraphics*[bb= 95 0 644 594,width=0.222\hsize]{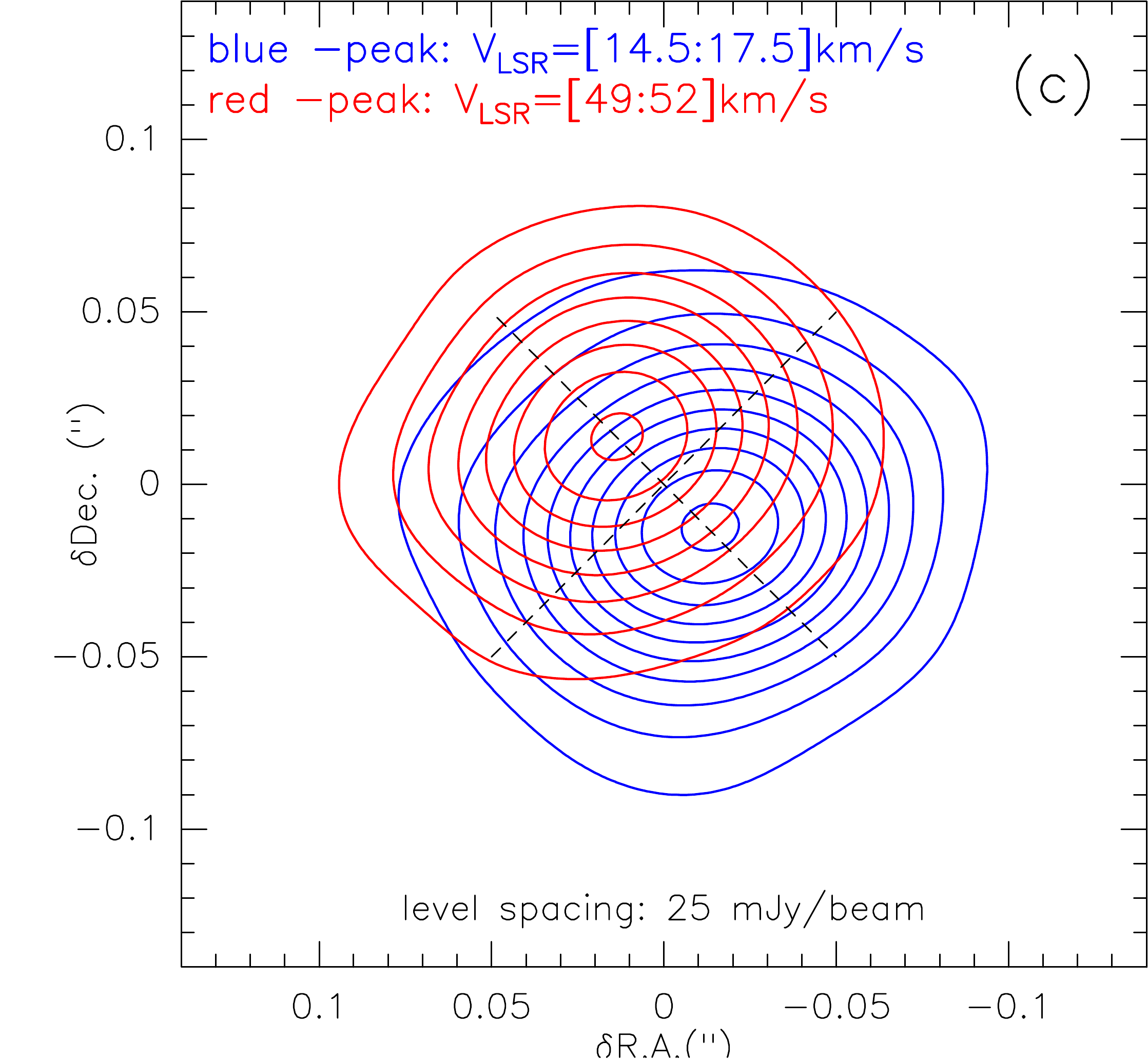}
   \includegraphics*[bb= 95 0 644 594,width=0.222\hsize]{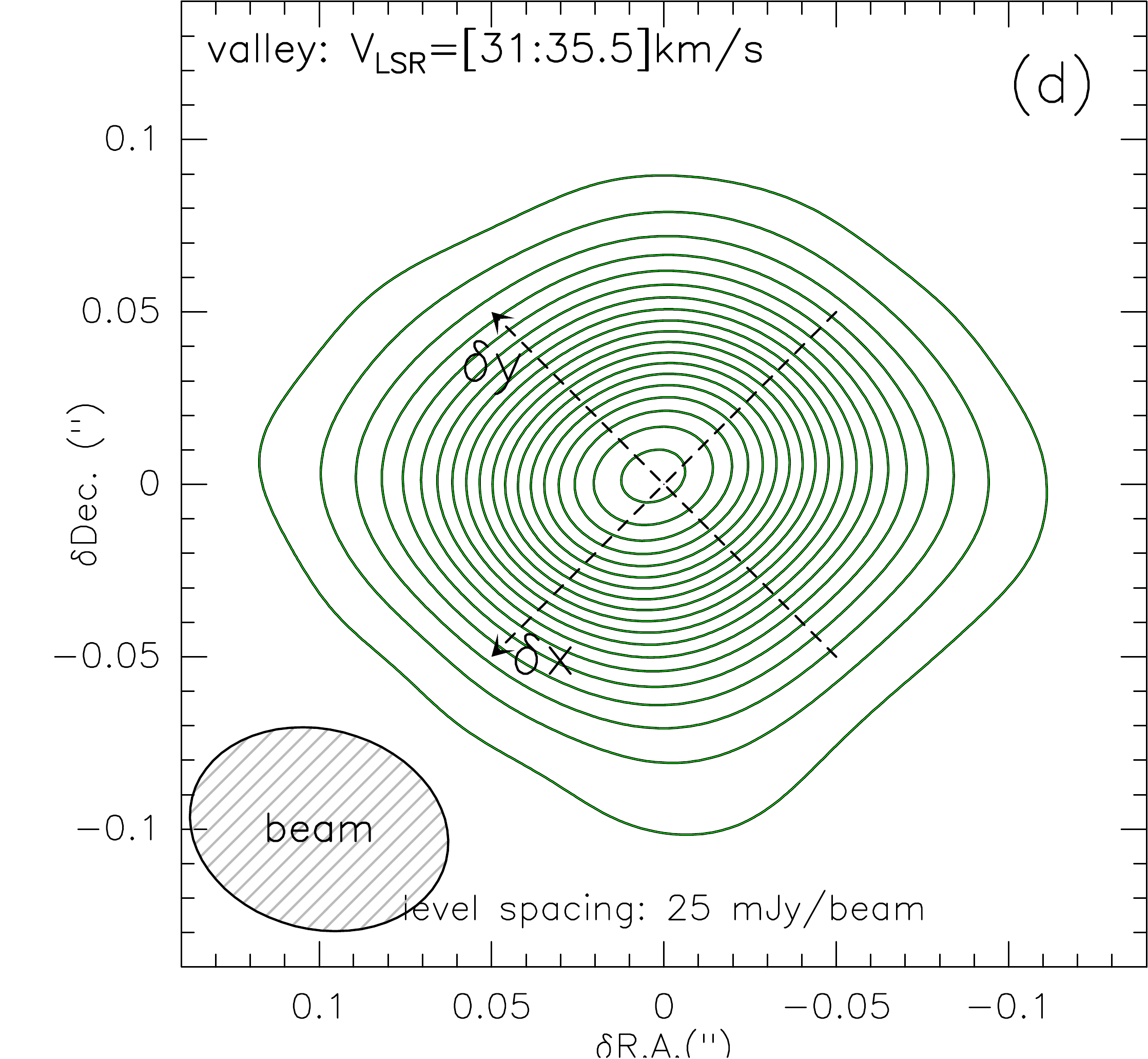} \\
   \vspace{0.5cm}
     \includegraphics[width=0.45\hsize]{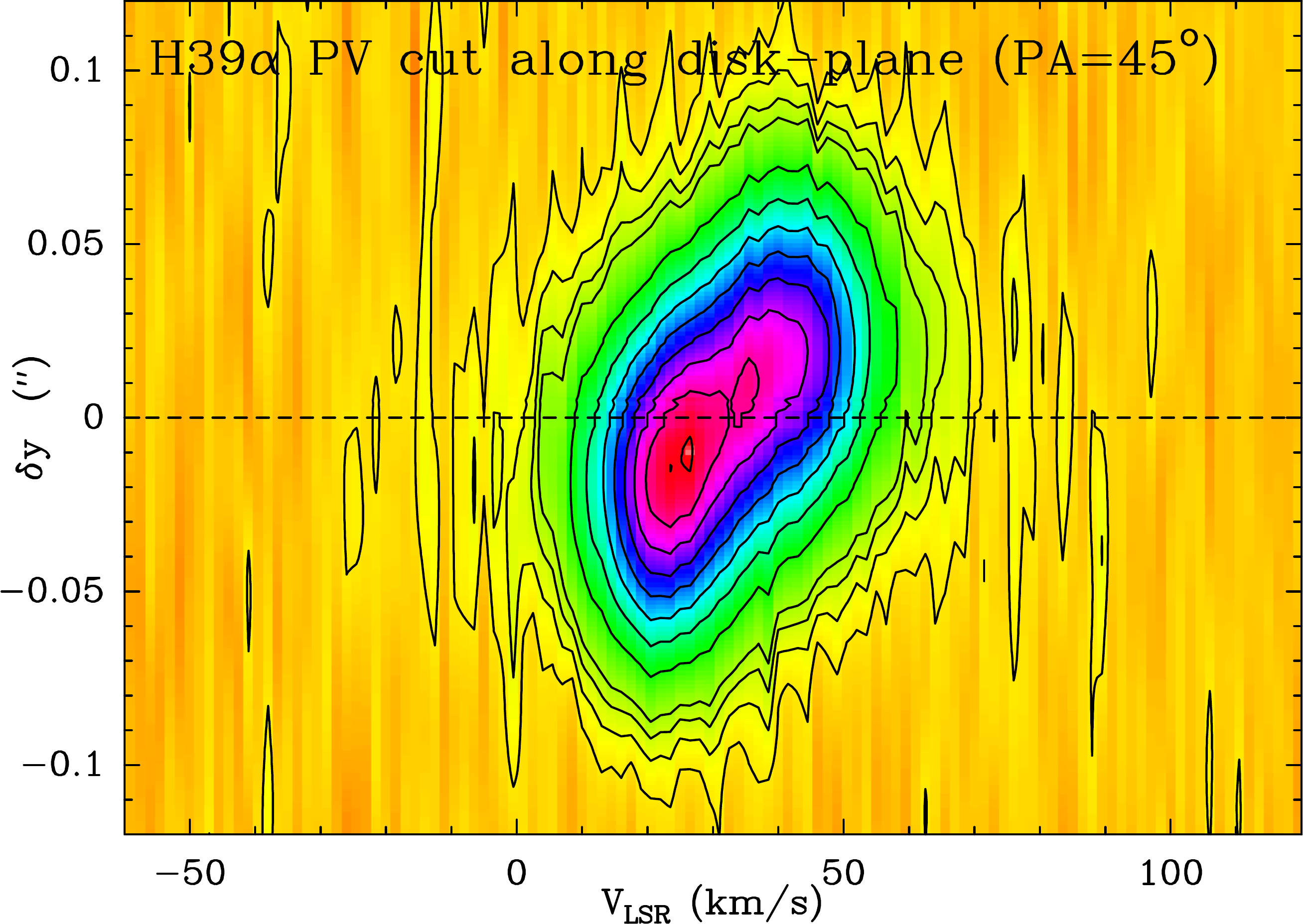}
     \includegraphics[width=0.45\hsize]{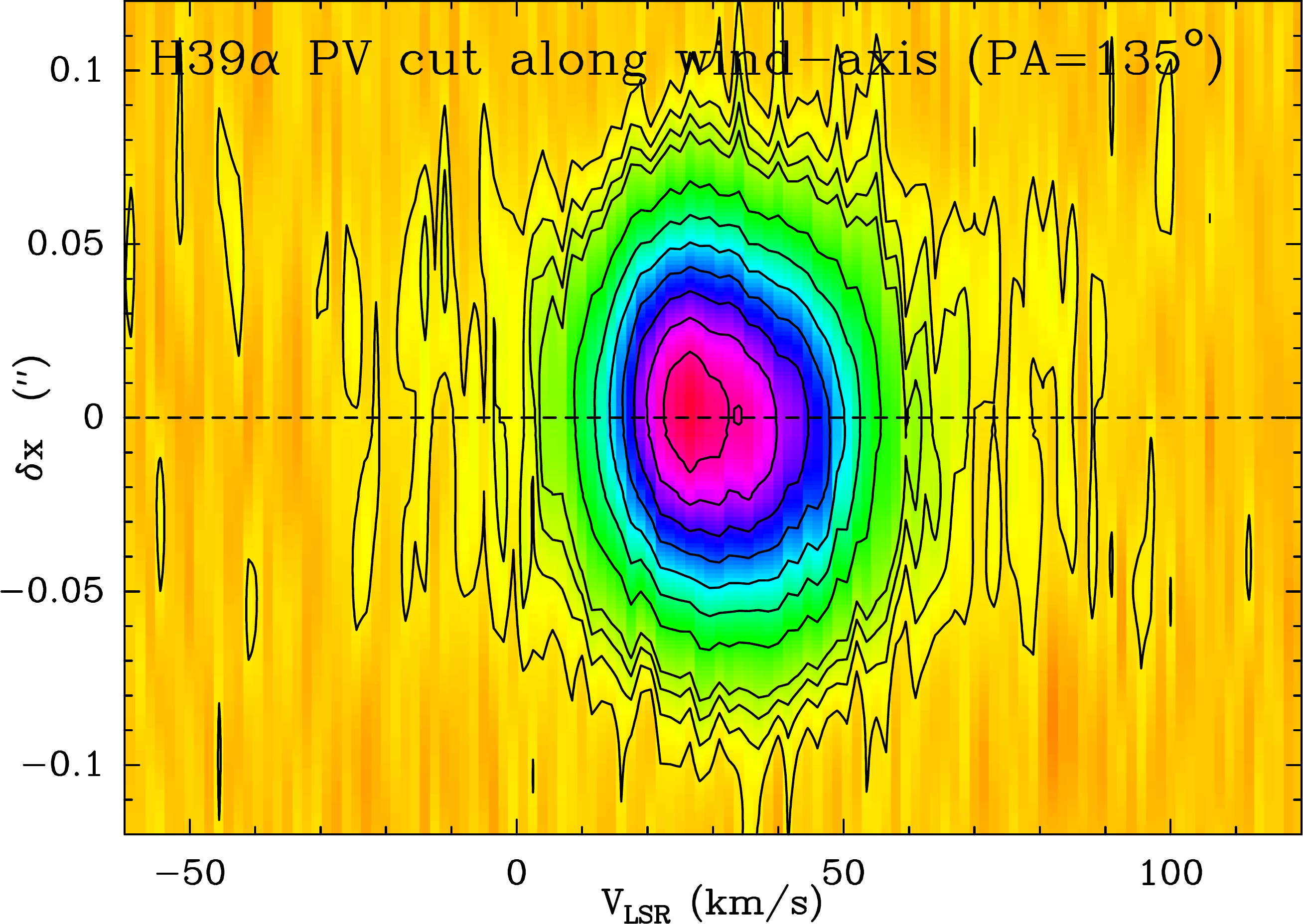} 
   
     \caption{Same as in Fig.\,\ref{f-h30a-all} but for the H39$\alpha$ line. Note the larger field of view and beam size in this case.
       Levels in the PV diagrams (bottom) are
     2.5, 5, 7.5, and from 10 to 90 by 10\,m\jb. 
          \label{f-h39a-all}}
   \end{figure*}

The ALMA observations of the H39$\alpha$ line are shown in
Figs.\,\ref{f-h39a-all} and \ref{f-cubeH39a}.  In contrast to the
H30$\alpha$ transition, the profile of the H39$\alpha$ line has a
nearly gaussian core, as already known from our previous single-dish
observations (CSC+17). The angular resolution at 3\,mm is a factor $\sim$2
poorer than at 1\,mm. In spite of this, the spatial emission
distribution of the H39$\alpha$ transition definitively corroborates
the presence of both the rotating disk and the fast bipolar outflow in
MWC\,922.

We detect with ALMA broad \htnal\ emission wings that extend over a
full velocity range of $\sim$140\,\kms.  The rotation of the fast wind
is not evident in the maps of the \htnal\ line wings
(Fig.\,\ref{f-h39a-all}b); in this case, the lower S/N ratio precludes
identifying the relative offset along the $y$ axis between the red-
and blue-wing emitting regions. The smaller width of the
\htnal\ wings compared with that of the \htal\ line is probably simply
a result of the lower S/N of the \htnal\ maps.  Alternatively or
additionally, this could be a sign of outward deceleration of the fast
wind since, due to opacity effects, the \htnal\ line is expected to
trace layers of the ionized gas that are slightly more distant from
the center than the \htal\ transition.
Higher sensitivity and angular-resolution

The comparison of the source-integrated H30$\alpha$ profile as
observed with ALMA and with the \iram\ single-dish antenna two years
earlier reveals apparent differences (Fig.\,\ref{f-h30a-all}a). In
addition to an overall brightening and slight profile broadening, we
observe a notable increase of the asymmetry between the blue and the
red peak, with the blue peak almost doubling the intensity of the red
one in the ALMA data.  These changes cannot be due to calibration errors
but rather point to variations with time of the physical properties
within the emitting volume. Given the maser nature of the
\htal\ transition, even small changes in the physical
conditions of the ionized core will be non-linearly amplified in a
significant amount.
Variability as well as strongly asymmetric profiles of maser mm-RRLs
have been reported before in other ultracompact \ion{H}{ii} regions
\citep{mar89b,thu92,gor00,jim13} 

Differences in the ALMA and \iram\ H39$\alpha$ profiles are also
perceived (Fig.\,\ref{f-h39a-all}a), e.g. the ALMA spectrum is slightly
more intense and wider than the \iram\ line. However, in this case,
these small discrepancies could be due, in principle, to pointing and
absolute flux calibration errors of the single-dish spectrum (of up to
$\sim$30\%, CSC+17).

Finally, as already mentioned, other non-$\alpha$ mm-RRLs lines have
been detected with ALMA, namely, \hcg, \hsd, and \hce. These
transitions are intrinsically weak and, therefore, the S/N of the obtained maps
is rather low, which prevents discerning structural or kinematical
details of the emitting region. Integrated intensity maps and
total emission profiles of these transitions are shown in the appendix, in 
Fig.\,\ref{f-otras}.

\section{Analysis: modelling of the emission}
\label{moreli}

We have modelled the free-free continuum and mm-RRL emission from the
ionized core around MWC\,922 using the non-LTE radiative transfer code
MORELI (MOdel for REcombination LInes), which is described in detail
in \cite{bae13}. We used MORELI before in CSC+17 to model MWC\,922 and
a small sample of pPNe candidates to constrain the structure, physical
conditions, and kinematics of the emerging \ion{H}{ii} regions in
these objects. The model presented in CSC+17 has been refined to
reproduce our ALMA maps. The reader is referred to CSC+17 for full
details on the parameters, major assumptions and general uncertainties
of our modelling approach. We offer a summary of the model basis in
the next paragraphs, which will be followed by a description of the
most important modifications made to the initial model in our current
study.

\begin{figure*}[ht!]
     \centering
     \includegraphics*[bb=112 70 755 575,width=0.5\hsize]{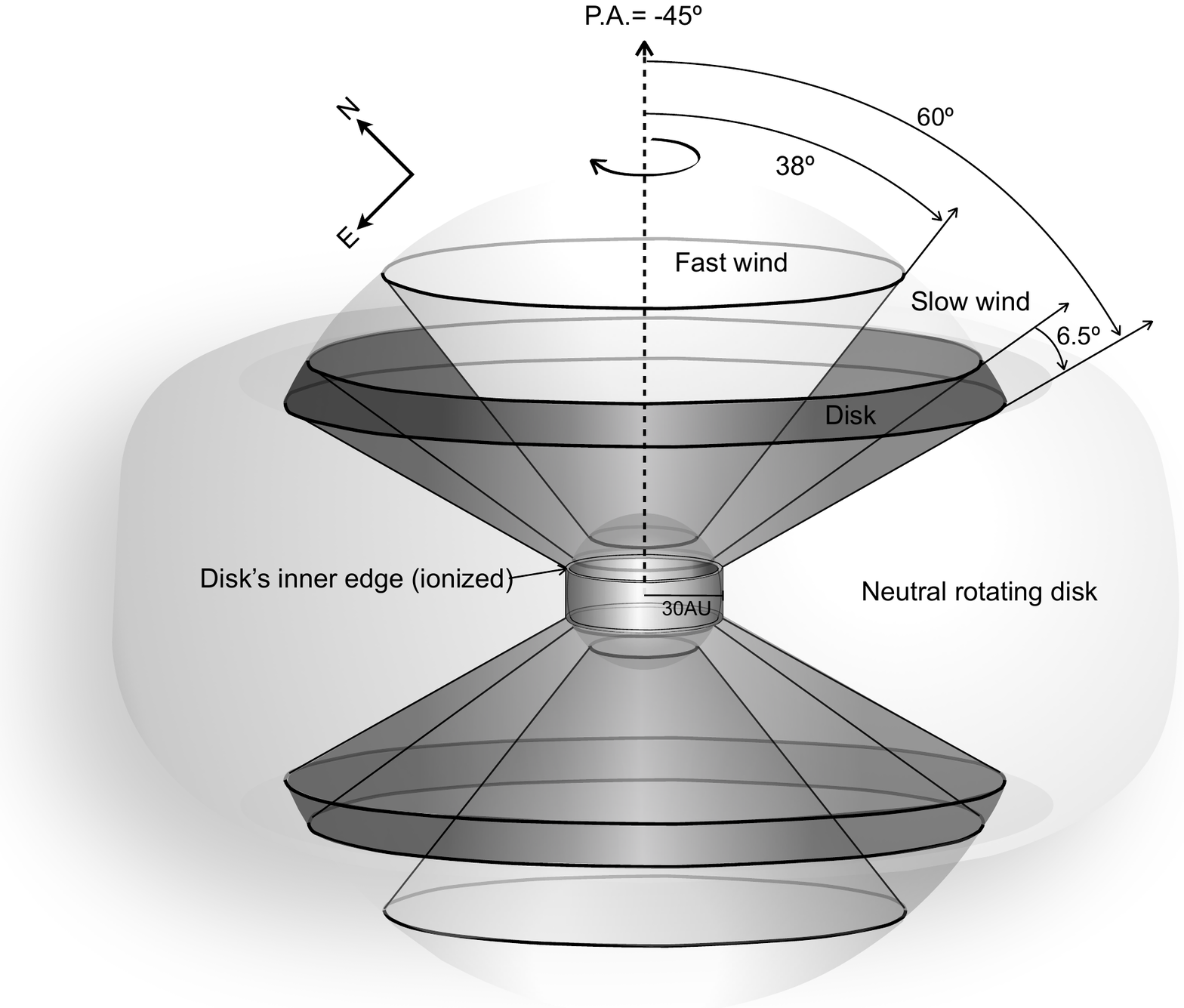}
     \includegraphics[width=0.45\hsize]{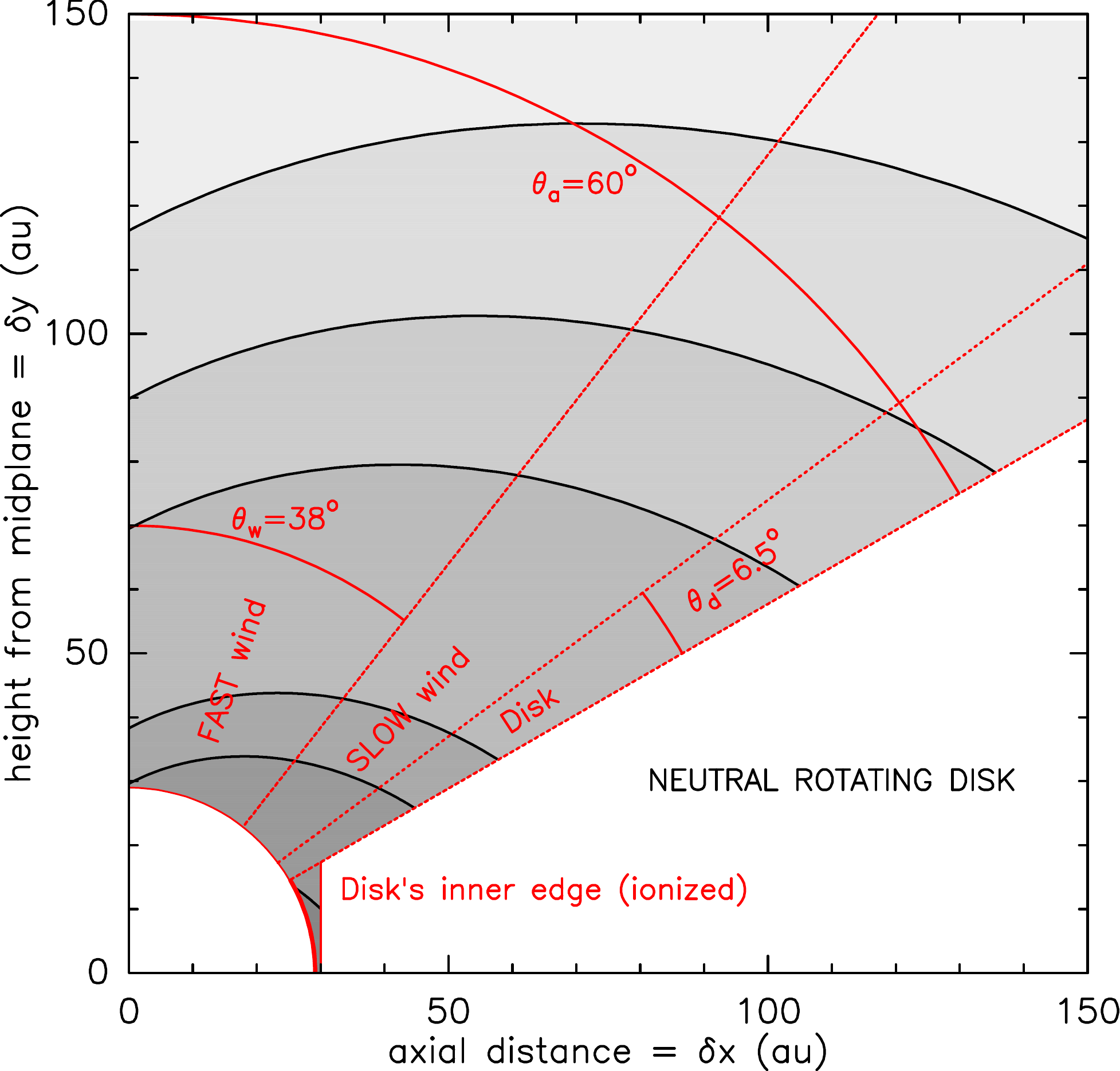}
   \caption{{\bf Left)} Sketch of the biconical geometry adopted in
     our model (\S\,\ref{moreli}). The equatorial plane is oriented in
     the sky along PA=45\degr\ and is seen edge-on. {\bf Right)}
     Two-dimensional representation of the density law adopted
     (Table\,\ref{t-moreli}). Countour levels are 2.5\ex{5}, 5\ex{5},
     10$^6$, 5\ex{6}, 10$^7$, and 5\ex{7}\,\cm3. Dotted lines delimit
     different model components. The neutral disk is not probed by our
     ALMA observations.
     \label{f-dens2D}}      
   \end{figure*}

The \ion{H}{ii} region around MWC\,922 has been represented by a conical
geometry, i.e., the ionized emission is assumed to arise in two
opposing conical structures that are coaxial with, and inscribed in,
an extended neutral rotating disk (Fig.\,\ref{f-dens2D}). The
outermost biconical layer, in contact with the neutral part of the disk,
simply represents the thin conical surface of the rotating disk that is
illuminated and photoionized by the central source. We refer to this
component as the ``ionized disk skin'' or simply the ``disk''
and its outermost boundary is characterized by a semi-opening angle \tha\, measured from
its revolution axis to the midplane. The disk has an angular width
\thd, that is, it extends from \tha\ to \tha$-$\thd.  The regions that
are inside of (surrounded by) the innermost boundary of the disk, at
colatitude $\theta$$<$\tha$-$\thd, form the ``outflow/wind''  -- see  Fig.\,\ref{f-dens2D}. 

The kinematics of the disk and the outflow are also as described in
CSC+17, i.e., the disk is assumed to rotate in a Keplerian
(\vrot$\propto$\,1/$\sqrt{r}$) fashion, and the outflow is modelled as a
rotating and radially expanding wind. As we will show below, in the new model the
outflow has been subdivided in a {\sl slow} and a {\sl fast} wind
component. The rotation in both the slow and fast wind components is
taken to be the same as in the disk.

The electron density (\dense) is assumed to vary throughout the disk
and outflow as a function of the radial distance, following a
power-law distribution ($\propto$$r^{-\beta}$), and also as a function
of the latitude, decreasing exponentially from the outer boundary of the disk toward the
poles ($\propto$\,$\exp{-(\frac{\theta_a-\theta}{\theta_0})}$) -- see
Fig.\,\ref{f-dens2D} and Table\,\ref{t-moreli}. For simplicity and to
restrict the number of model input parameters, we adopt a constant
value of the electron temperature (\te) within the disk and within the
outflow.

The major modifications made on our base model
for MWC\,922 in this work are:

   \begin{enumerate}
   \item We introduce a new component, the fast wind, to explain the broad emission
     wings of the \htal\ and \htnal\ lines. The spatial distribution
     of the line wing emission denotes the presence of a fast 
     bipolar wind orthogonal to the disk
     (Fig.\,\ref{f-h30a-all}b and \ref{f-h39a-all}b).  This
     fast outflow is represented in our model as a bicone with a semi-opening angle
     \thw\ measured from the revolution axis (Fig.\,\ref{f-dens2D}).
   \item In order to recreate the X-shape morphology of the ionized core around
     MWC\,922 implied by the ALMA data, a larger value of the disk
     semi-opening angle (\tha$\sim$60\degr) is now adopted. 
   \item We increase the size of the central cavity
     (\rin$\sim$29\,au). The reason is twofold. First, a \rin$\sim$29\,au
     cavity best reproduces the separation between the two relative
     maxima of the red- and blue-wing emitting regions of the
     \htal\ line (Fig.\,\ref{f-h30a-all}b). Second, a larger cavity is
     needed to explain the flattening of the free-free continuum
     emission observed towards 1\,mm (Fig.\,\ref{f-sed}). (Note that
     in CSC+17, the slope of the continuum could not be measured
     separately at 1\,mm and 3\,mm, which led us to adopt a unique
     common value of the spectral slope consistent with optically
     thick free-free emission in the whole 1-to-3\,mm range.)
   \item We now consider the contribution to the total emission by a
     thin layer of ionized gas located immediately beyond the cavity,
     at the inner edge of the rotating neutral disk (referred to as
     the ``Disk's inner edge'' in Fig.\,\ref{f-dens2D}). This
     component is needed to enhance the \htal\ emission from low-latitude
     regions and, thus, to bring the emission from the red- and
     blue-peaks closer to the disk midplane as observed (see, e.g.,
     Fig.\,\ref{f-h30a-all}c).
   \item We consider a non-zero, yet small, value of the turbulence
     velocity dispersion ($\sigma_{\rm turb}$$\sim$2-3\,\kms) as an
     additional component of line broadening\footnote{Turbulence was
       considered to be negligible in our pilot study, CSC+17.}  (coupled with
     thermal, pressure, and macroscopic motion
     broadening).
    \end{enumerate}

   The code MORELI produces synthetic images of the free-free
   continuum and mm-RRLs with a spatial resolution that is equal to
   the size of the grid cells used to perform the radiative transfer
   computations (typically of $\sim$1-2\,au). To simulate the
   observations with ALMA, we used our input model and the same UV
   sampling and weighting of our ALMA observations to generate the
   synthetic or model visibilities. This has been done using the task
   {\tt uv\_model} of MAPPING (GILDAS). After a model UV table has
   been created, image restoration and cleaning of the latter has been
   performed using exactly the same tasks and parameters as for the
   ALMA data (\S\,\ref{sec-obs}). This way an accurate model-observations
   comparison is guaranteed. [For example, although it is not the case
     in MWC\,922, there would be no problem with large scale
     structures filtered out in the data as they will be equally
     filtered by the UV coverage in the restoration of the model image.]
   
   \subsection{Model results}

%
\begin{table*}[ht!]
\small
\caption{Model parameters used to reproduce the observations of MWC\,922 (\S\,\ref{moreli}).}
\label{t-moreli}      
\centering                          
\begin{tabular}{l c c}        %
  \hline\hline                      
Parameter & value & value \\ 
\hline
Distance ($d$) & 1700 pc & 3000\,pc \\ 
LSR Systemic velocity (\vsys) & $+$33\,\kms  & -- \\ 
Inclination ($i$) & 0\degr & -- \\ 
Ionized disk's semi-opening angle (\tha) & 60\degr  & -- \\
Angular width of the ionized disk (\thd) & 6.5\degr  & -- \\
Fast outflow's semi-opening angle (\thw) & 38\degr  &--  \\
Inner radius (\rin)  & 29\,au  & 51.5\,au \\
Radius of the disk's inner edge (ionized) & 30\,au & 53\,au \\ 
Electron density (\dense) & 4.3\ex{7}($\frac{r}{30{\rm au}}$)$^{-2.7}$exp($-(\theta_a-\theta)/40\degr$)\,\cm3\  &
1.5\ex{8}($\frac{r}{30{\rm au}}$)$^{-2.7}$exp($-(\theta_a-\theta)/40\degr$)\,\cm3\  \\
Electron temperature (\te) & 8750\,K  & 9000-11000 \\
Slow outflow's velocity (\vexslow)  & 9 \,\kms\ ({\sl radial expansion})  & -- \\
Fast outflow's velocity (\vexfast)  & 95\,\kms\ ({\sl radial expansion})  & -- \\ 
Keplerian rotation velocity (\vrot) & 36.4/$\sqrt{r/6.7 {\rm au}}$\,\kms\  &  48.5/$\sqrt{r/6.7 {\rm au}}$\,\kms\\  
Central mass ($M_{\rm c}$) & 10\,\msun\ & 18\,\msun\ \\ 
Mass-loss rate of the slow wind (\mloss$_{\rm  slow}$) &  3\ex{-7}\,\my\   &  6\ex{-7}\,\my\\   
Mass-loss rate of the fast wind (\mloss$_{\rm  fast}$) &  2\ex{-6}\,\my\   &  4\ex{-6}\,\my\\   
Ionized mass ($M_{\rm ion}$) & 2\ex{-5}\msun\  &  9\ex{-5}\msun\ \\   
\hline 
\end{tabular} \\
\tablefoot{The geometry adopted is depicted in
  Fig.\,\ref{f-dens2D}. The mass-loss rates given in the table
  represent the values at the inner radius of the disk (\rin). The
  value of the ionized mass is that in the modelled region, within
  $\sim$150\,au, adopting $d$=1.7\,kpc ($\sim$300\,au, adopting $d$=3\,kpc).  The
  departure coefficients $b_{\rm n}$ used are from \cite{sto95}. \\
}
\end{table*}
%


   
   \begin{figure*}[ht]
     \centering
     \includegraphics[width=0.34\hsize]{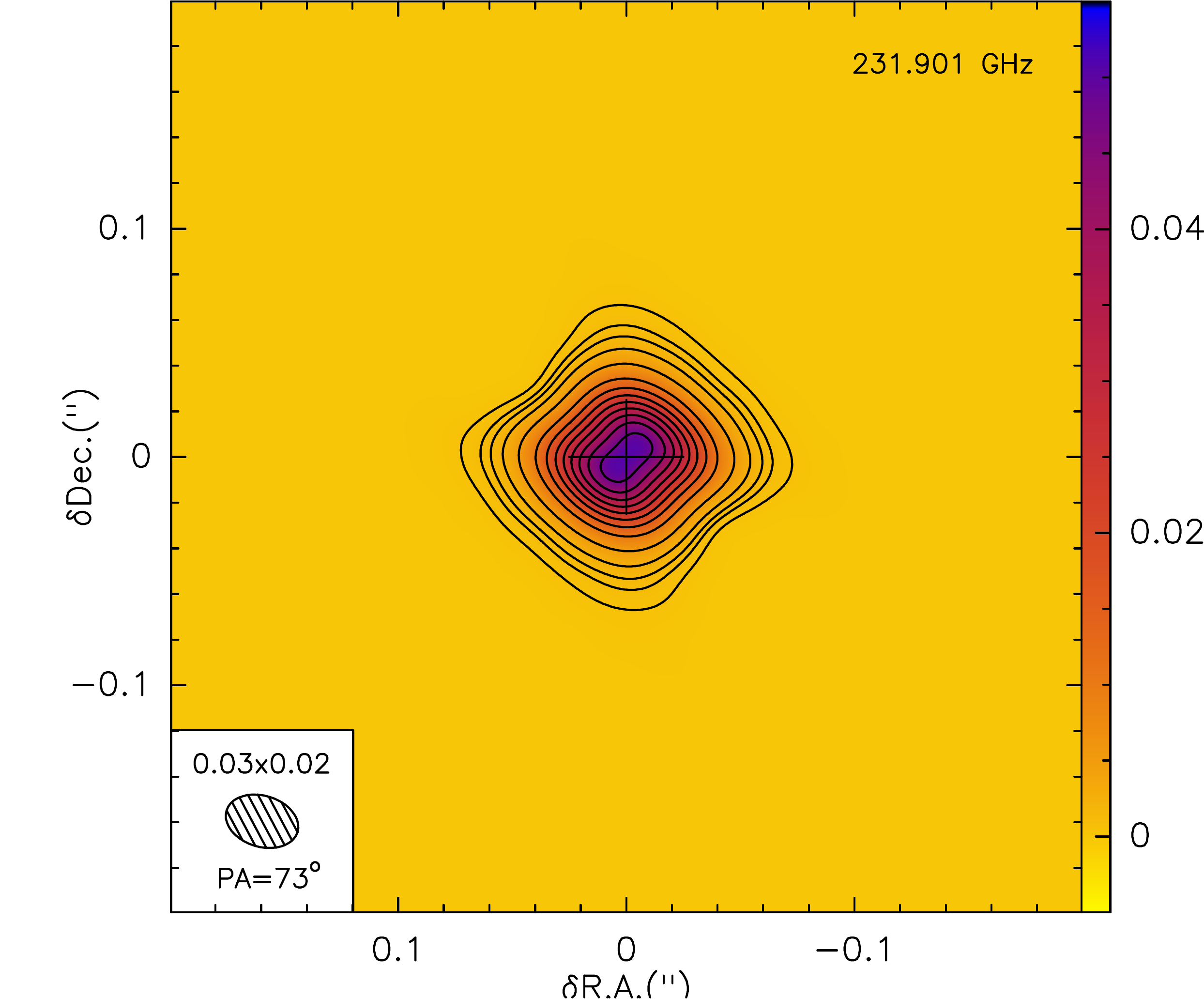}
     \includegraphics[width=0.34\hsize]{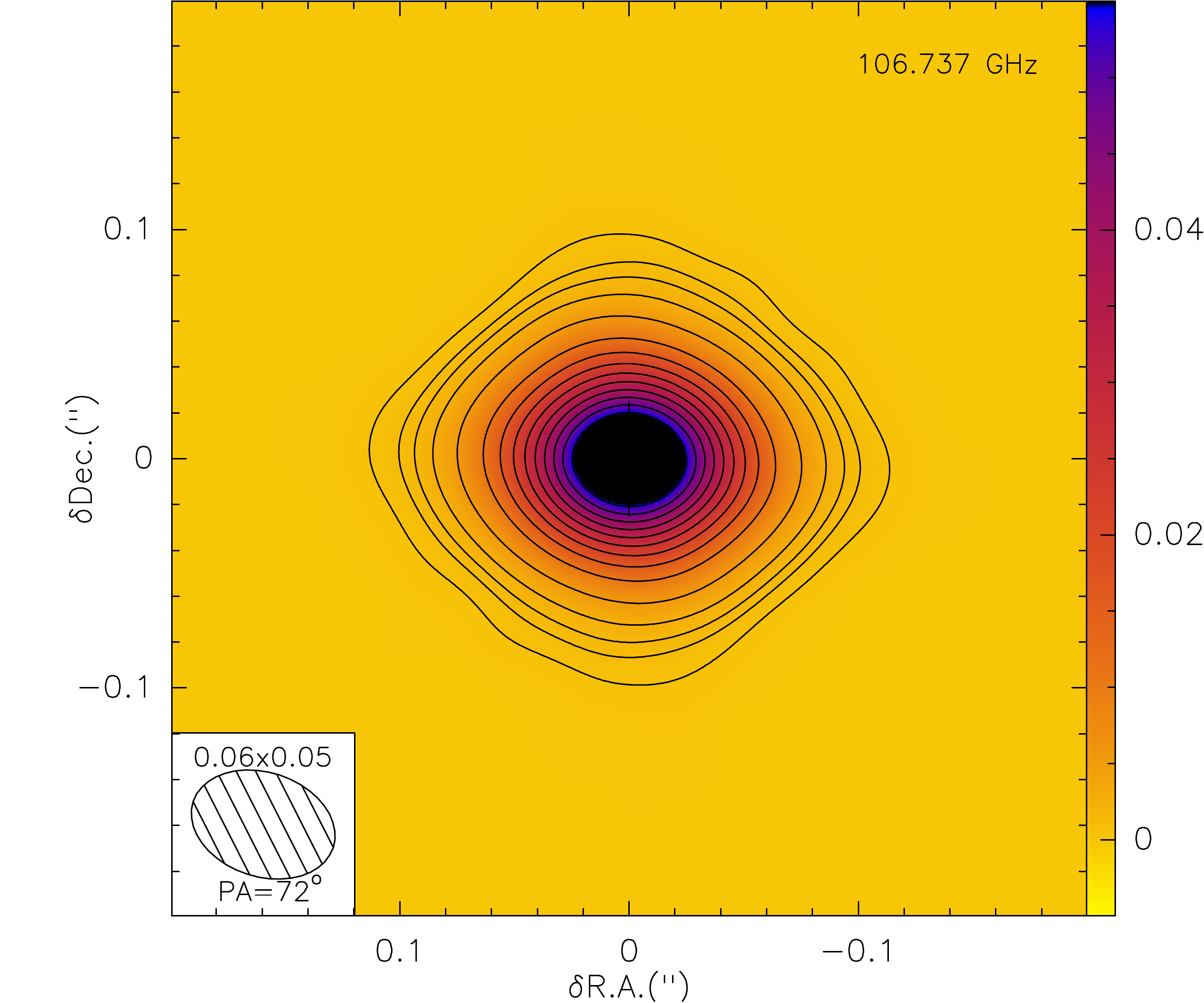}  
     \includegraphics[width=0.29\hsize]{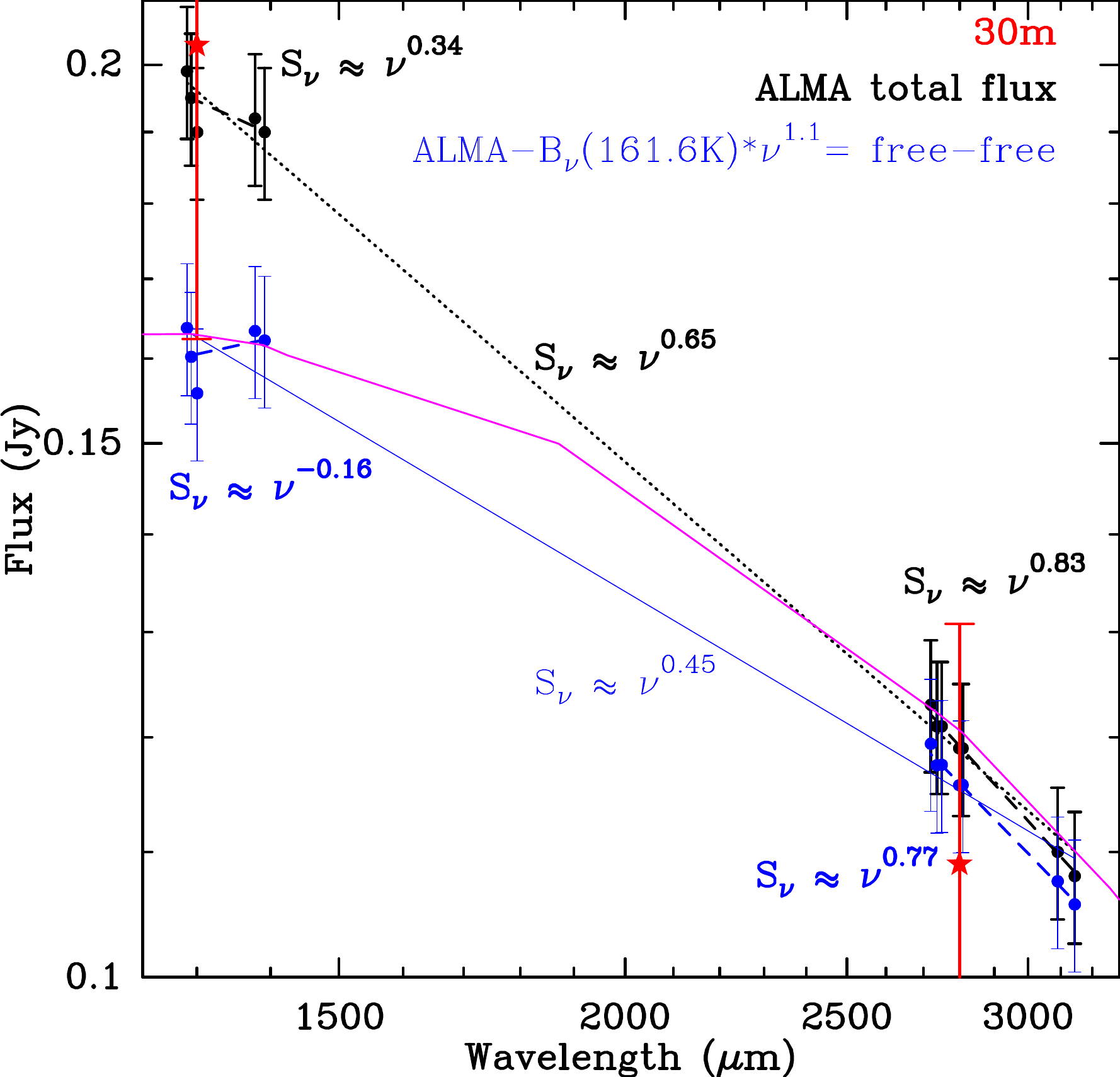} \\ 
     \caption{Model predictions of the free-free continuum of MWC\,922
       (\S\,\ref{moreli}, Table\,\ref{t-moreli}) . {\bf Left \&
         Middle)} 1 and 3\,mm continuum maps as in, and to be compared with, 
       Fig.\,\ref{f-cont-acc}.  {\bf Right)} SED as in, and to be compared with, 
       Fig.\,\ref{f-sed}b. Model free-free continuum integrated fluxes
       are indicated by the pink line.  
   \label{f-cont-moreli}}
   \end{figure*}
 
In Table\,\ref{t-moreli} we give the input parameters of a model that
successfully reproduces our ALMA images and line profiles of the
mm-RRLs and the free-free continuum emission in MWC\,922.

As we can see in Fig\,\ref{f-cont-moreli}, this model predicts the
free-free continuum fluxes observed at 1 and 3\,mm and the flattening
of the emission at intermediate wavelengths
(Fig.\,\ref{f-cont-acc}). The morphology of the predicted and observed
continuum emission maps is also very similar, even with
super-resolution (Figs.\,\ref{f-cont-HiRes} and
\ref{f-cont-HiRes-model}). In particular, our model adequately
emulates the X-shaped brightness distribution denoted by the 3\,mm
continuum maps. 
At 1\,mm, data-model discrepancies are expected since the observations
also include the dust emission contribution, which is not modelled and
could represent a $\sim$15-20\%\ of the total flux observed (and not
resolved out). We believe that the faint, extended emission along the
disk equator observed at 1\,mm, but not reproduced by our model, is
probably due to dust settled in the disk's midplane.

\begin{figure*}[htbp!]
     \centering
     \includegraphics[width=0.28\hsize]{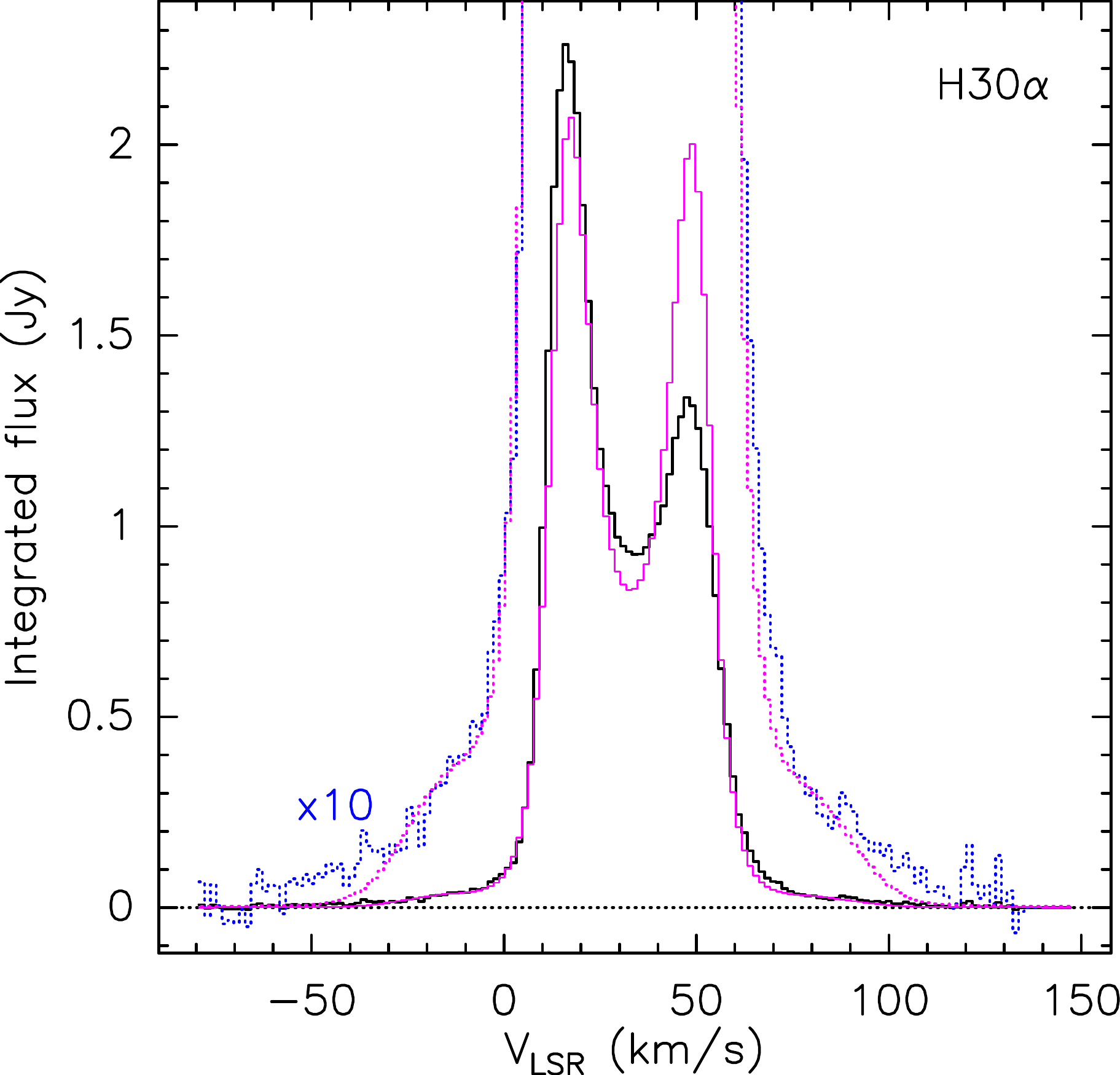}
     \includegraphics*[bb= 0 0 644 594,width=0.255\hsize]{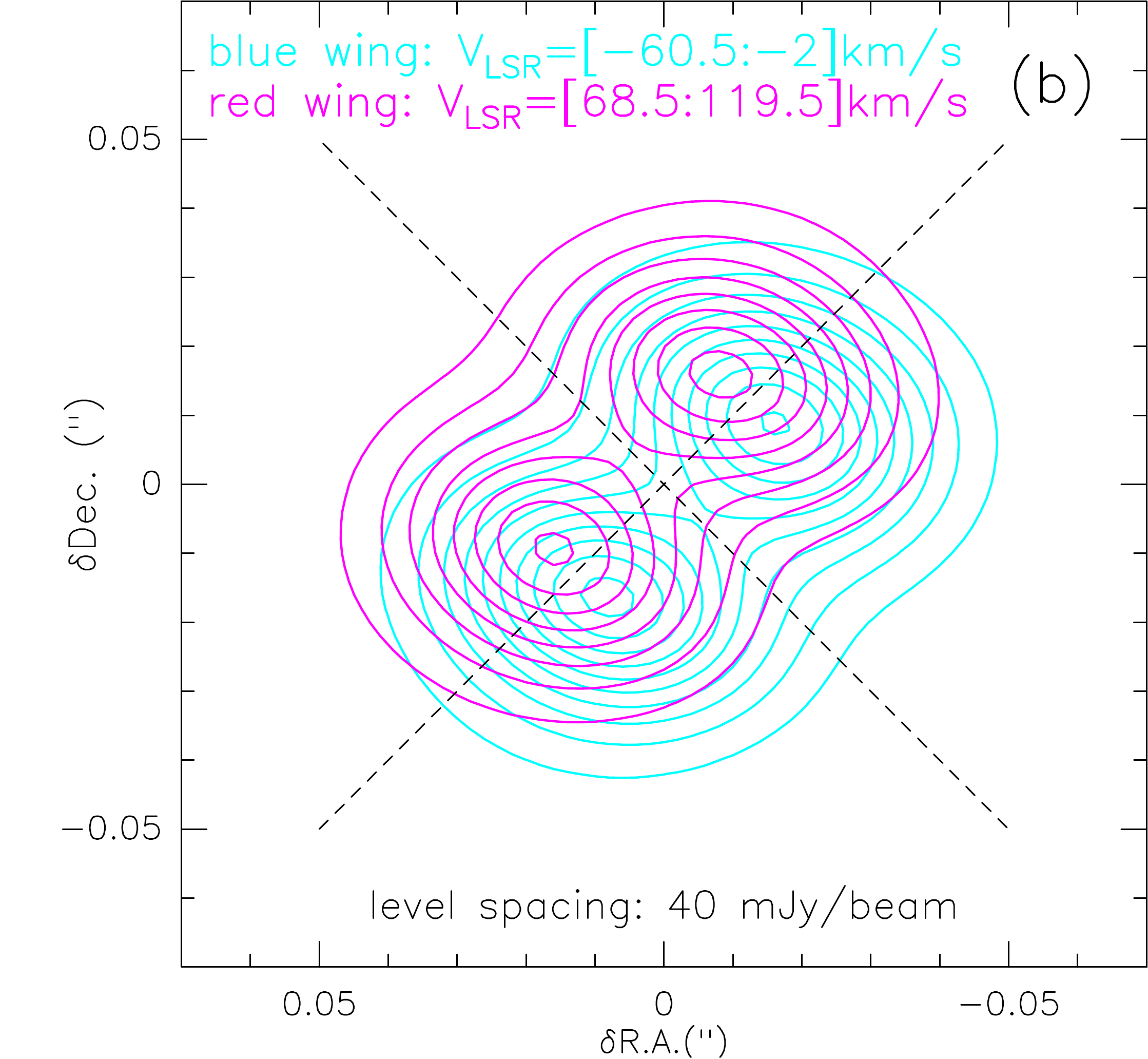}
     \includegraphics*[bb= 95 0 644 594,width=0.222\hsize]{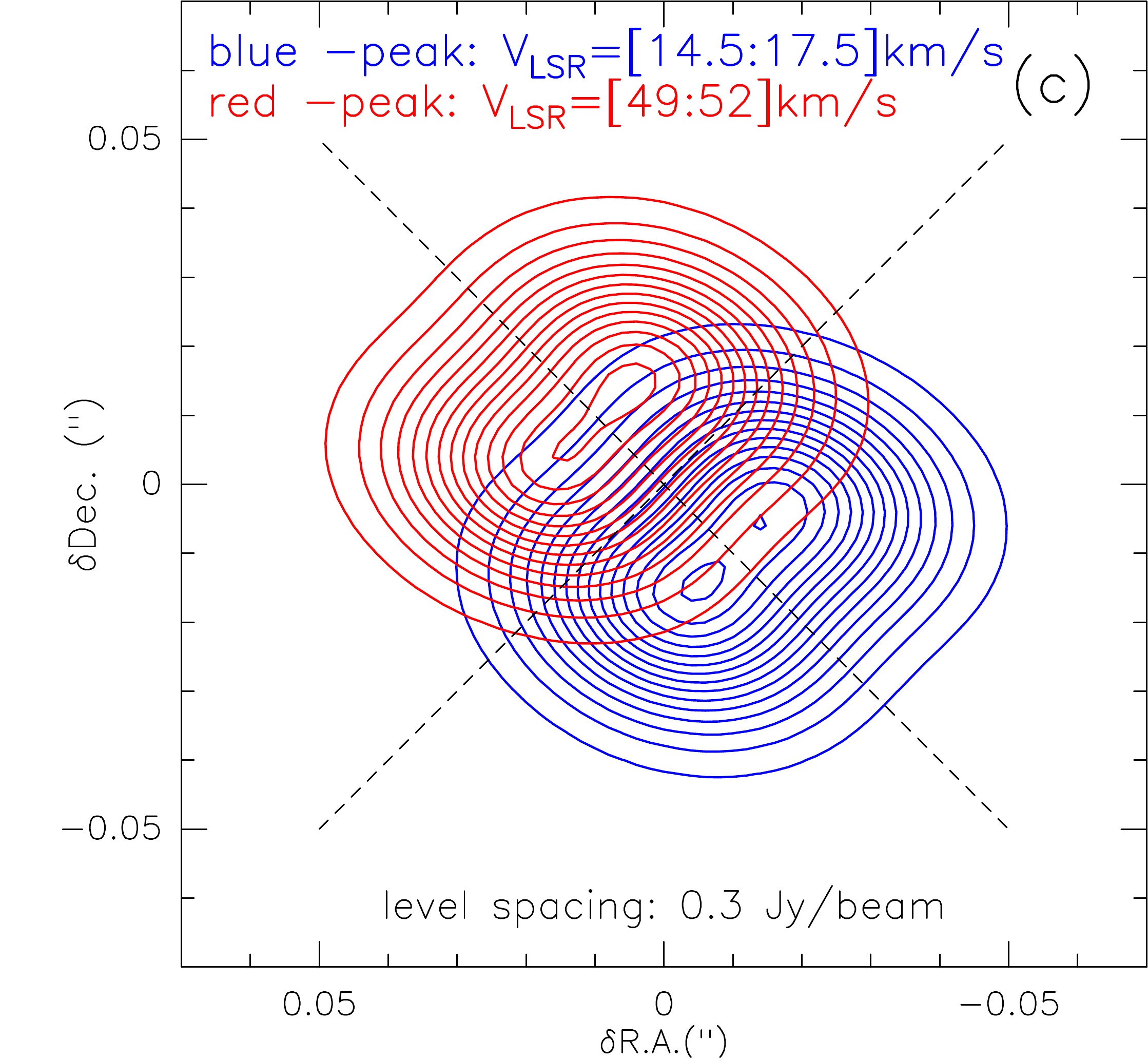}
     \includegraphics*[bb= 95 0 644 594,width=0.222\hsize]{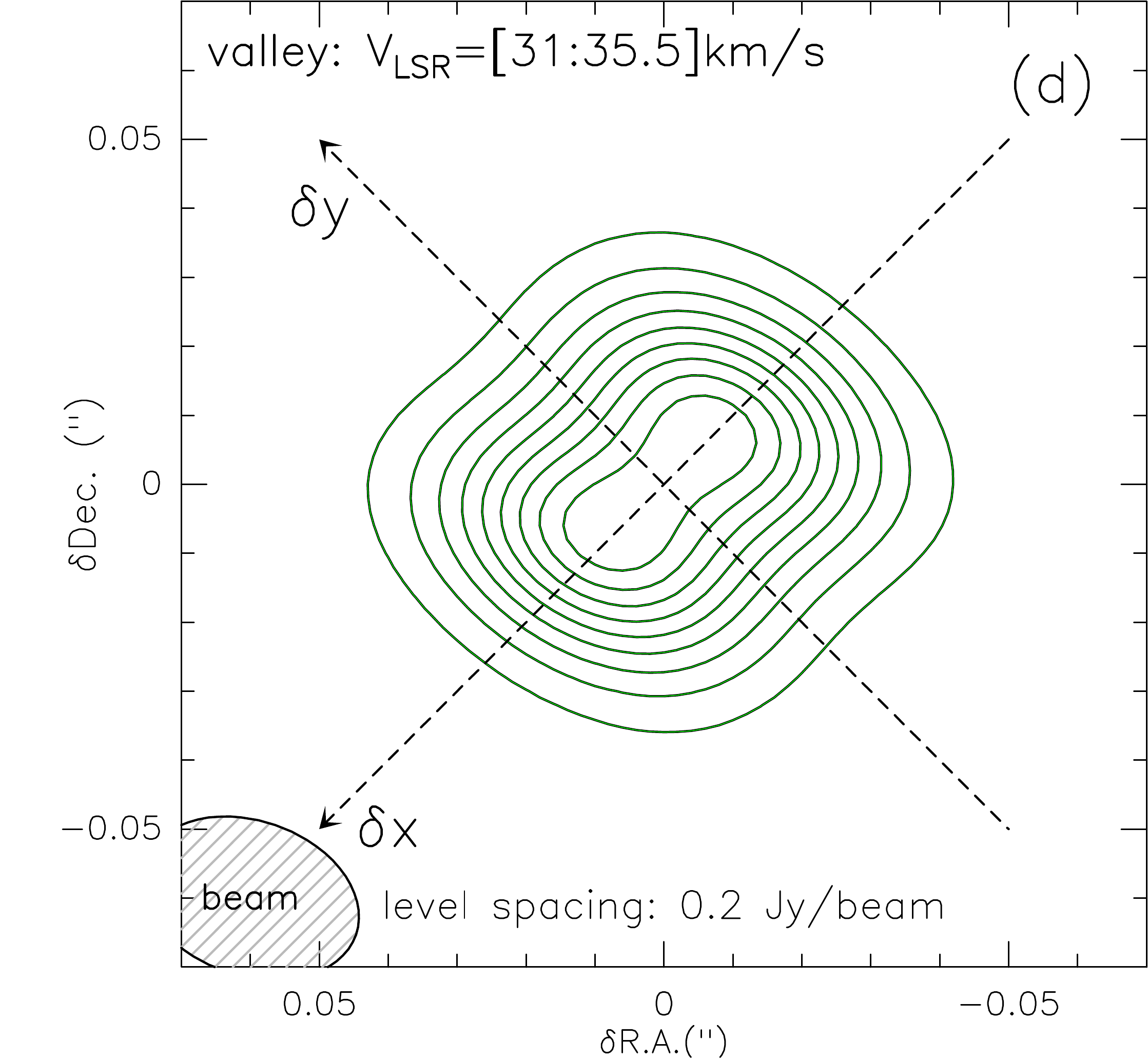}
     \\
 \vspace{0.5cm}
     \includegraphics[width=0.45\hsize]{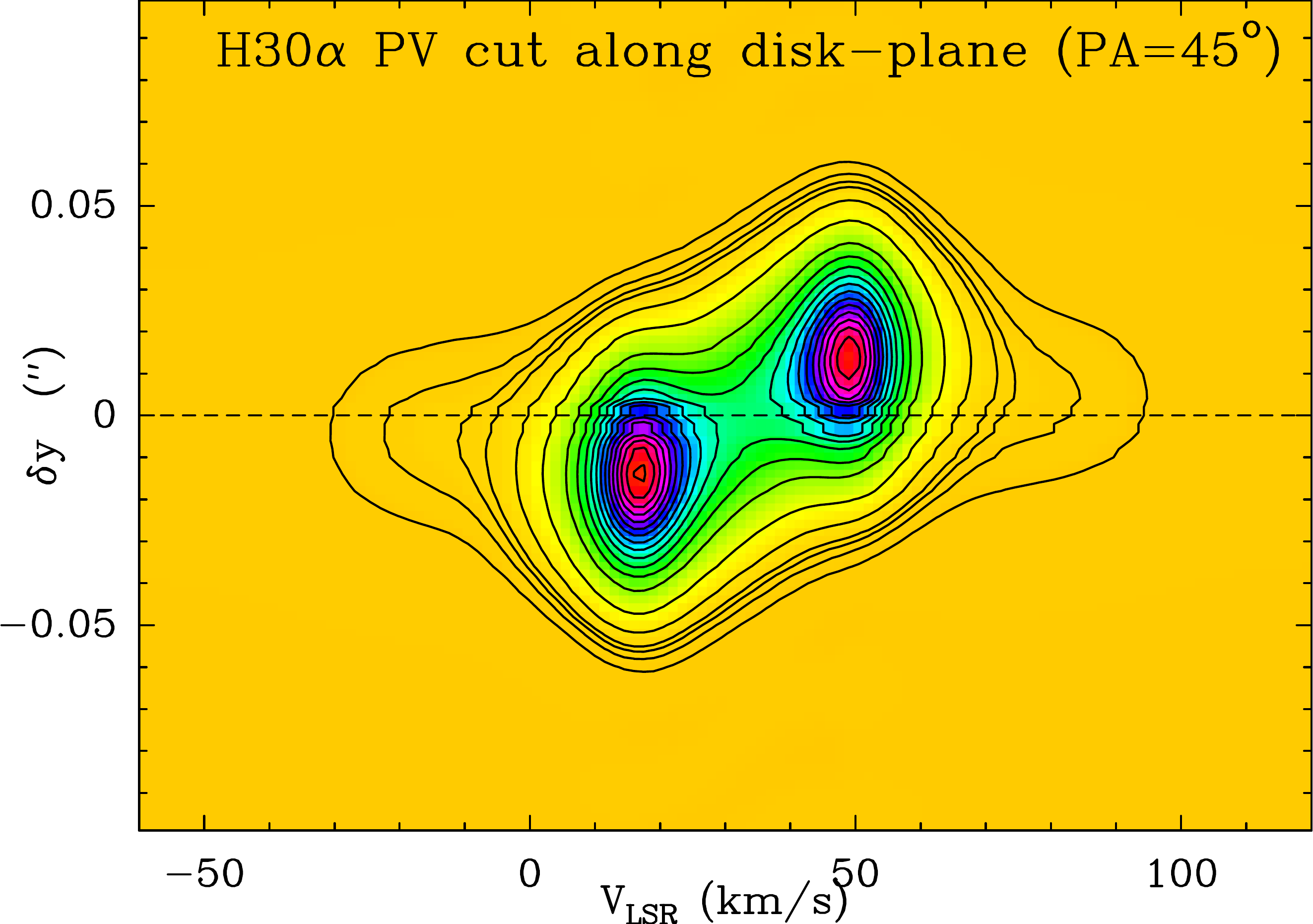}
     \includegraphics[width=0.45\hsize]{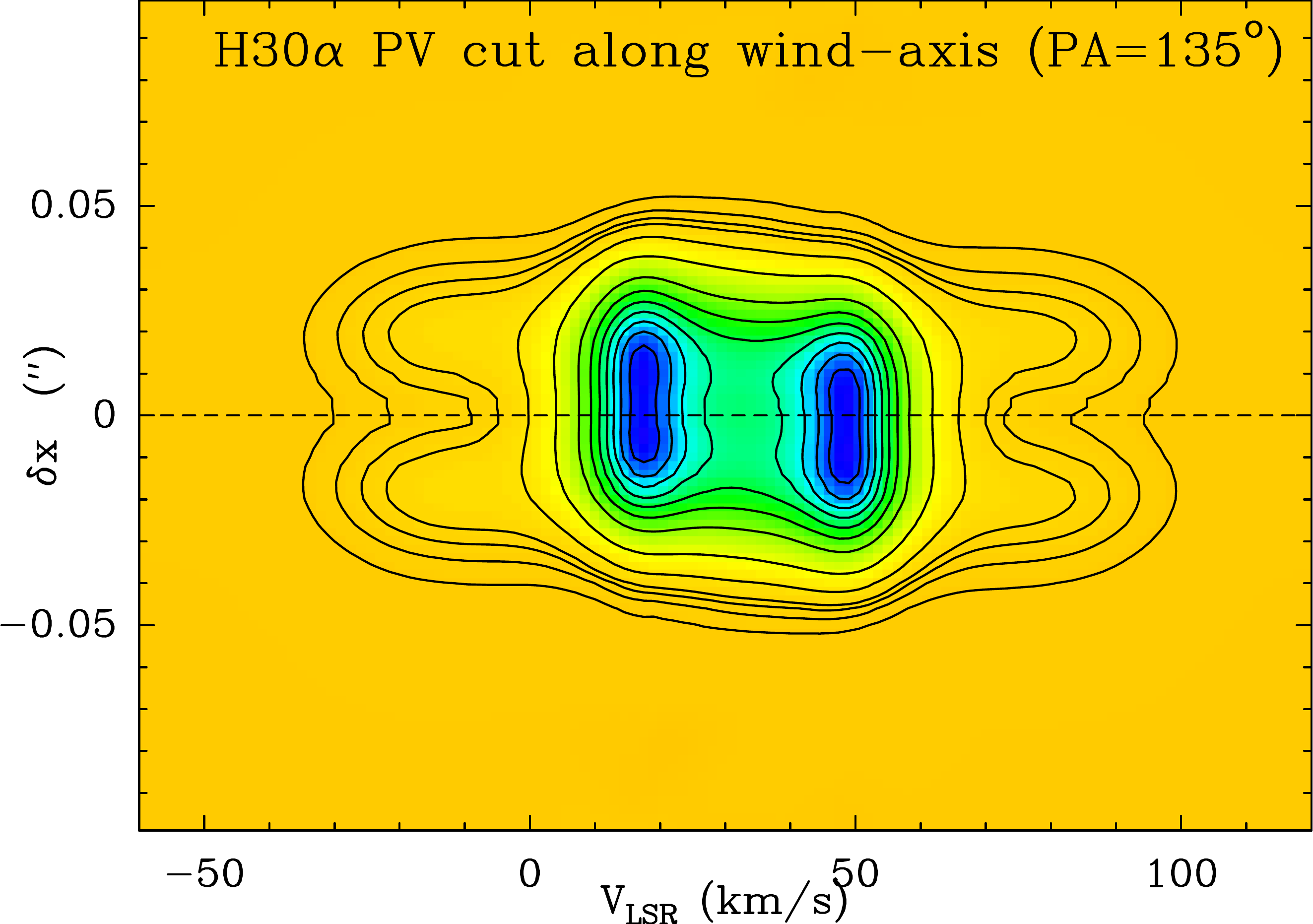}
     \caption{Same as in Fig.\,\ref{f-h30a-all} but showing the predictions of our model (\S\,\ref{moreli}, Table\,\ref{t-moreli}). 
        \label{f-h30a-all-moreli}}
   \end{figure*}

\begin{figure*}[htpb!]
     \centering
\includegraphics[width=0.28\hsize]{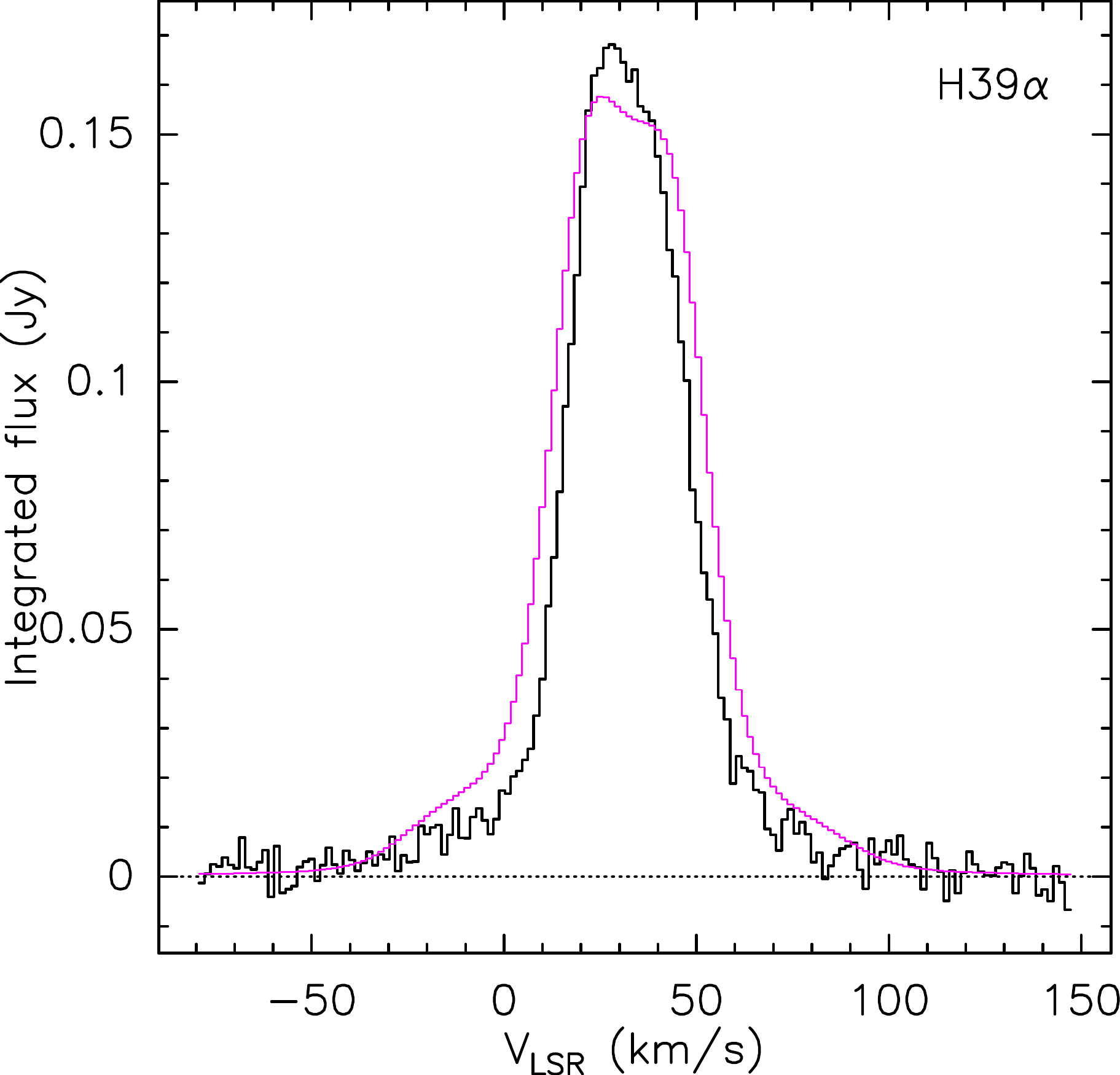}
\includegraphics*[bb= 0 0 644 594,width=0.255\hsize]{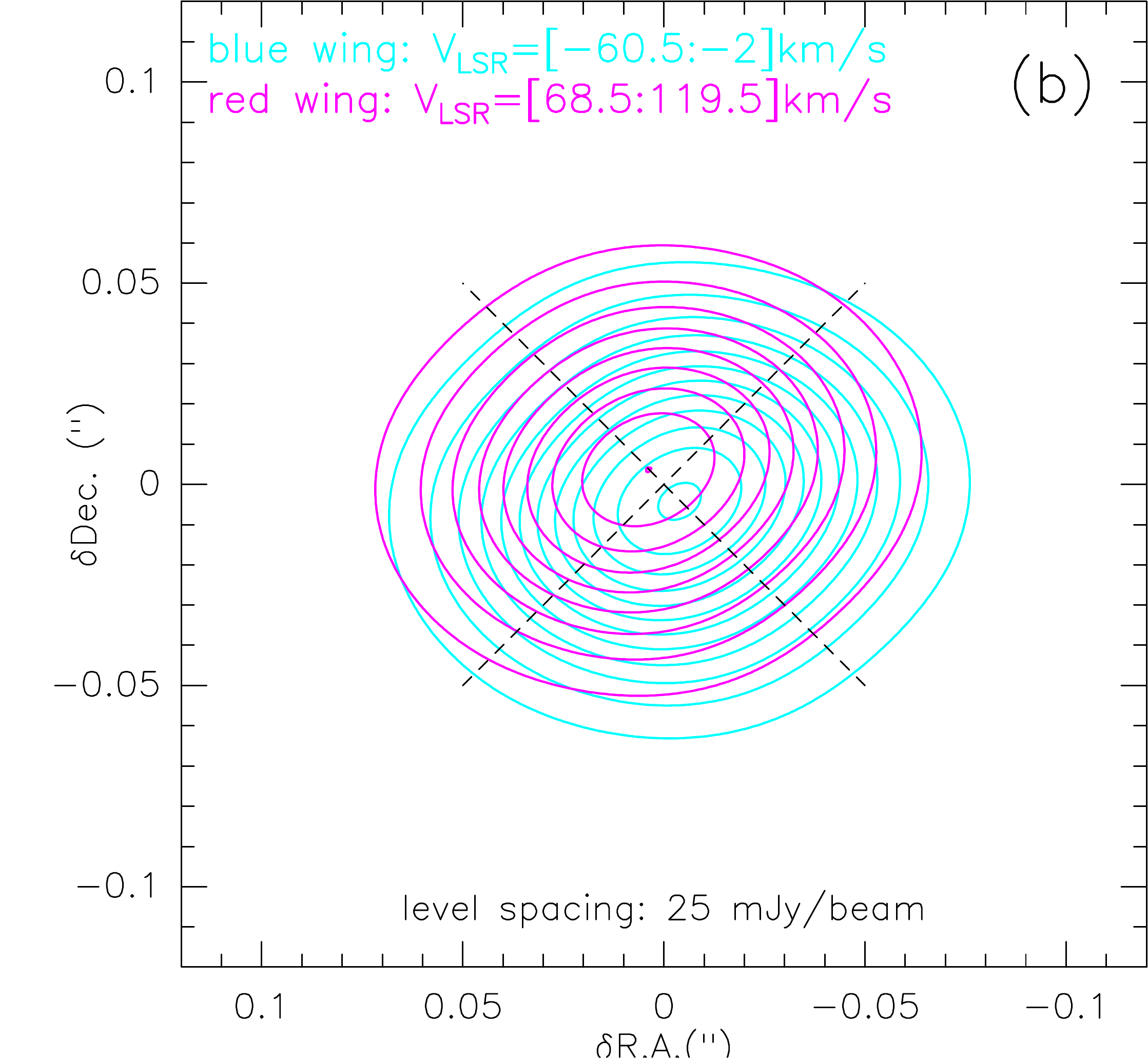}
\includegraphics*[bb= 95 0 644 594,width=0.22\hsize]{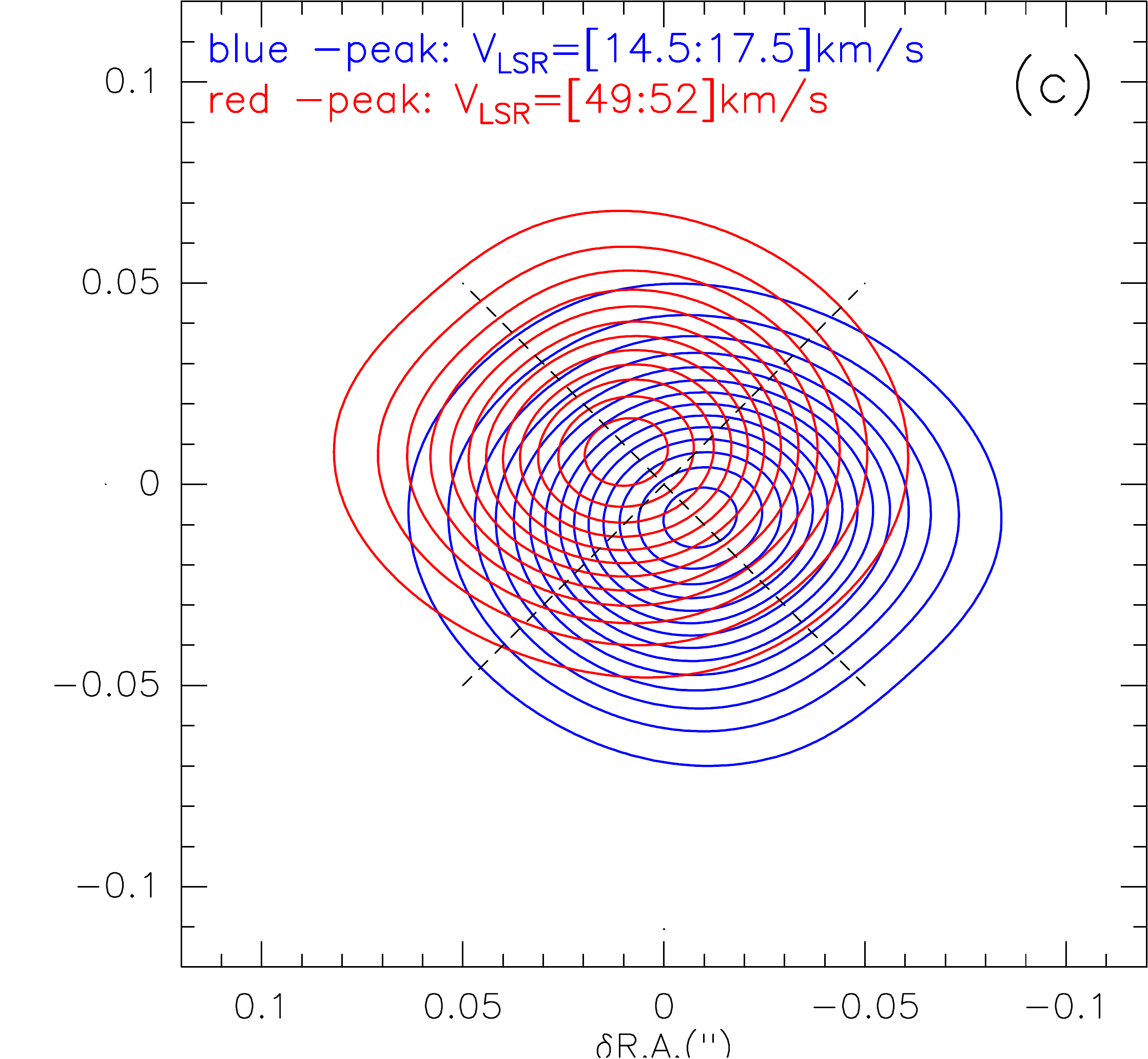}
\includegraphics*[bb= 95 0 644 594,width=0.22\hsize]{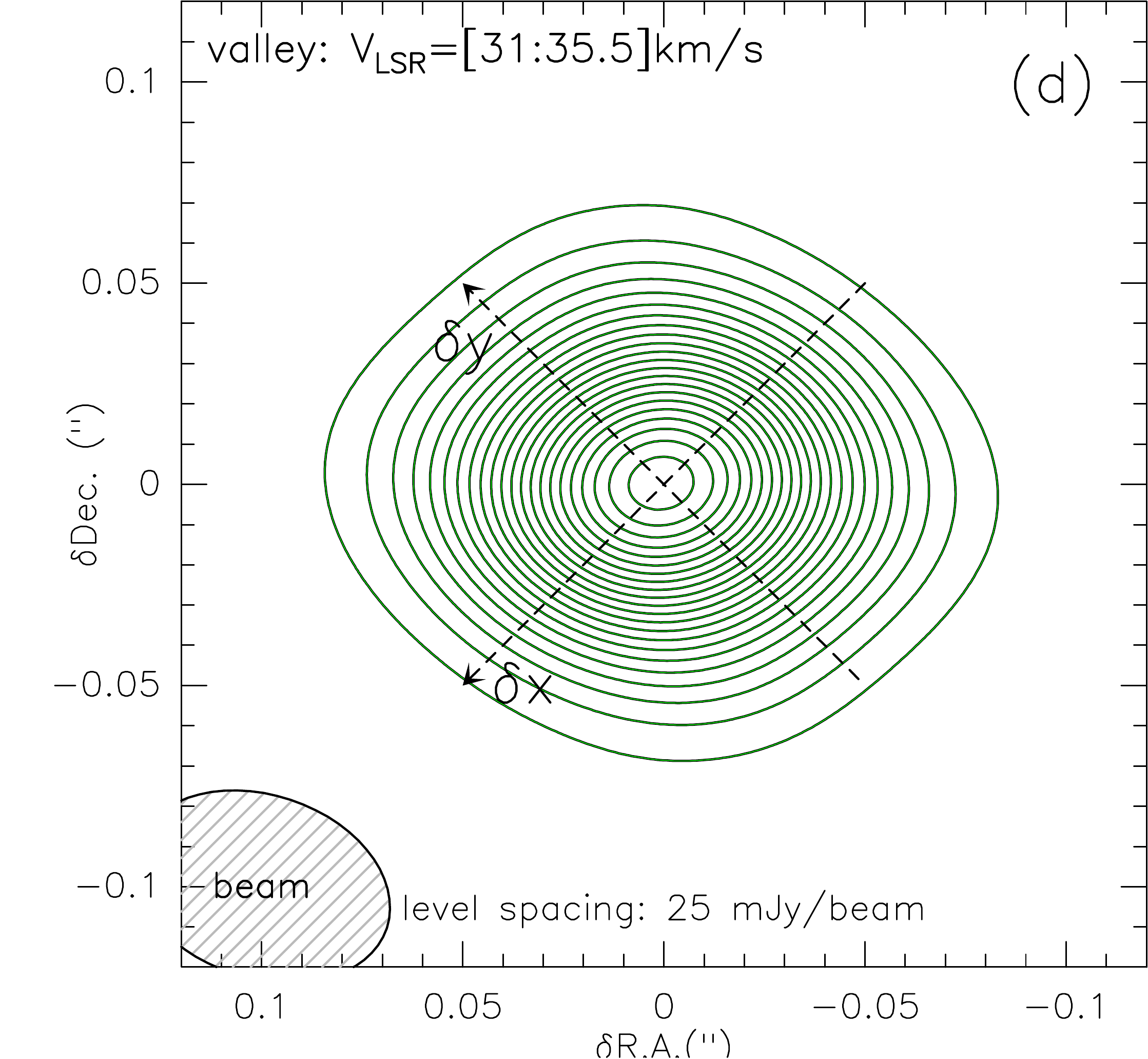}
\vspace{0.5cm}
\includegraphics[width=0.45\hsize]{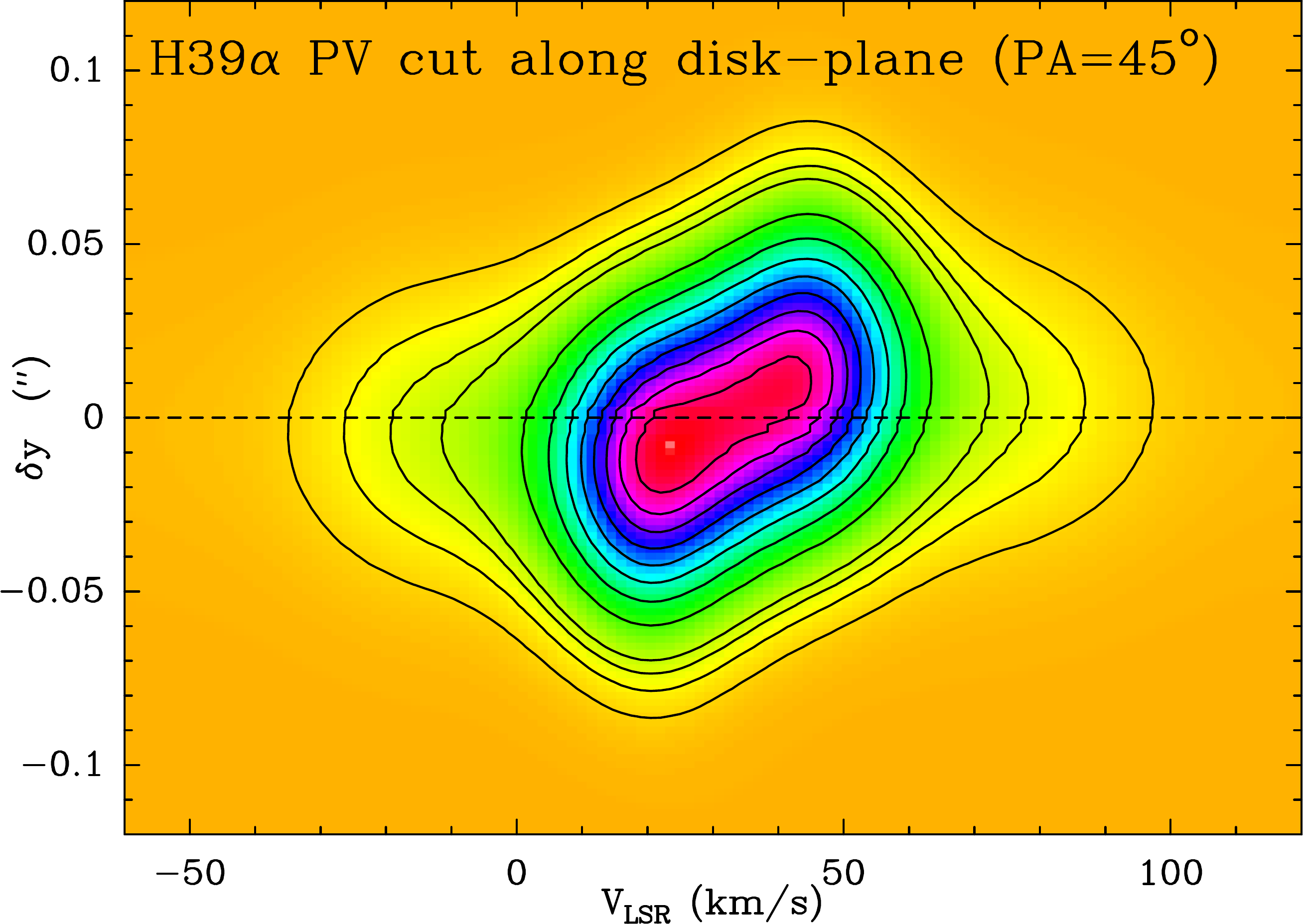}
     \includegraphics[width=0.45\hsize]{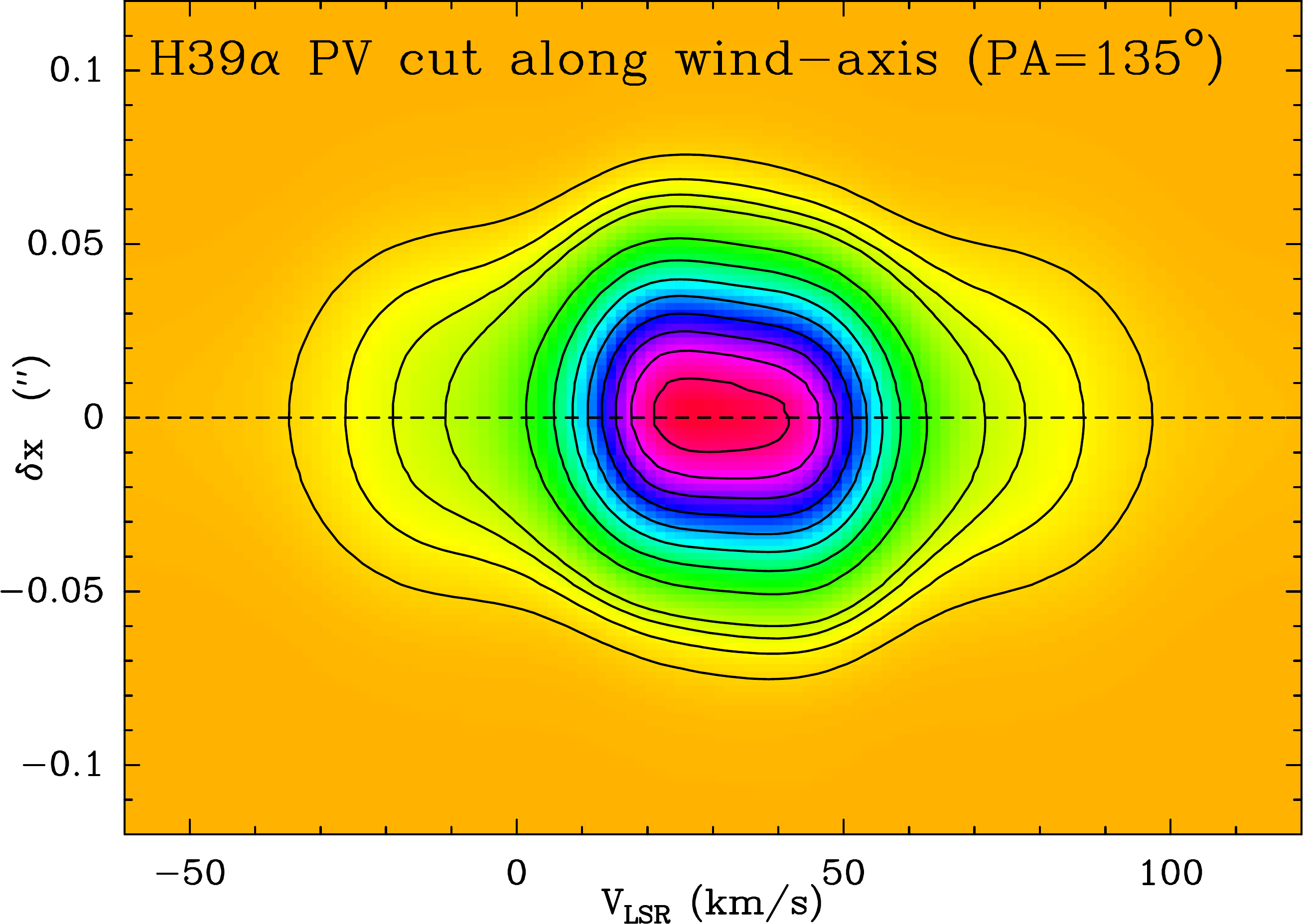}
     \caption{Same as in Fig.\,\ref{f-h39a-all} but showing the predictions of our model (\S\,\ref{moreli}, Table\,\ref{t-moreli}) \label{f-h39a-all-moreli}}
     \end{figure*}

The profile and spatial distribution of the mm-RRLs are also
reasonably well reproduced by our model
(Figs.\,\ref{f-h30a-all-moreli}, \ref{f-h39a-all-moreli} and
\ref{f-otras}). In the case of the $\alpha$-transitions, the model
predicts the change from a single-peaked profile at 3\,mm (\htnal) to
a double-horn profile at $\sim$1\,mm (\htal), as well as the overall
intensity and shape of both profiles, including the broad wings.
As expected, our axially symmetric model is not able to reproduce the
larger intensity of the \htal\ blue peak, relative to the red one, as
observed with ALMA. The line asymmetry, which has changed with time as
deduced from our observations with the \iram\ antenna (Fig.\,\ref{f-h30a-all}), could
be mimicked by adding one or more dense ionized clumps in the
approaching side of the disk. 
Given the time-variable profile and the non-linear
response of the \htal\ maser line to small changes in the physical
properties in the emitting region, such a model is out of the scope of this paper. 

Our model confirms the maser nature of the \htal\ transition, with a
minimum optical depth at the line (blue) peak
$\tau_{30\alpha}$$\sim$$-$6, which explains the large intensity of
this line as well as its time-varying and double-horn profile.  Our
model indicates that, although in a smaller amount, the \htnal\ line
also shows maser amplification ($\tau_{H39\alpha}$$\sim$$-$2). The
rest of the mm-RRLs reported here (\hce,\hcg, and \hsd) are
predominantly thermal lines, with a minor contribution by stimulated
emission ($|\tau|$$<<$1) and, thus, with intensities close to those
expected in LTE.


Concerning the surface brightness distribution of the
$\alpha$-transitions, we are able to emulate the shape and position of
the emission wings, the angular separation of the red- and blue-peak
emitting regions along the disk equator (i.e., the $y$-axis), and the
morphology at the systemic velocity (panels b, c, and d of
Figs.\,\ref{f-h30a-all-moreli} and \ref{f-h39a-all-moreli}). Our
model, in general, best reproduces the \htal\ maps. In the case of the
\htnal\ transition, our model slightly overestimates the width of the
profile and is not able to explain the emission dip at the center of
the emission wings (in contrast to \htal). These two facts suggests
that the \htnal\ emission arises in regions slightly more distant from
the center than predicted by the model.

The \htal\ red- and blue-peak emitting regions appear slightly more
extended in the direction perpendicular to the disk (i.e.\,the $x$-axis)
in the model than in the observations (Fig.\,\ref{f-h30a-all}, panel c and bottom-right panel). 
The ALMA super-resolution maps show that, as predicted by our model,
both the red- and blue-peak \htal-emitting regions are actually
elongated along the $x$-axis and hint at two local maxima, which are
$\delta$x$\sim$0\farc02 apart (Fig.\,\ref{f-h30a-all-moreli-hires}c).
In the data, the asymmetry of these regions about the equator is
notable, with the SE clump being brighter than the NW one. This is not
reproduced by our model, which is symmetric about the disk equatorial
plane. Note that, in any case, this asymmetry is not apparent in the
\htnal\ maps (Fig.\,\ref{f-h39a-all-moreli-hires}c), which indicates
that the differences in the physical conditions of the disk regions
above and below the equator are probably relatively small but 
are exponentially amplified by the \htal\ maser.

The \htal\ red- and
blue-peak emission clumps lie closer to the disk equator in the data
than in the model (most remarkably, for the intense blue-peak).  We
include in our model the emission from the inner wall/edge of the
neutral rotating disk, presumably ionized by the star
(Fig.\,\ref{f-dens2D}), to enhance the \htal\ emission from the
midplane of the disk, although our model still slightly underestimates
the amount of ionized gas in these very low-latitude
($\theta$\raw90\degr) disk regions.

Our model confirms the kinematic structure of the ionized \ion{H}{ii}
region around MWC\,922 inferred directly from the line maps. The
rotation of the gas at low-to-mid latitudes in the disk is well
described by a Keplerian law (\vrot$\propto$1/$\sqrt{r}$). The central
mass is about $\sim$10\,\msun\ (adopting $d$=1.7\,kpc), although the
uncertainty of this value is relatively high, of up to 25\%, as a
consequence of the uncertain \bn\ departure coefficients (as discussed
already in CSC+17) and the high sensitivity of the maser line
intensity to the latter. We have also run a model adopting an
angular-momentum conservation rotation-law (\vrot$\propto$1/$r$, not
shown). However, the angular resolution of the maps does not enable to
discern if there is a significant improvement when using this
law.

As in our initial model presented in CSC+17, in addition to rotation,
the mid-latitude regions of the ionized core lying just above the
thin
ionized surface of the rotating disk include radial expansion at low
velocity. This component, which is referred to as the slow wind (as
already mentioned, Fig.\,\ref{f-dens2D}), has been introduced in our
model because it is found in most systems where rotating disks have
been spatially resolved to date, including the Red Rectangle
and MWC\,349, which resemble MWC\,922 in many aspects
\citep[e.g.][]{mar11,bae13,buj13,buj17,buj18}.  However, based on our current ALMA data, we cannot
firmly establish the presence of a slow wind in MWC\,922 since
different models with radial expansion, or even infall motions, with
velocities in the range $|$\vexslow$|$$\sim$0-10\,\kms\ lead to
synthetic maps that are equally satisfactory, within
uncertainties. In a similar way, we cannot rule out some
expansion (at low velocity, $\lsim$10-15\,\kms) in the so-called disk
component (which has been assumed to have a width of \thd=6.5\degr),
where pure rotation is adopted for consistency with our original
model presented in CSC+17, which was inspired in that performed for MWC\,349A
\citep{mar11,bae13}.


One of the novelties and, probably, most relevant discovery in this
paper is the presence of a fast bipolar wind emerging from MWC\,922
and the fact that this wind is rotating. The expansion velocity
adopted here is \vexfast$\sim$95\,\kms, although slightly faster
motions are possible given that the full width of the \htal\ line
wings are somewhat underestimated by our model. We use a conservative
value of \vexp\ that simultaneously explains the profiles of all the
mm-RRLs detected (note that the \htnal\ wings are rather well
emulated).  The rotation of the fast wind, in the same sense as the
disk, reveals itself in the data and in the model as a small offset
about the disk's revolution axis (i.e., the $x$-axis) of the blue- and
red-wings emission clumps; this is best seen in the intense
\htal\ transition (Figs.\,\ref{f-h30a-all}b and
\ref{f-h30a-all-moreli}b). The offset is also visible in the PV
diagram along the $y$-axis: note that the red-wing (blue-wing) overall
emission lies above (below) the $\delta y$=0\arcsec\ line (bottom-left
panels of Figs.\,\ref{f-h30a-all} and \ref{f-h30a-all-moreli} and
Fig.\,\ref{f-pvoffsets}). To corroborate the rotation in the fast
wind, we show in the appendix (Fig.\,\ref{f-h30a-all-moreli-hires})
the predictions of a model in which rotation has been suppressed in
the fast wind: as expected, this model is not able to reproduce the
observed off-axis distribution of the \htal\ red- and blue-wings
clumps, which overlap on top of the disk/wind revolution axis.

\subsection{Model caveats and uncertainties}
\label{f-caveats}

Our model is a relatively simple representation of the
spatio-kinematic structure and physical conditions of the emerging
\ion{H}{ii} around MWC\,922 that is able to describe reasonably well
the dominant features of our ALMA observations. As already noted in
CSC+17, we stress the extremely high sensitivity of the \htal\ maser
line intensity (but also of the \htnal\ transition) not only to small
changes in the local physical conditions but also to the uncertain
\bn\ departure coefficients. Thus, the biggest challenge is always to
reproduce simultaneously the maser and non-maser RRL transitions. In
this section, we discuss the uncertainties of some of the model
parameters given in Table\,\ref{t-moreli}.

{\sl The central cavity.} As shown by the data, and corroborated by
our model, most of the free-free continuum and RRL emission at
mm-wavelengths arise in a compact ionized region within
$r$$\la$80-90\,au (adopting $d$=1.7\,kpc). \citep[The outermost least dense
layers of the \ion{H}{ii} region, which are best traced at
cm-wavelengths, reach out to larger distances of $\sim$200-300\,au as
shown by][]{rod12}. The radius of the central cavity is uncertain,
with values smaller than \rin$\sim$25-30\,au making it difficult to
simultaneously reproduce the mm-SED (in particular, the free-free
continuum becoming optically thin near $\sim$\,1\,mm) and the maps (in
particular, the spatial separation of the red- and blue-wing emitting
clumps of the \htal\ transition). This is true even considering density gradients slightly
shallower than the one adopted in our preferred model (Table\,\ref{t-moreli}).

We note that, in any case, the spherical cavity used in our model
could well be an oversimplified representation of the density
distribution of the most inner regions in the core of MWC\,922, which
could be a porous/sponge-like (or, reversely, clumpy) structure, or could
accommodate one or more non-spherical cavities (e.g. finger-like
holes/voids excavated in a denser environment maybe by the action of
jets), etc. Additional observations of mm-RRLs at higher frequencies
(H$n\alpha$ with $n$$<$30), ideally with high angular resolution, are
needed for an accurate characterization of the innermost ($\la$30\,au)
regions of the core of MWC\,922.

{\sl The temperature and density laws.} For simplicity, we have used
an isothermal model, i.e., with no significant variation of \te\ across the
different kinematic components of the ionized core. Under this assumption, the
X-shape morphology of the free-free continuum maps is
dictated by the adopted density law, which is also the same for the
disk and the wind(s) and has been chosen to vary as a uniform function
of the radial distance to the center and the latitude. We have found
satisfactory models using a range of radial power-law indexes,
$\beta$$\sim$2.5-2.9, and a more or less steep density decrease with
the latitude, $\theta_0$$\sim$20\degr-60\degr.

Once a suitable density law has been established, we have varied
\te\ to achieve a good compromise between the model predictions
for the free-free continuum and the mm-RRLs, which are highly
dependent on \te. The value of \te\ adopted here ($\sim$8000-9000\,K)
is slightly larger than that used in CSC+17 ($\sim$7000\,K). We stress
that \te\ is especially uncertain and this difference has
not to be interpreted as a real time-variation of \te, but just as a
consequence of the slightly different values of certain model
parameters, particularly of the density distribution and the geometry.
Our model indicates that the typical electron densities in the
emitting layers are in the range \dense$\approx$10$^6$-10$^7$\,\cm3,
with little uncertainties.

{\sl Disk and fast-wind aperture and kinematics}.  The opening angle
or aperture of the disk is relatively well constrained by the X-shape
morphology of the continuum source deduced directly from the maps,
with a range of values \tha$\sim$50\degr-70\degr\ yielding acceptable
model predictions. The aperture of the fast wind is, however,
relatively uncertain because this component is compact (particularly,
across the wind) and is not fully spatially resolved.  The final
aperture adopted (\thw$\sim$38\degr) is optimal for the simple density
law adopted in our model and reproduces reasonably well the shape and
position of the peanut-like \htal\ wing-emitting regions. The ALMA
maps of the \htnal\ emission wings, although noisier, also show a
peanut-like shape that is not well reproduced by our model
(Fig.\,\ref{f-h39a-all}b and Fig.\,\ref{f-h39a-all-moreli}b). This
could suggest that the density fall becomes shallower in the outer
regions of the ionized core under study, which are better traced by
\htnal\ than by \htal\ due to opacity effects.

As already mentioned, the presence of radial expansion at
mid-latitudes of the ionized core of MWC\,922 is expected (based on what is
observed in similar objects) but it is not
firmly established from our data and model.  This implies that the
disk angular width is highly unconstrained, with any value from
\thd$\sim$2\degr\ to $\sim$20\degr\ being possible.  Our model
indicates that, if the slow wind exists, then its average velocity is
\vexslow$\la$15\,\kms, while the average expansion velocity of the
fast wind is about 100\,\kms. Obviously, the different model
components we have used (the disk, the slow wind, and the fast wind) are
most likely not separated by sharp boundaries. This means that there
is most likely a gradual transition from pure rotation in the disk to
fast expansion plus rotation in the high-velocity wind, maybe through
a slow-expansion and rotation transition region at mid-latitudes.
Based on our data, a radial velocity gradient in the fast wind is not
ruled out either but, in principle, it is not necessary to explain the
observations, which in any case trace a rather thin
($\sim$30-40\,au-wide) layer of ionized gas beyond the cavity.

{\sl The distance to the source}.  The distance to MWC\,922 is
probably the most uncertain parameter that affects the interpretation
and modelling of the data.  By default and for a better direct
comparison with the model presented in CSC+17, we adopt a value for
the distance of $d$=1.7\,kpc (\S\,\ref{intro}). We also ran our model
using the near kinematic distance, $d$=3\,kpc, as estimated in
CSC+17. The model parameters that are different in case $d$=3\,kpc are
specified in Table\,\ref{t-moreli}.  Note that given the non-linear
behaviour of the \htal\ maser, not only the density, the size of the
cavity, and the rotation velocity field have been modified, but also
the electron temperature (\te) has been slightly readjusted to produce
satisfactory predictions.  As we will show in \S\,\ref{dis-hrd}, a
value of the distance $\sim$3\,kpc enables a better agreement between
the mass of the central star and the expected location of MWC\,922 in
the HR diagram (HRD) based on current single star post-main sequence
evolutionary tracks.

\section{Discussion}
\label{dis}

\subsection{The mass in MWC\,922's close environment}
\label{dis-mass}

As we have seen, dust produces a dominant emission component in the
far-IR and it has a non-negligible contribution of $\sim$15-20\%\ at
1\,mm (Fig.\,\ref{f-sed}). Since there is no missing flux in our ALMA
data, the bulk dust thermal emission must arise in a rather compact
region within a few hundred au from MWC\,922, given the maximum
recoverable scale of the ALMA configuration used in this work at 1\,mm
($\sim$0\farc35, \S\,\ref{sec-obs}). The fit to the far-IR SED of
MWC\,922 by a modified black-body (\snu{2.1}) is consistent with a
major component of optically thin \td$\sim$160\,K dust emission with a grain
emissivity frequency dependence as $\sim$$\nu^{1}$. Using the
formulation described, e.g., in \cite{san98}, and adopting a dust
emission flux at 230.9\,GHz of $\sim$35\,mJy, we estimate that the
total dust mass 
 in MWC\,922 is \md$\sim$8\ex{-5}\,\msun\ (at $d$=1.7\,kpc). 
For a typical gas-to-dust mass ratio of $\sim$100, we deduce a total
mass of \mtot$\sim$8\ex{-3}\,\msun\ in the proximity of MWC\,922.  We
note that the accuracy of our dust mass determination is limited by
the uncertainties in the grain emissivity (to a factor of a few),
which depends on the unknown composition and structure of the grains.

The total mass of ionized gas implied by the free-free continuum
emission at mm-wavelengths is
\mhii$\sim$2\ex{-5}\,\msun\ (Table\,\ref{t-moreli}, $d$=1.7\,kpc),
which is two orders of magnitude lower than \mtot\ deduced above.
This indicates that there may be a predominantly neutral gas component
around MWC\,922.

Part of this neutral gas component is indeed detected by means of
first and second overtone CO transitions in the near IR \citep{weh17}.
\cite{weh17} estimate a gas temperature of about 3000\,K in the CO
emitting region and conclude that this molecule lies in a
disk/ring-like structure, surviving dissociation due to dust in the
inner disk blocking the stellar UV radiation by the B[e] central
star. These authors deduce a relatively low CO column density of
$\sim$1.7\ex{15}\,\cm2, which implies a H$_2$ column density of
$\sim$8.5\ex{18}\,\cm2 adopting a typical CO-to-H$_2$ fractional
abundance of X(CO)=2\ex{-4}. \cite{weh17} do not attempt to compute
the total mass of hot molecule rich gas traced by NIR CO emission
(\mco) probably because the dimensions of the CO emitting volume are
unknown. In the following paragraph, we provide a crude estimation of
the mass that suggests that, as \mhii, \mco\ probably represents only
a very small fraction of the mass traced by the mm-continuum dust
emission.

Adopting an average H$_2$ density of $\approx$10$^4$-10$^5$\,\cm3 in
the hot CO emitting region
and using the column density estimated by \cite{weh17} we deduce that
the size of the intervening CO-emitting region along the line of sight
is probably $\Delta r$$\approx$1-10\,au. Note that the density of the
hot CO-emitting gas is plausibly intermediate between that in the
central \ion{H}{ii} region studied here ($\sim$10$^6$-10$^7$\,\cm3,
\S\,\ref{moreli}) and that of the more extended surrounding nebulosity
traced by numerous forbidden and permitted emission lines in the
optical \citep[in the range $\approx$10-10$^4$\,\cm3,
  see][]{weh17,bal19}.  If we assume that the hot CO gas is located in
a hollow cylindrical structure with a representative inner radius of
$R$, and a thickness ($\Delta r$) and a scale height $h$$\sim$$\Delta
r$, then we deduce a total mass in this component of only
\mco$\approx$[10$^{-7}$-10$^{-8}$]$\times$$\frac{R}{1000\,au}$\,\msun.

The radius of the hot CO-ring is unknown since none of the slits used
by \cite{weh17} were oriented along the nebula equator.  However, we
believe that values much larger than $\approx$1000\,au are unlikely
since it would imply that the hot CO emission is reaching to, and
coexisting with, the less dense and relatively extended optical/NIR
nebulosity where the gas is known to be atomic and begins to be dissociated/ionized by the
external (ISM) UV radiation field
\citep[e.g.][]{tut07,weh17,bal19}. The CO-ring scale height, $h$, is also
unlikely to be significantly larger than $\Delta r$ since a very long,
thin-walled tubular structure would hardly remain stable. 
A severe
upper limit of
$h$$<$1\arcsec\ ($<$1700\,au at $d$=1.7\,kpc) is deduced from the
observations, since the CO emission is not claimed to be spatially resolved by \cite{weh17}.
Even in the case that our crude estimate of \mco\ is underestimated
by 1-2 orders of magnitude, the total mass in the hot CO-ring is
probably smaller than, or at most comparable to, the mass of the \ion{H}{ii} region. 

We then conclude that the dust mm-continuum emitting structure and the
gas presumably contained in it\footnote{Note that it is not clear the
  existence of this gas since the conditions in the disk are unknown.},
$\mtot$$\sim$8\ex{-3}($\frac{d}{1.7\,kpc}$)$^2$\,\msun, is the
dominant mass component in the close environment of MWC\,922.  As we
show in the next subsection, this mass could give some
insights into the evolutionary stage of this enigmatic source, in
particular, if it is $i)$ a young (pre-main sequence) star still
enshrouded by its natal could or $ii)$ a main-sequence or post-main
sequence star experiencing mass transfer onto a companion.

\subsection{The nature of MWC\,922}
\label{dis-nat}

Establishing the evolutionary stage of IR excess B[e] stars like
MWC\,922 is difficult attending only to their \teff\ and \lstar, since
B[e] stars in the main-sequence and in pre- and post-main sequence
stages 
cluster around the same region of the HR diagram.  Here, we list a
number of arguments related to the amount of matter around the star that, in our opinion, make a pre-main
sequence nature in the early stages of evolution improbable.

$i)$ First, the amount of material in the close environment of
MWC\,922 (\mtot$\sim$8\ex{-3}($\frac{d}{1.7\,kpc}$)$^2$\,\msun,
\S\,\ref{dis-mass}) is too low compared to what it is generally
observed in intermediate- and high-mass young stellar objects (YSOs)
\citep[see e.g.\,the review by][]{bel16}.  In particular, for a
$\sim$10\,\msun\ YSO the average disk mass is $\approx$1\,\msun\, with
the least massive disks being above 0.1\,\msun\ in all cases reported
to date (see Fig.\,13 in Beltran and Wit).
If MWC\,922 is located at a larger distance, e.g.\ $d$=3\,kpc, the
mass in its close environment and at the center of the rotating disk would
be \mtot$\sim$0.025\,\msun\ and $\sim$18\,\msun, respectively.  Such a
massive YSO, however, again will be expected to have a more massive
disk than observed, typically from a few solar masses to
$\sim$100\,\msun.  The value of \mtot\ computed for MWC\,922 would
only be consistent with a low-mass YSO of less than 2-3\,\msun. However, the luminosity
for such a low-mass YSO, log(\lstar/\ls)$\sim$1.5-2, would be orders of magnitude lower than
the luminosity of MWC\,922.

$ii)$ Second, as shown by our ALMA data and other previous works,
MWC\,922 is totally deprived of line emission from any molecular
species, except for CO \citep[to date only detected in the NIR,][]{weh17}. This is again in
marked contrast to what is observed in YSOs, which generally display
a very rich molecular chemistry that includes, among many other
species, complex organic molecules such as CH$_3$OH, CH$_3$CN, etc
(e.g.\,Beltran and Wit, \citeyear{bel16}, and references therein). The large variety of
molecular species detected toward most YSOs is due to the fact that
they are still enshrouded in the giant molecular clouds, cold and
dense, where they were formed. However, there is no trace of a
significant amount of cold and dense material surrounding MWC\,922. In
fact, the recent detection of a parsec-scale jet orthogonal to the
equator at optical wavelengths \citep{bal19} indicates that such a jet
has been travelling for $\sim$3000\,yr without any sign of strong
interaction with ambient material \citep[e.g.\,this represents yet another
dissimilarity with typical YSOs, in which jets surrounded by swept out
molecular cavities are commonly observed; see e.g.][]{arc06}.

$iii$) As we will see in \S\,\ref{dis-winds}, the mass-accretion rate
inferred from the mass-loss rate of the fast bipolar wind (assuming
that typically $\sim$10\%\ of the accreted material is injected into
the fast wind) is \mloss$_{\rm accr}$$\approx$1\ex{-5}\,\my, which is
significantly lower (by 1-2 orders of magnitude) than that expected in
a massive YSO (typically 10$^{-4}$-10$^{-3}$\,\my, see e.g. Fig.\,10 of
Meyer et al., \citeyear{mey19}, and also Fig.\,17 in Beltr\'an \& de Wit, \citeyear{bel16}). However, the
mass-loss rate of the slow and fast winds deduced in this work are
consistent within uncertainties with the values generally observed in
blue giant stars with a spectral type and luminosity comparable to
that of MWC\,922, especially, if it is located at $d$=3\,kpc, which would
imply \lstar$\sim$5.9\ex{4}\,\ls\ \citep[e.g.\,][see the blue-shaded region in their
Fig.\,1]{meyn15}. 

\subsubsection{Position in the HR diagram}
\label{dis-hrd}

   \begin{figure}[htbp!]
     \centering
      \includegraphics*[bb=130 15 640 517,width=0.99\hsize]{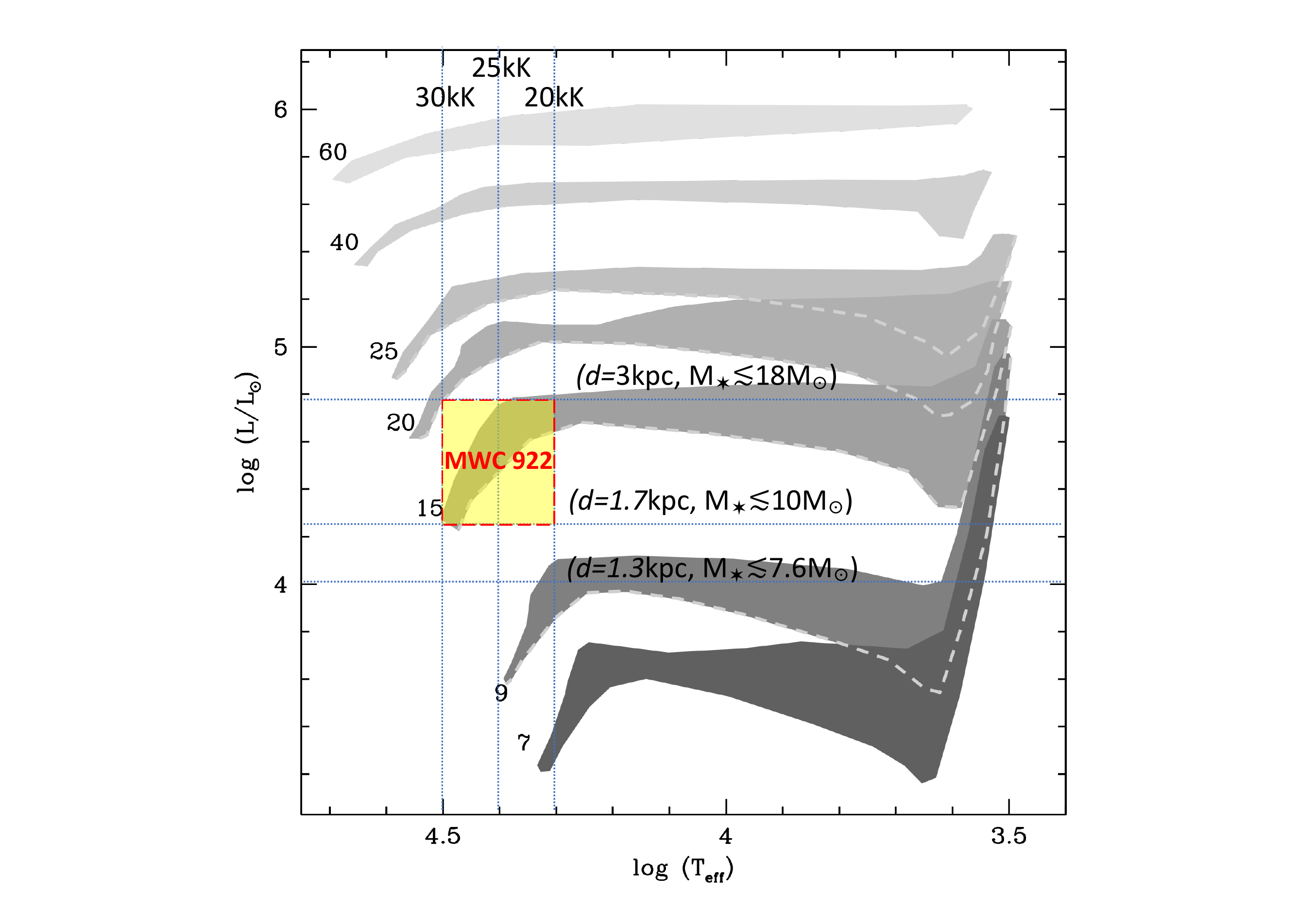}
      \caption{Figure adapted from Fig.\,6 in \cite{mar05} showing the
        position of MWC\,922 in the HR diagram (inside the yellow box)
        together with the envelopes of evolutionary paths for
        \mstar=7, 9, 15, 20, 25, 40, and 60\,\msun\ stars taking into
        account the different codes studied by these authors. The horizontal dotted lines show
        the luminosity of MWC\,922 adopting different values of $d$; the upper limit to the stellar mass deduced from
        the central mass inside of the rotating disk at different $d$ is also indicated. 
          \label{f-hrd}
        }
   \end{figure}

One crucial parameter to constrain the nature of MWC\,922 is the mass
of the object at the center of the rotating disk. Although the
effective temperature and luminosity of MWC\,922 are uncertain, we can
compare the location of this object in the HR diagram (HRD) with the
post-main sequence evolutionary tracks of massive stars. In
particular, we aim to test whether such a position is consistent with
a stellar mass \mstar$\lsim$10 or $\lsim$18\,\msun\ adopting $d$=1.7
and 3\,kpc, respectively, as deduced from the ALMA data
(\S\,\ref{moreli}).  As already mentioned, the upper limit arises
because the central source can be a binary or multiple system.

We have used the set of different evolutionary tracks compiled and
jointly evaluated by \cite{mar05}, which are reproduced in
Fig.\,\ref{f-hrd}. As we can see in this
figure, evolutionary tracks for 
$<$10\,\msun\ stars lie well below the HRD box defined by the probable
ranges for the effective temperature and luminosity of MWC\,922. This
makes improbable a mass of \mstar$\lsim$10\,\msun\ and, thus, a
distance $d$$\sim$1.7\,kpc for MWC\,922. Adopting a lower value of the distance to match
the luminosity of a less massive star (e.g.\,a $\sim$9\,\msun\ with
log(\lstar/\ls)$\sim$4, which would imply $d$$\sim$1.3\,kpc) does
not help to explain this inconsistency because the upper limit to the
stellar mass consistent with the rotating disk observed will be correspondingly smaller 
(\mstar$<$7.6\,\msun\ at $d$$\sim$1.3\,kpc), implying again
evolutionary tracks shifted downwards from MWC\,922. 

As we see in Fig.\,\ref{f-hrd}, the HRD position of MWC\,922 is most
consistent with a \mstar$\sim$15\,\msun\ star, which is 
fully compatible with the upper limit to the stellar mass derived
adopting the near kinematic distance, \mstar$\lsim$18\,\msun. For a
distance $d$$\sim$3\,kpc, the luminosity of MWC\,922,
log(\lstar/\ls)$\sim$4.8, would be most consistent with a star that
has recently left the main-sequence and is entering the blue giant
stage. This would imply an age of $\sim$10\,Myr.  If this is correct,
MWC\,922 could be very similar to the B[e] star MWC\,137, a slightly
more massive B[e] supergiant with a confirmed evolved nature based on
the \trecem/\docem\ ratio \citep{oks13}, that also shows similar
properties to MWC\,922, including a parsec-scale jet emerging from the
center \citep{meh16}. The morphology of the circumstellar nebulosity
around MWC\,922 (squared or biconical) and MWC\,137 (roughly elliptical) is however
quite different, which is probably related with the different
properties of the central binary system and the way the stars interact
and transfer mass.

Assuming a post-main sequence nature, the rotating disk and fast
bipolar ejections found almost certainly require the presence of a
companion to MWC\,922. Therefore, since MWC\,922 could not be
following the evolutionary tracks of single stars, a different stellar
mass cannot be totally ruled out. Also, the mass of the central object
deduced from the disk rotation is also somewhat uncertain (by
20\%-30\%, if our hypothesis of Keplerian rotation is true or larger
if it is not).  Therefore, we are cautious on the conclusion about the
stellar mass presented above. For this reason, throughout this paper
we show the results from the two most likely values of the distance,
namely, $d$=1.7\,kpc, commonly accepted in the literature and used in
our original model reported in CSC+17, and $d$=3\,kpc, the near
kinematic distance to MWC\,922, which enables a best reconciliation of
its HRD position with the upper limit to \mstar\ inferred by us.

Considering the uncertainties in the evolutionary models \citep{mar05}
and in the values of \teff, \lstar, and \mstar, we do not rule out
MWC\,922 to be a member of the 'FS CMa' class, as proposed by
\cite{mir07} for other unclassified B[e] stars. FS CMa stars are
believed to be binary systems currently undergoing a phase of rapid
mass exchange composed of one main-sequence star and a compact
companion. More recently, \cite{Fue15} have proposed that FS CMa stars
could be post-merger products that still retain part of the dusty
ejection in the shape of a disk.

Finally, in CSC+17 we showed that the \teff\ and \lstar\ of MWC\,922
could be, in principle, consistent with a post-AGB star evolved from a
massive progenitor ($>$5-10\,\msun\ in the main sequence).  However,
in view of the little amount of material in the stellar surroundings
deduced in this paper (\S\,\ref{dis-mass}), we now believe this is
improbable. This is because in a post-AGB stage the star should have
already ejected most of its initial mass (mainly during the AGB phase
in the form of a slow, dense wind). Since for massive stars the
AGB-to-PN transition happens very quickly, possibly in a few decades
and certainly in less than a few hundred years \citep[e.g.][]{ber16},
we should observe at least a fraction of the few \msun\ expected in
the circumstellar envelope. A significant amount of molecular gas is
indeed found even in well-developed, $\sim$few$\times$10$^3$\,yr old PNe
with hot (\teff$\approx$10$^5$\,K) central stars, e.g.\,NGC\,7027 or
NGC\,6302 \citep{hua10,sg17}.

\subsection{The ionizing central source}
\label{dis-ion}

Our code MORELI computes the number of Lyman continuum photons emitted
per second (\nlyc) needed to reproduce the free-free and mm-RRL
emission observed. The value found, log\nlyc$\sim$46.9\,s$^{-1}$, has been
compared with those obtained from the grid of model atmospheres of
early B-type stars by \cite{lan07} to investigate the central ionizing
source. We find that a \teff$\sim$30\,kK star (with \rs$\sim$5\rsun,
adopting \lstar$\sim$1.8\ex{4}\,\ls\ at 1.7\,kpc)
is able to produce the ionization observed even if it is near the main
sequence (i.e.\, if we adopt the values of $q_0$=log\nlyc\ for a star
with a surface gravity of log$g$=4 cm/s$^2$). If the star is a blue giant
star (log$g$=3 cm/s$^2$) then the \nlyc\ inferred can also be
accounted for at lower stellar temperatures, \teff$\gsim$25\,kK
(\rs$\lsim$7\,\rsun).

Assuming that the mass at the center of the rotating disk ($\sim$10\,\msun)
is that of the star, then the stellar surface gravity is
log$g$$\sim$3.7-4.0 for the range of temperatures \teff$\sim$25-30\,kK
(and adopting the stellar radius \rs\ computed from
\lstar=4$\pi$\rs$^2$$\sigma$\teff$^4$).  The gravity values obtained
suggest a star that is in the main sequence or has just left it
(i.e.\,a young post-main sequence star).

We have done the same exercise adopting a distance to the source
of $d$=3\,kpc, for which the central mass inferred from the observations
and the model is $\sim$18\,\msun. In this case, a \teff$\sim$25-30\,kK
star (with \rs=13-9\,\rsun\ and \lstar=5.8\ex{4}\,\ls), with
log$g$$\sim$3.5-3.8, will also be able to produce the amount of
\nlyc\ needed. The values of the gravity deduced in this case suggest
a post-main sequence nature.

We note, in any case, that there are several uncertainties in the
estimations presented above.  First, the central mass inside the
rotating disk is an upper limit to the mass of the ionizing star since
there could be binary (or even multiple) system at the core. Indeed,
the presence of a binary is the most likely scenario to explain the
formation of a rotating disk if MWC\,922 is not a YSO
(\S\,\ref{dis-mass}). In case of a lower mass for the central star,
then the gravity would be also lower, with the subsequent increase of
$q_0$=log\nlyc\ from the model atmospheres even for stars with
slightly lower values of \teff. Note that the distance to MWC\,922,
and thus its luminosity, is rather uncertain.  Also, we cannot rule
out that the B[e] star MWC\,922 is slightly cooler than
\teff$\sim$25-30\,kK (as deduced above) since part of the ionization
could be produced by shocks resulting from the interaction between the
fast $\sim$100\,\kms\ bipolar wind with the ambient material.

For these reasons, we still consider in our discussion a conservative
range of \teff$\sim$20-30\,kK for MWC\,922.

\subsection{The winds emerging from MWC\,922}
\label{dis-winds}

{\sl The fast bipolar wind}. The presence of rotation in the fast bipolar wind emerging from
MWC\,922 is one of the most interesting results from this work. To our
knowledge, there are no other fast ($\gsim$100\,\kms) bipolar outflows
with compelling observational evidence for rotation. Rotation in {\em
  slower} winds, however, has been found in a handful of objects
\citep[e.g.][]{lee08,mar11,bur15,buj16,hir17}.  Among these, MWC\,349A is the
source with the fastest rotating wind, with a radial expansion
velocity of $\sim$60-70\,\kms\ \citep{mar11}.

The rotation in the fast wind of MWC\,922 takes place in the same
sense as in the disk, which suggest $i$) that there is a causal
connection between the two (e.g. that the wind is launched via a
disk-mediated mechanism from the rotating disk) or $ii$) that both
components have a common underlying origin.

The launching radius of the fast wind is constrained by our ALMA maps
to $\lsim$29\,au (adopting $d$=1.7\,kpc). This value is smaller than
the gravitational radius for a central mass of \mstar,
$r_g$$\sim$84\,au$\times$(8750\,K/\te)($M_*$/10\msun), which implies
that the fast wind is not simply photoevaporating from the inner rim
of the disk.
(The same conclusion holds adopting $d$=3\,kpc after appropriately scaling of the launching radius and $r_g$.) 
This result was somewhat expected since the high velocity of the fast wind is also not
contemplated in case of photoevaporation, which would happen with a
velocity comparable to the sound speed of the gas in the disk ($c_{\rm
  s}$$\lsim$10\,\kms, for the values of \te\ deduced,
Table\,\ref{t-moreli}).
The large velocity of the fast wind suggests that 
it is launched much closer to the central system, well inside the inner rim of the ionized
disk traced by our observations. Since the velocity of the wind is
typically similar to (or slightly larger than) the escape velocity
at the launching-site, the launching radius is probably 
$\lsim$2\,au ($\lsim$3.5\,au) for a central mass of $\sim$10\,\msun\ ($\sim$18\,\msun), adopting $d$=1.7\,kpc ($d$=3\,kpc). 

We cannot rule out that the rotating disk observed represents the
outer counterpart of a more compact disk component with a
dominant role in the launch of the fast wind, maybe by a
magneto-centrifugal launching mechanism, as it has been proposed for, 
e.g., MWC\,349A \citep{mar11} and the massive YSO candidate Orion
Source I \citep{hir17}. However, we believe that it is unlikely that the disk layers
that are actively participating in this process are the ones probed by
our ALMA observations, which are relatively distant from the center
($>$[30-120]$\times$$\frac{d}{1.7 \rm{kpc}}$\,au).

The fast bipolar wind
and the rotating disk of MWC\,922 could both be the outcomes of a common
underlying physical process (scenario $ii$ above). 
For example, in the case of a
mass-exchanging binary system composed of a mass-losing blue
giant (presumably the B[e] star MWC\,922) and a compact
companion, the rotating $\sim$30-100\,au-scale disk observed could
represent a {\sl circumbinary} disk formed as a result of angular momentum 
transfer from the binary orbital motion to the slow wind of the mass-losing star. This is indeed the case of the Red Rectangle
\citep{buj13} and all post-AGB stars with rotating disks spatially and spectrally
resolved to date \citep[][and references therein]{buj13,buj17,buj18,con18}. In this
scenario, the fast wind would be launched not from the surface of the
circumbinary disk but from the (unseen) accretion disk around the
companion. Circumbinary disks and accretion disks (around the
secondary) are conspicuously found in numerical simulations of 
interacting binary systems \citep[e.g.][and references
  therein]{mas98,liu17,che17,sal19}.

The mass-loss rate in the fast wind is uncertain since this component
is barely spatially resolved and, thus, its detailed morphology and
density distribution are not optimally constrained.  From our model,
we deduce a mass-loss rate of \mloss$_{\rm
  fast}$$\sim$[2-4]\ex{-6}\,\my\ (adopting $d$=[1.7-3]\,kpc, Table\,\ref{t-moreli}). Assuming that the
accretion rate is typically a factor $\sim$10 larger than the
mass-loss rate of the wind \citep[e.g.][]{bel16,mey19}, matter could
be falling into the compact companion at a rate of \mloss$_{\rm
  accr}$$\approx$10$^{-5}$\,\my.
For an average expansion velocity of the fast wind of $\sim$95\,\kms,
the dynamical age of its innermost layers is only $\sim$1.5 or 2.6
years (at $d$=1.7 or 3\,kpc, respectively).

{\sl The slow wind.} Although radial expansion at intermediate
latitudes (between $\sim$30\degr\ and 52\degr) is not strictly
necessary to reproduce our ALMA observations, our data are consistent
with the presence of a slow ($\lsim$15\,\kms) wind lifting off the
ionized disk, as observed in the Red Rectangle and other objects with
rotating disks \citep[e.g.][]{buj13,buj17,buj18}.
Assuming that a slow wind is present, we deduce a mass-loss rate of
\mloss$_{\rm slow}$$\sim$[3-6]\ex{-7}\,\my\ in its innermost layers, at
$\sim$[29-51]\,au (Table\,\ref{t-moreli}), ejected only $\sim$[15-27]$\times$$\frac{9\,{\rm
    km\,s^{-1}}}{\vexslow}$ yr ago.

Since the total mass in the close environment of MWC\,922 obtained
from the dust mm-continuum emission is about
\mtot$\sim$8\ex{-3}$\times$$(\frac{d}{1.7 \rm{kpc}})^2$\,\msun, the reservoir of material in the stellar
surroundings will probably dissipate
in $\sim$[3-4]\ex{4}\,yr (or less if the mass-loss rate
continues increasing as suggested by the density law inferred, which is steeper than an
inverse square function). If the aperture of the slow wind is larger than
adopted, for example, if it reaches down to the surface of the neutral
disk (i.e. \thd=0\degr, Fig.\,\ref{f-dens2D}) then the mass-loss rate
needed to explain the observations is $\sim$25\% larger, implying a
correspondingly lower disk dissipation time. 
The lifetime of the disk is moderately short but this type of sources are rare,
so there is no contradiction. Moreover, if there is a continuated matter supply
into the disk (for example, from a mass-losing giant star in a
mass-exchanging binary system), the dissipation time could be longer.

\section{Conclusions}
\label{conclusions}
In the fall of 2017, we mapped with ALMA the inner layers
($r$$\lsim0\farc06$, i.e.\,within one or two hundred au) of the
ultracompact \ion{H}{ii} region around the far-IR excess B[e] star
MWC\,922. We imaged for the first time the continuum and the
\htal\ and \htnal\ line emission in bands 6 and 3 with spatial
resolution down to $\sim$0\farc03 and $\sim$0\farc06,
respectively. The high sensitivity of ALMA also led to the first
detection of \hce, \hcg, and \hsd\ emission in this source. We have
modelled our data using the non-LTE radiative transfer code MORELI,
which has enabled us to constrain the morphology, kinematics and
physical conditions at the core of this outstanding object. In the
following, we summarise our observational and modelling results:

   \begin{itemize}
      \item Our continuum maps show a
        pinched-waist squared-like morphology reminiscent of the
        NIR nebula observed at much larger angular scales. This
        shape is consistent with the free-free emission arising in a
        biconical shell inscribed in a larger, and predominantly
        neutral, rotating disk that is illuminated and photoionized by
        the central source (as hypothesized in CSC+17). Our current
        model of the core of MWC\,922 corroborates that, indeed, the
        shape of the continuum maps stands for an underlying X-shaped
        morphology.

      \item We have computed the slope of the continuum spectrum 
        in the 1\,mm and 3\,mm bands using data from individual
        spectral windows. This suggests a previously unreported
        flattening of the free-free continuum spectrum around 1\,mm (after
        dust emission subtraction) that would imply a transition from
        optically thick to optically thin free-free emission near
        $\sim$2\,mm.

      \item As expected, the \htal\ and \htnal\ line emission arise
        from a region that is comparable, in shape and dimensions, to
        the continuum-emitting area. The pronounced double-peaked
        profile and large intensity of the \htal\ transition denotes the
        maser nature of this line, as already known (CSC+17).  We
        observe an overall brightening, a slight profile broadening,
        and a notable increase of the asymmetry between the blue and
        the red peak of the \htal\ profile relative to our single-dish
        observations of this transition performed 2 years earlier
        (CSC+17). The single-peaked profile of the \htnal\ transition,
        with a much less prominent maser amplification, has not varied
        significantly. The changes in the \htal\ profile can be
        attributed to variations with time of the physical properties
        within the emitting volume, which are non-linearly amplified
        in a significant amount given the maser nature of this line.

      \item Our ALMA line maps have provided direct observational
        confirmation of rotation in a near edge-on disk. The inner rim of the disk is at
        $\sim$29$(\frac{d}{1.7\,\rm{kpc}})$\,au where the gas is
        rotating at a velocity of $\sim$18\,\kms. Assuming Keplerian
        rotation, we infer a central mass inside the disk of
        $\sim$10 and $\sim$18\,\msun\ adopting a distance to
        MWC\,922 of $d$=1.7 and 3\,kpc, respectively.

      \item Our data unveil the presence of a fast ($\sim$100\,\kms)
        bipolar ejection orthogonal to the disk. In the \htal\ maps,
        with the highest S/N, we identify a velocity gradient
        perpendicular to the outflow axis, which is a clear sign for
        rotation in the fast wind. This is confirmed by our model. To
        our knowledge, there are no other fast ($\gsim$100\,\kms)
        bipolar outflows with compelling observational evidence for
        rotation in the literature.

      \item We present in \S\,\ref{moreli} and Table\,\ref{t-moreli} a
        model for the ionized core of MWC\,922 that satisfactorily
        reproduces our ALMA continuum and mm-RRL maps. In addition to
        the rotating disk and the fast bipolar outflow, there may be a
        slow ($\lsim$15\,\kms) wind lifting off the disk. The presence
        of a central cavity with radius $\sim$29-51\,au
        ($d$=1.7-3\,kpc) is needed in our model to reproduce the
        free-free continuum spectrum flattening near 2\,mm and the
        surface brightness depression at the center in the continuum
        and mm-RRL maps.

      \item We estimate average electron temperatures of
        \te$\sim$8000-11000\,K and densities in the range
        \dense$\approx$10$^7$-10$^6$\,\cm3 in the ionized core of
        MWC\,922. A density radial law \dense$\propto$$r^{-2.7}$, and
        a smooth variation with the latitude (decreasing from the disk
        outer boundary to the poles) is inferred. The density radial law
        could be slightly shallower than adopted, as suggested by the
        \htnal\ cubes, which show that the emission from this
        transition arises at slightly larger distances from
        the center than predicted by our model.

        \item We deduce a mass-loss rate of the fast bipolar wind of
          \mloss$_{\rm fast}$$\sim$few\ex{-6}\,\my. Assuming that a
          slow wind is also present at intermediate latitudes
          (i.e.\,bridging the ionized surface of the rotating disk and the fast
          bipolar wind), we obtain a mass-loss rate of \mloss$_{\rm
            slow}$$\sim$few\ex{-7}\,\my. The dynamical age of the
          winds traced by our observations is at most of a few
          decades, with the innermost layers of the fast wind (at
          $\sim$29-51\,au) ejected just $\sim$1.5-2.6 years ago.

        \item The total ionized mass deduced from
          the free-free continuum emission at mm-wavelengths is
          \mhii$\sim$2\ex{-5}$(\frac{d}{1.7\,\rm{kpc}})^2$\,\msun. The
          dominant mass component in the close environment of MWC\,922
          is probably that probed by the $\sim$160\,K dust
          mm-continuum emission, from which we deduce
          \mtot$\sim$8\ex{-3}$(\frac{d}{1.7\,\rm{kpc}})^2$\,\msun.

        \item In \S\,\ref{dis} we discuss the evolutionary status of
          MWC\,922 from different perspectives and provide a number of
          arguments that favour a post-main sequence stage. We believe
          that MWC\,922 could be a $\sim$15\,\msun\ blue giant (with
          \teff$\sim$20-30\,kK and \lstar$\sim$5.9\ex{4}\,\ls) located
          at $d$$\sim$3\,kpc that has recently left the main-sequence
          ($\sim$10\,Myr ago) and is probably undergoing a phase of
          mass exchange with an unseen companion.

   \end{itemize}

\begin{acknowledgements}
We thank the referee, Eric Lagadec, for his comments.  This paper
makes use of the following ALMA data: ADS/JAO.ALMA\#2016.1.00161.S and
ADS/JAO.ALMA\#2017.1.00376.S.  ALMA is a partnership of ESO
(representing its member states), NSF (USA) and NINS (Japan), together
with NRC (Canada), MOST and ASIAA (Taiwan), and KASI (Republic of
Korea), in cooperation with the Republic of Chile. The Joint ALMA
Observatory is operated by ESO, AUI/NRAO and NAOJ. The data here
presented have been reduced using CASA (ALMA default calibration
software; {\tt https://casa.nrao.edu}); data analysis was made using
the GILDAS software ({\tt http://www.iram.fr/IRAMFR/GILDAS)}. This
work has been partially supported by the Spanish MINECO through grants
AYA2016-75066-C2-1-P, AYA2016-78994-P, ESP2015-65597-C4-1-R, and
ESP2017-86582-C4-1-R, and by the European Research Council through ERC
grant 610256: NANOCOSMOS. This research has made use of the The JPL
Molecular Spectroscopy catalogue, The Cologne Database for Molecular
Spectroscopy, the SIMBAD database operated at CDS (Strasbourg,
France), the NASA’s Astrophysics Data System and Aladin.

\end{acknowledgements}

%
%

\appendix

\section{Additional Figures}
\label{ap-figures}

\begin{figure*}[ht!]
   \centering 
   \includegraphics[width=0.465\hsize]{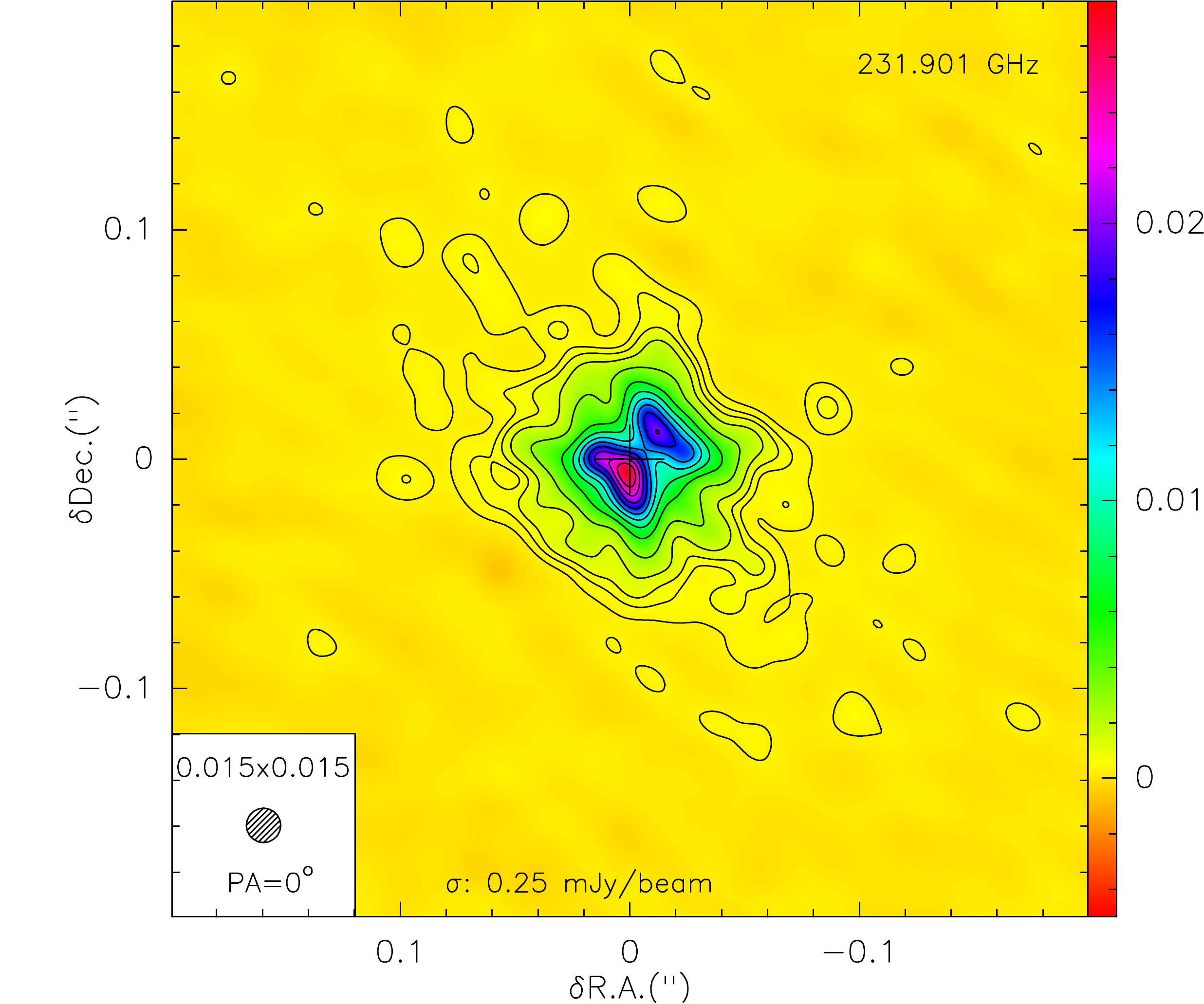} 
         \includegraphics[width=0.465\hsize]{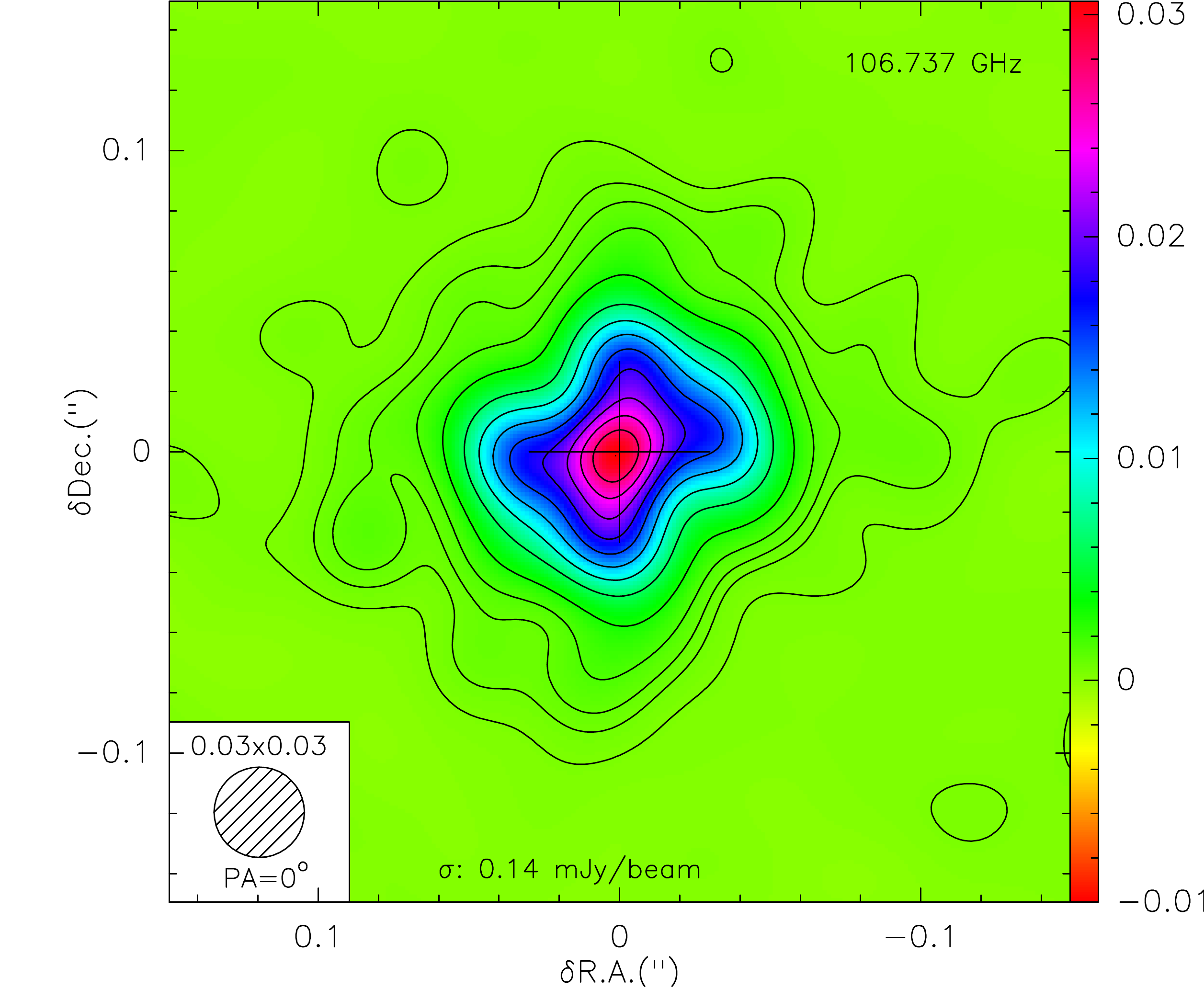} 
    \caption{Maps of the continuum emission adjacent to the H30$\alpha$
      (231.9\,GHz) and H39$\alpha$ (106.7\,GHz) lines with super-resolution. Contour levels are at 
      1\%, 2\%, 3\%, 5\%, and from 10\% to 100\% by 10\%\ of the peak (28.5 and 30.6\,m\jb\ at 1 and 3\,mm, respectively).
   \label{f-cont-HiRes}}      
   \end{figure*}

\begin{figure*}[ht!]
   \centering 
   \includegraphics[width=0.465\hsize]{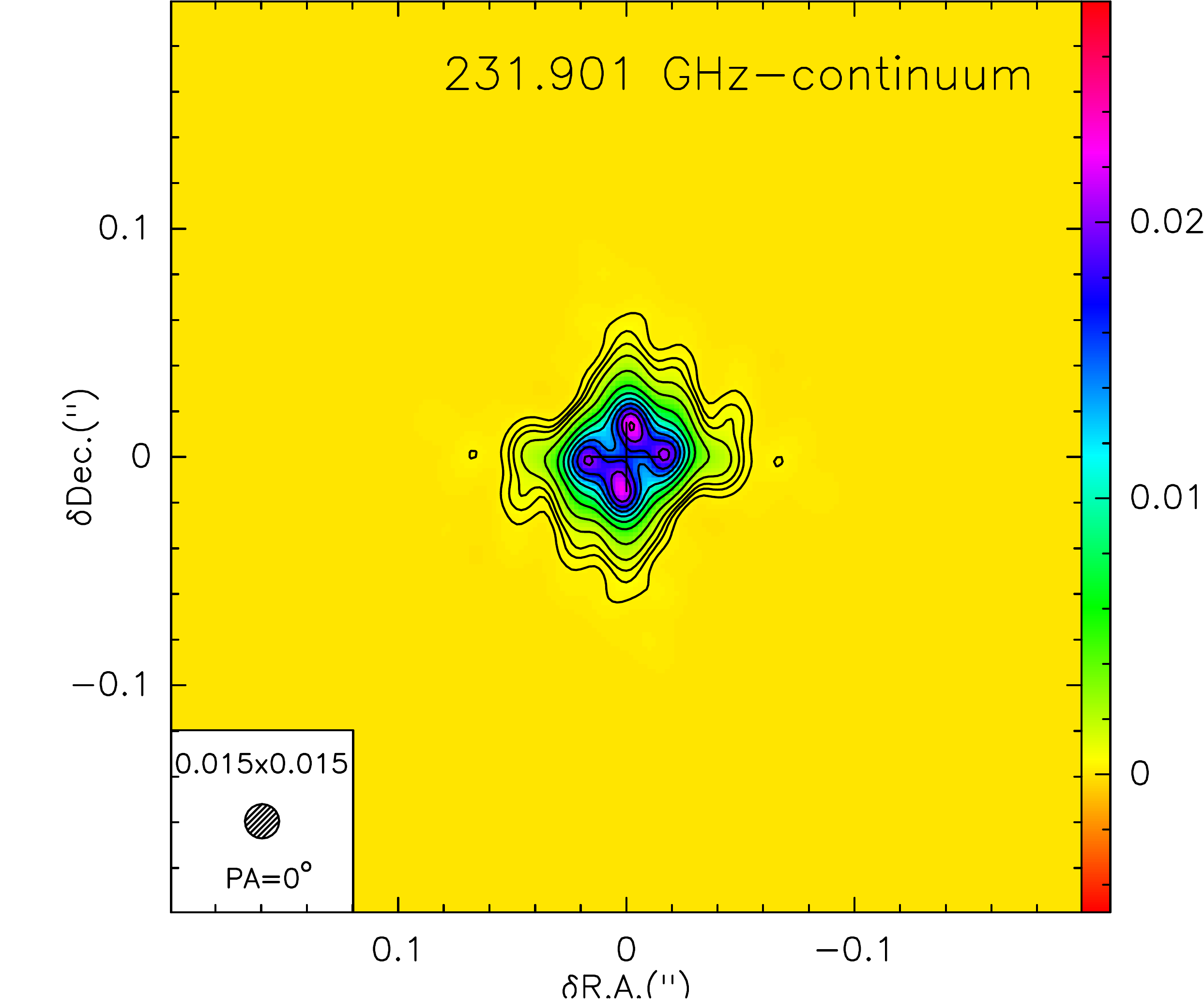}
   \includegraphics[width=0.465\hsize]{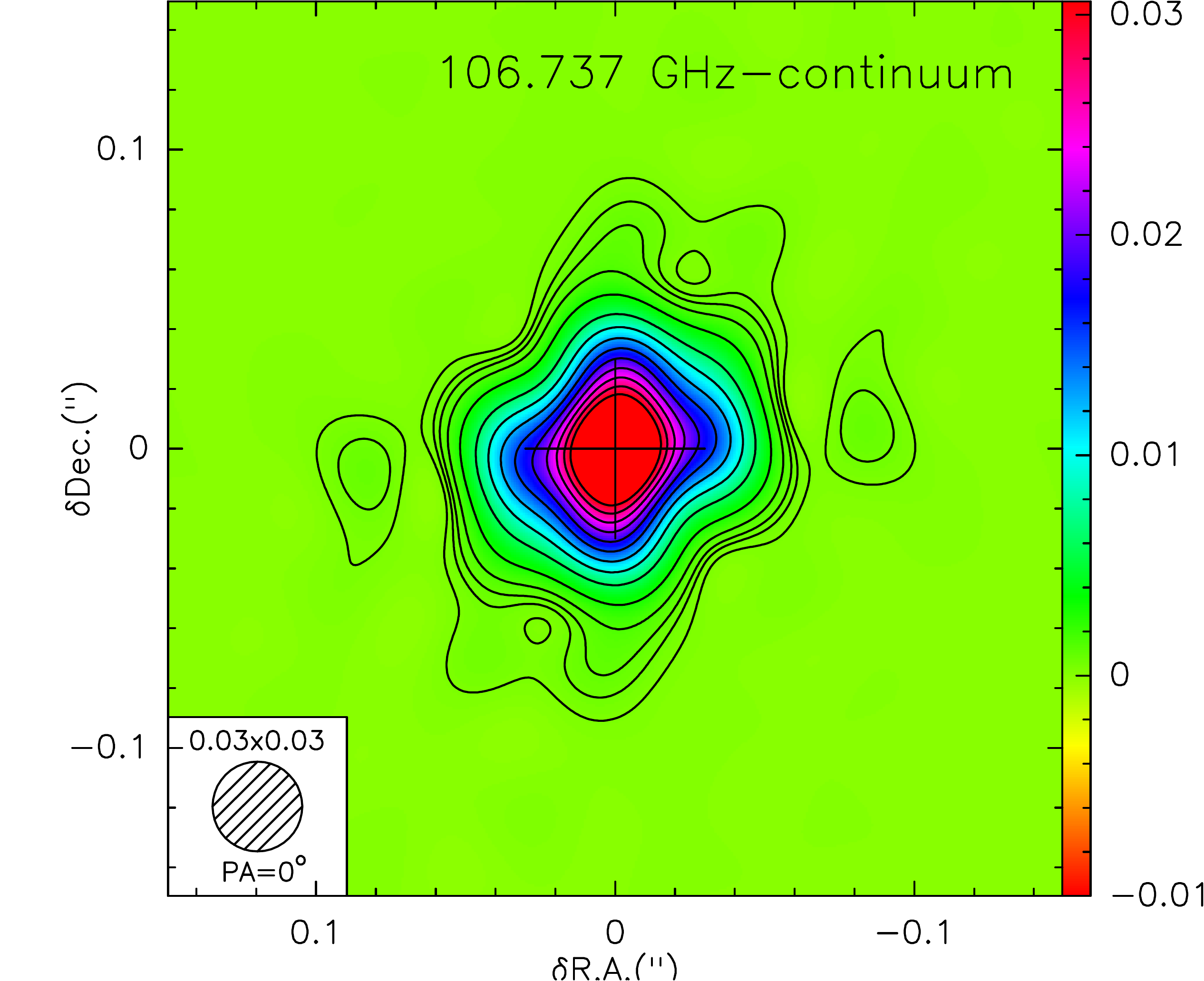} \\
   \caption{Same as in Fig.\,\ref{f-cont-HiRes} but for the MODEL. 
   \label{f-cont-HiRes-model}}      
   \end{figure*}

The Band 3 and \mbox{Band 6} continuum emission maps of MWC\,922
obtained with 15 and 30\,mas-resolution, respectively, are shown in
Fig.\,\ref{f-cont-HiRes}. Note the X-shaped morphology, reminiscent
of that of the NIR and MIR nebulosities around MWC\,922
(\S\,\ref{intro}). The predictions from our model, which is described
in \S\,\ref{moreli} and is schematically represented in
Fig.\,\ref{f-dens2D}, are show in Fig.\,\ref{f-cont-HiRes-model} with
the same angular resolution. The equatorial plane of the disk is
oriented along PA$\sim$45\degr.

\begin{figure*}[ht!]
   \centering 
   \includegraphics[width=0.99\hsize]{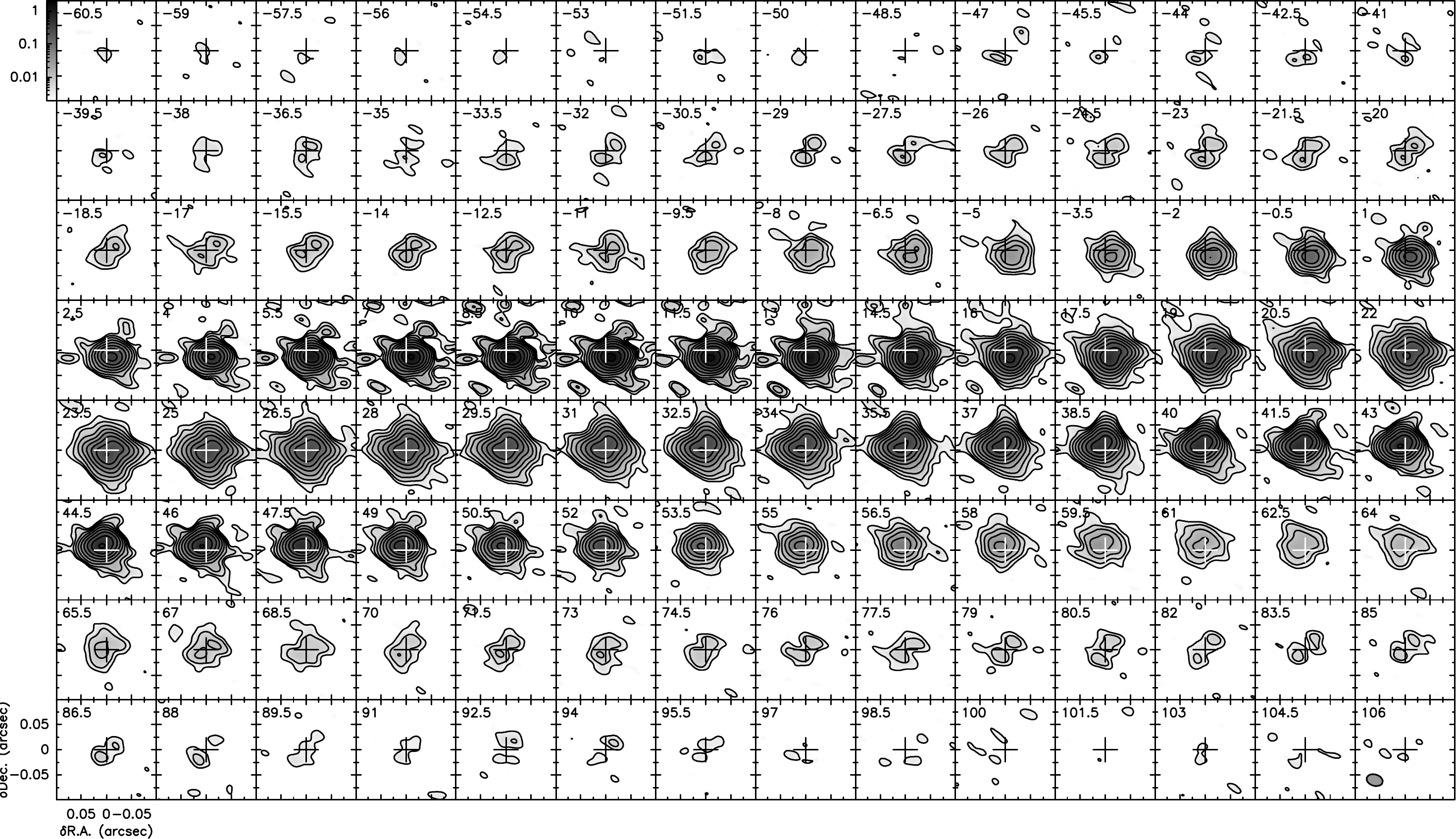} 
   \caption{ALMA velocity-channel maps of the H30$\alpha$ line in
     MWC\,922. The beam is 0\farc032$\times$0\farc022,
     PA=73\degr\ (bottom-left corner of the last panel). Level
     contours are 2.5$\times$2$^{(i-1)}$\,m\jb, for $i$=1,2,3,... 
          \label{f-cubeH30a}
        }
   \end{figure*}

\begin{figure*}[ht!]
   \centering 
   \includegraphics[width=0.99\hsize]{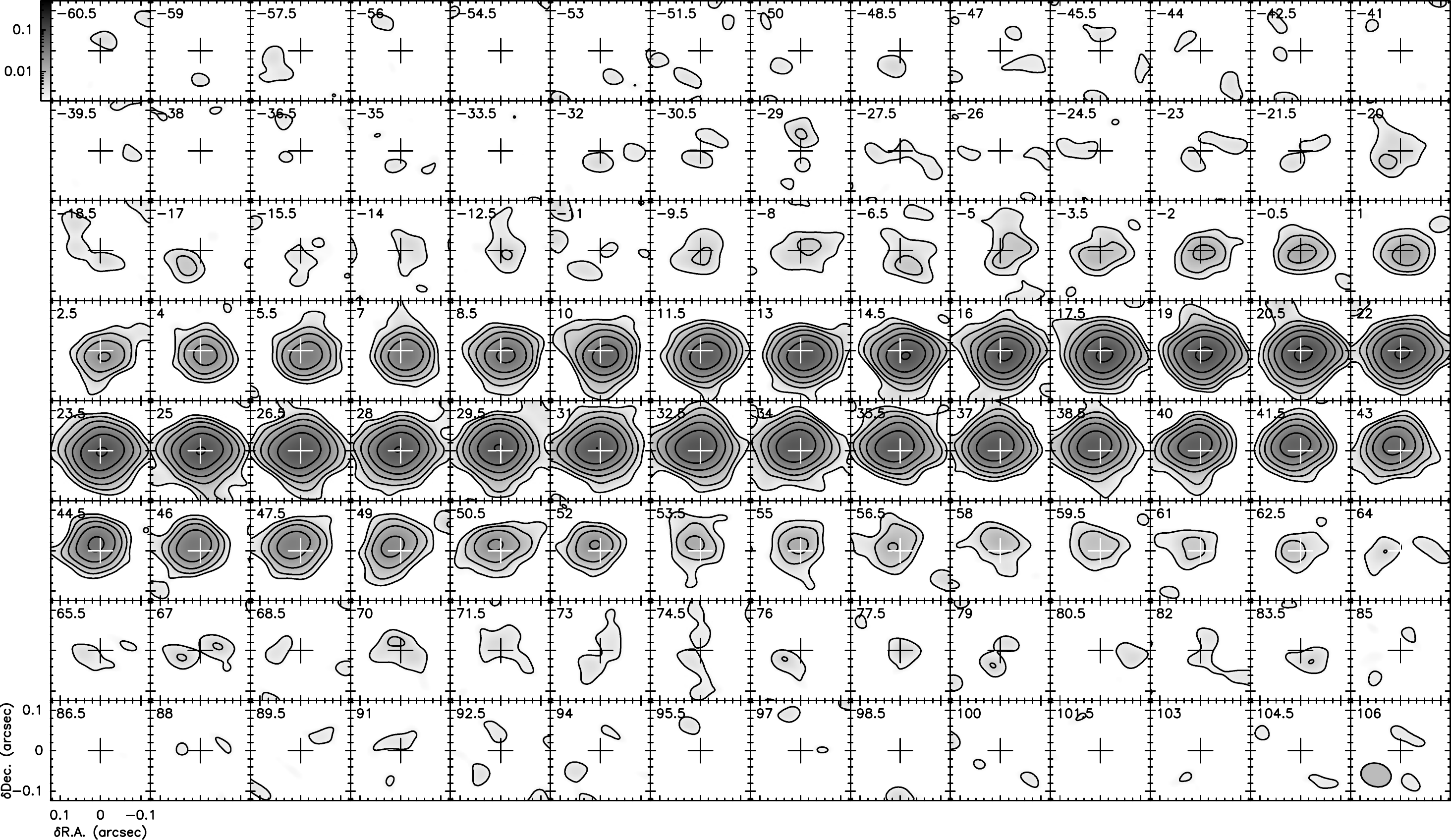} 
   \caption{ALMA velocity-channel maps of the H39$\alpha$ line in
     MWC\,922. The beam is 0\farc076$\times$0\farc058,
     PA=$-$104\degr\ (bottom-left corner of the last panel). Level
     contours are 2.5$\times$2$^{(i-1)}$\,m\jb, for $i$=1,2,3,... 
          \label{f-cubeH39a}
        }
   \end{figure*}

The velocity-channel maps of the H30$\alpha$ and H39$\alpha$ lines
towards MWC\,922 are presented in Figs.\,\ref{f-cubeH30a} and
\ref{f-cubeH39a}, respectively.

\begin{figure*}[htbp!]
     \includegraphics*[bb=30 1 723 591,width=0.30\hsize]{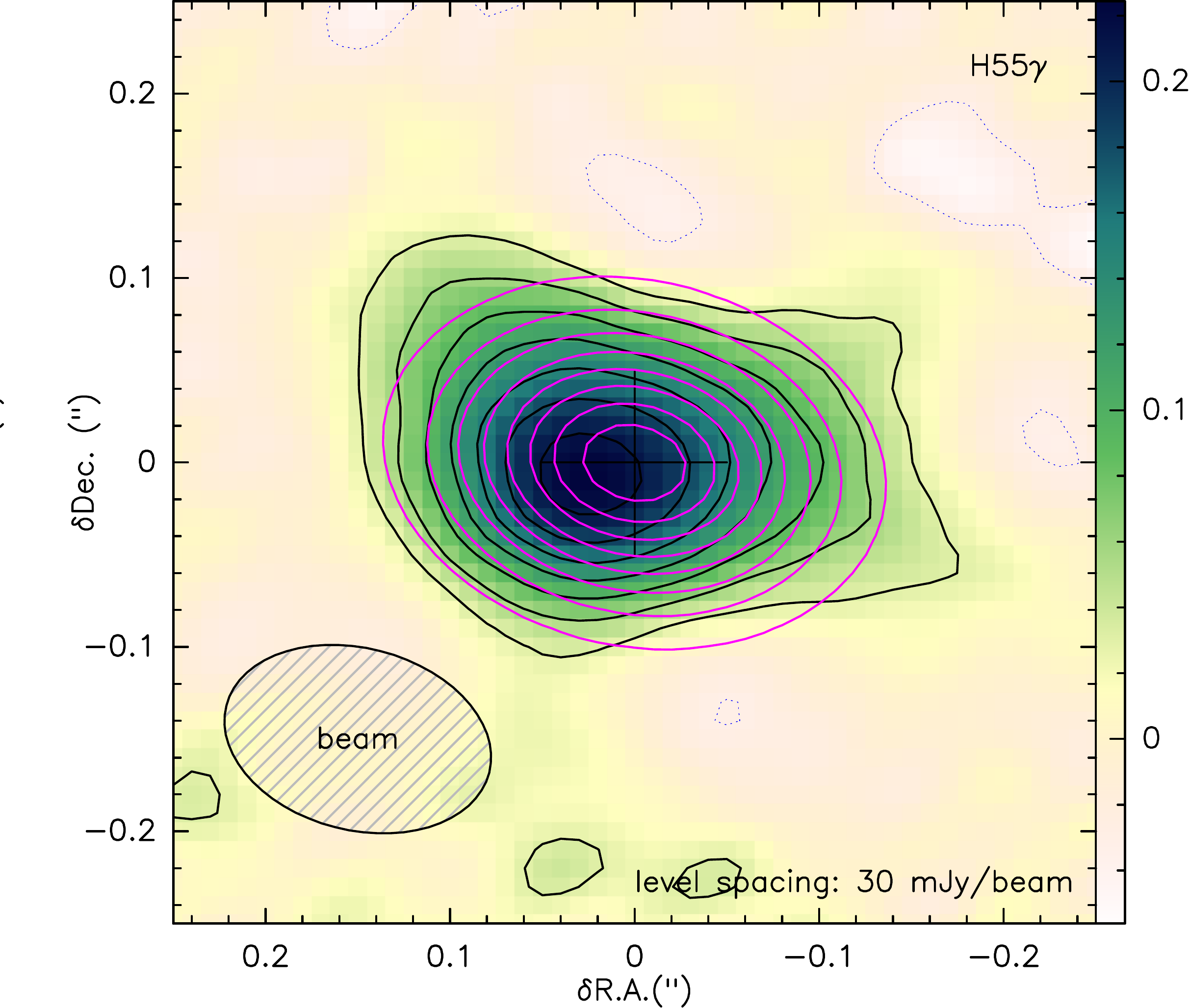}
     \includegraphics*[bb=30 1 723 591,width=0.30\hsize]{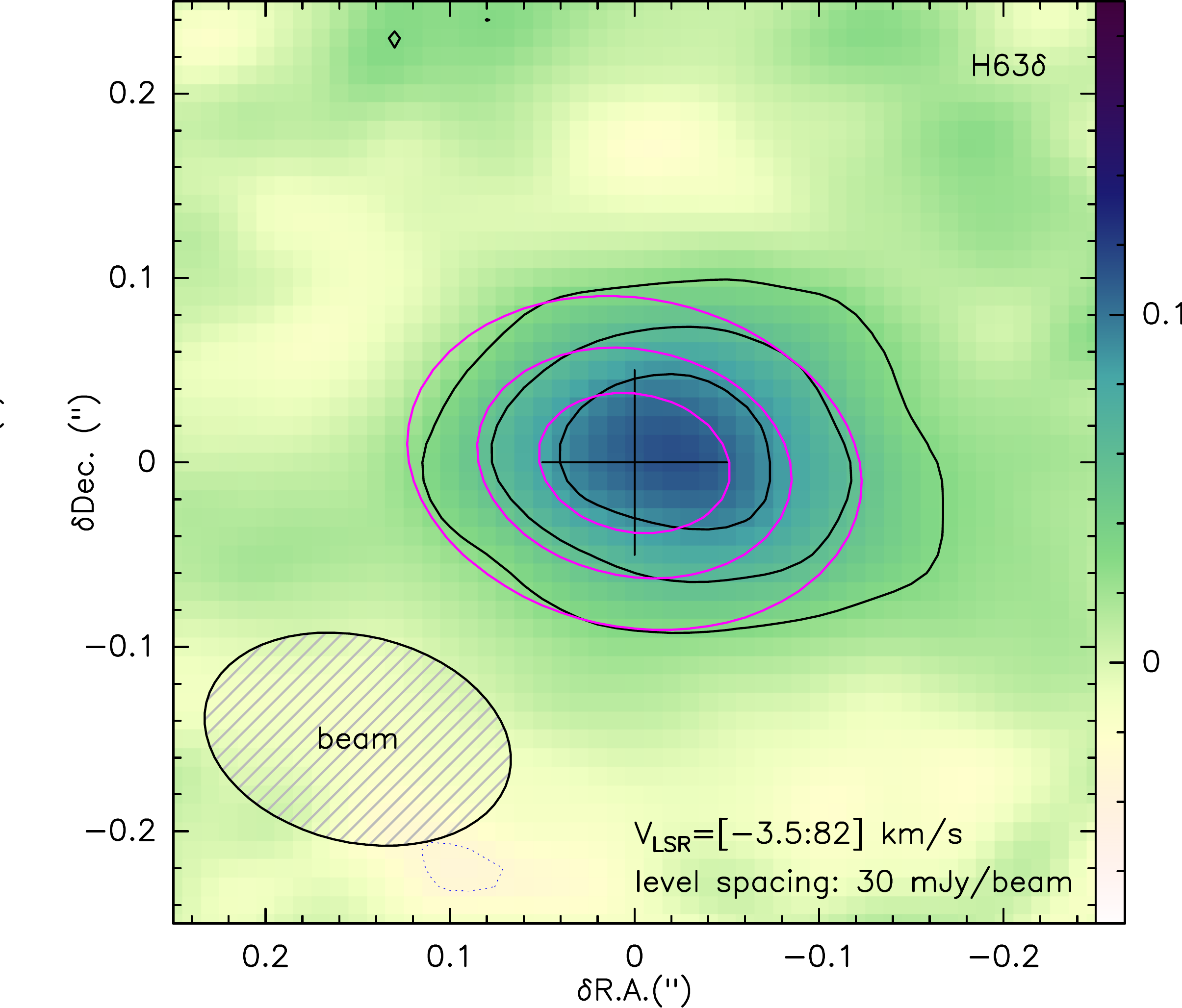}
     \includegraphics*[bb=30 1 723 591,width=0.30\hsize]{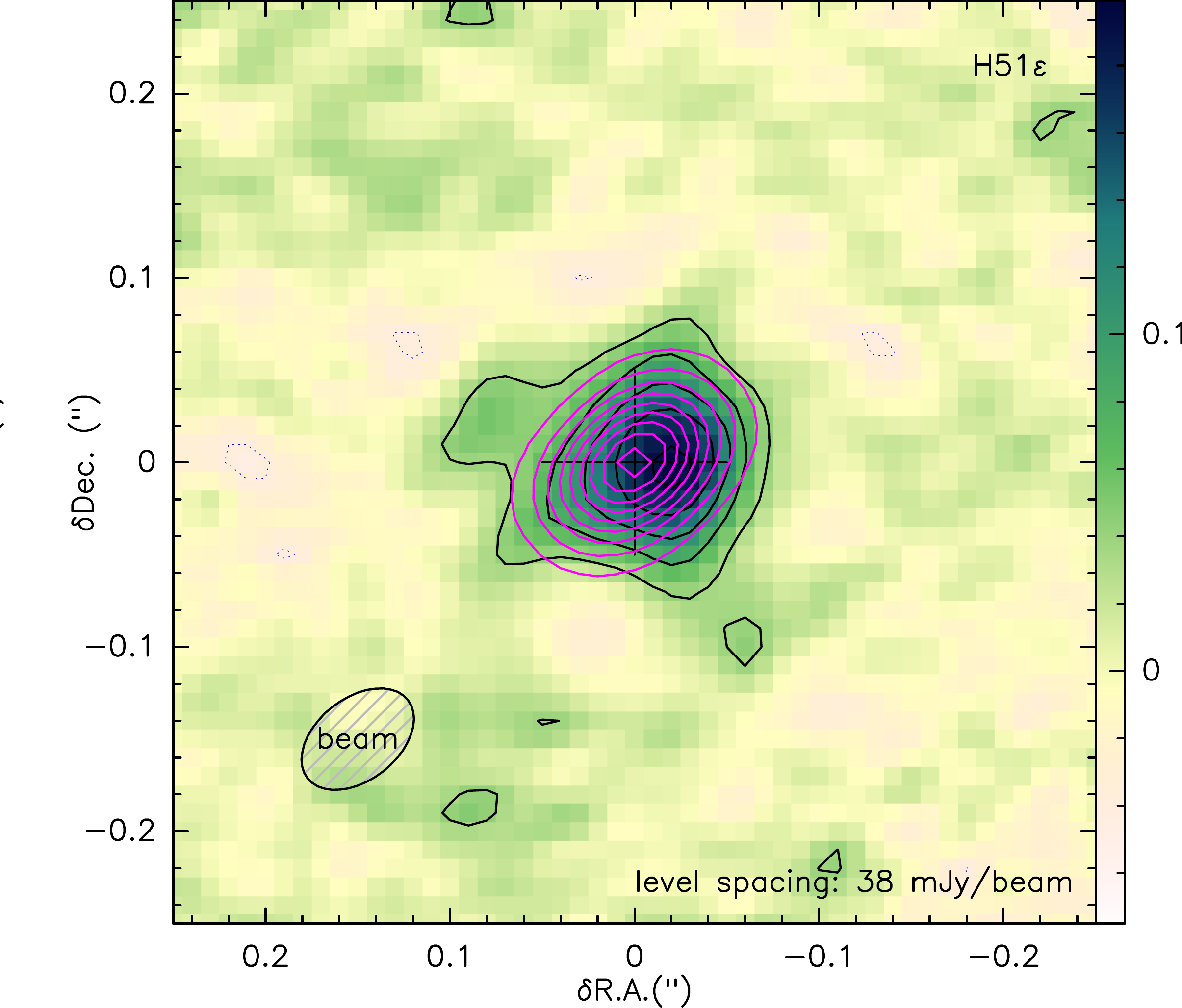}
     \includegraphics[width=0.267\hsize]{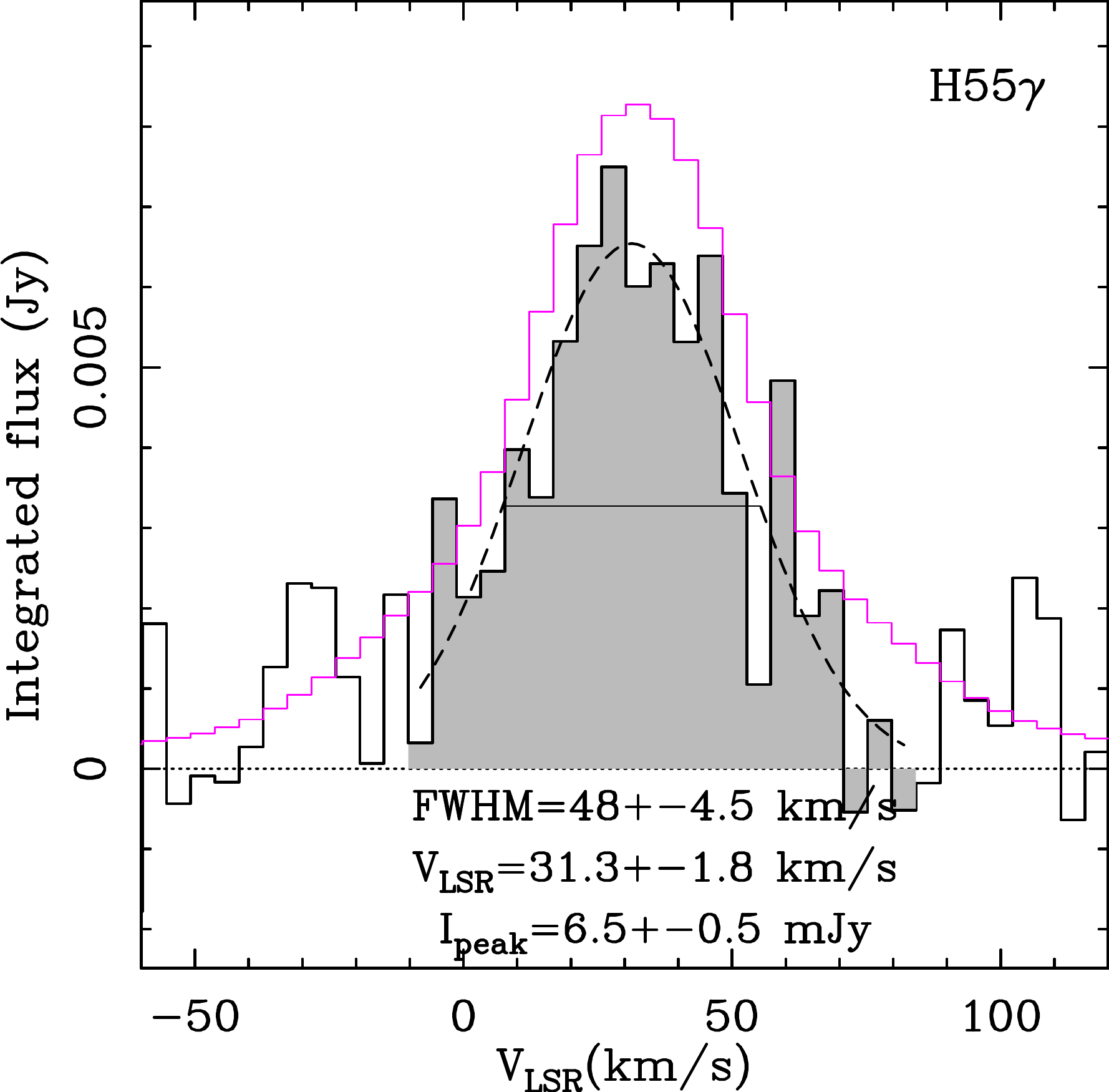}       
     \hspace{1.0cm}
     \includegraphics[width=0.267\hsize]{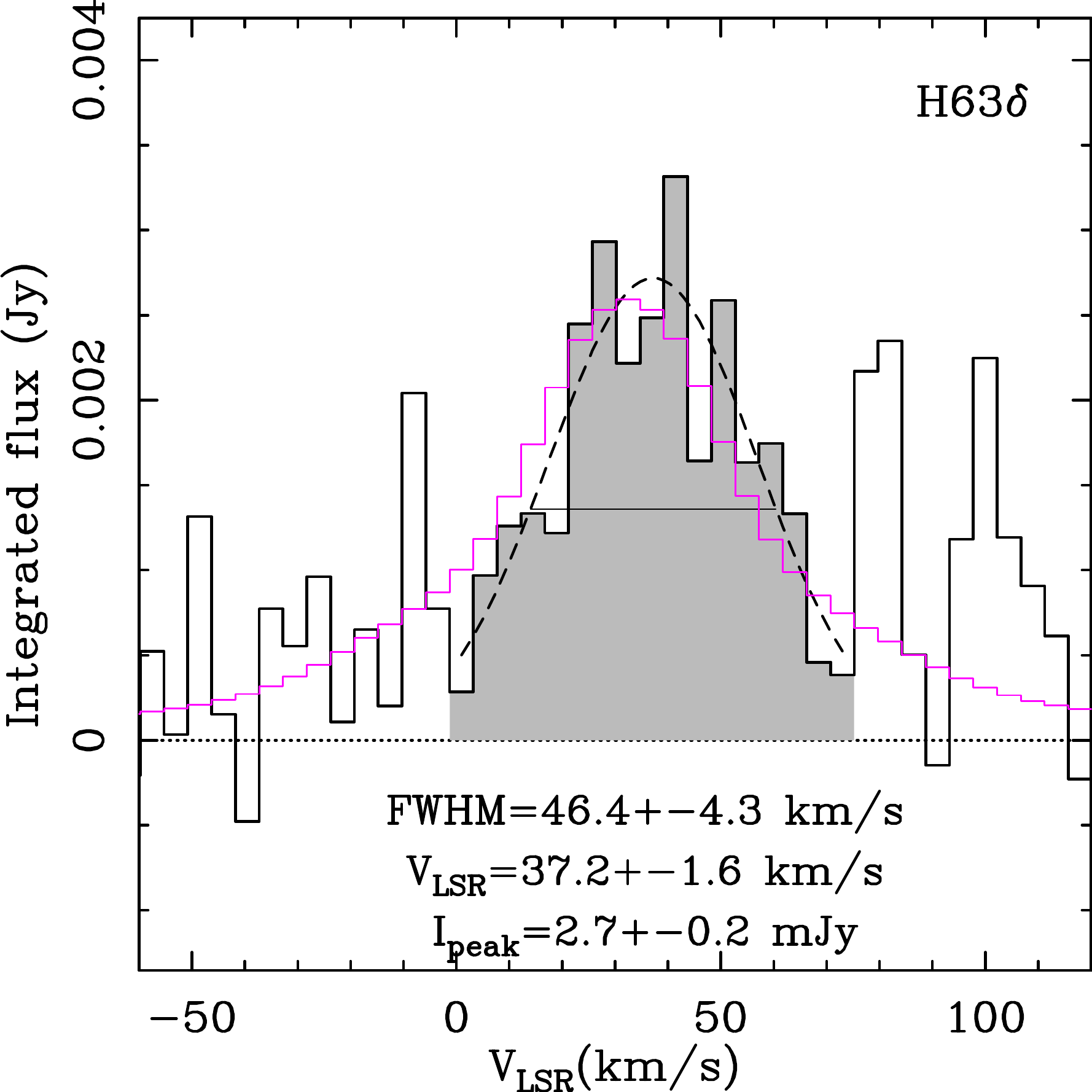}       
     \hspace{1.0cm}
     \includegraphics[width=0.267\hsize]{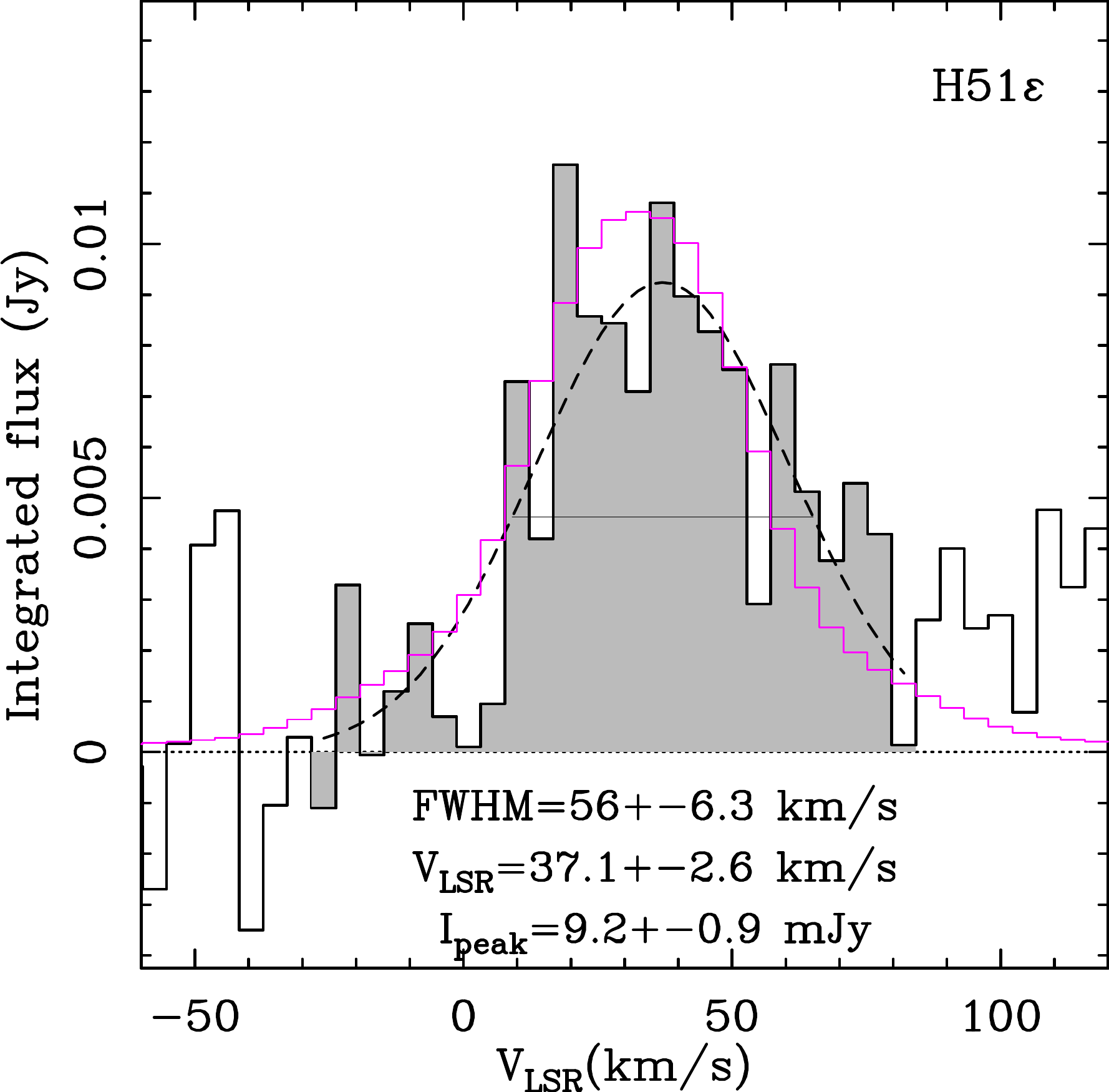}
     \caption{Other mm-RRLs detected with ALMA: H55$\gamma$,
       H63$\delta$, and H51$\epsilon$. Maps (natural weighting)
       integrated over the \vlsr=[$-$3.5:82]\,\kms\ range (top) and
       spectral profiles (bottom) are shown together with the predictions (in pink) from our
       model (\S\,\ref{moreli} and Table\,\ref{t-moreli}).
     \label{f-otras}}      
   \end{figure*}

In Fig.\,\ref{f-otras} a summary of the observational results for the
weak non-$\alpha$ mm-RRLs detected in MWC\,922 together with our model
predictions is offered.


\begin{figure*}[htbp!]
     \centering
     \includegraphics*[bb= 0 0 644 594,width=0.26\hsize]{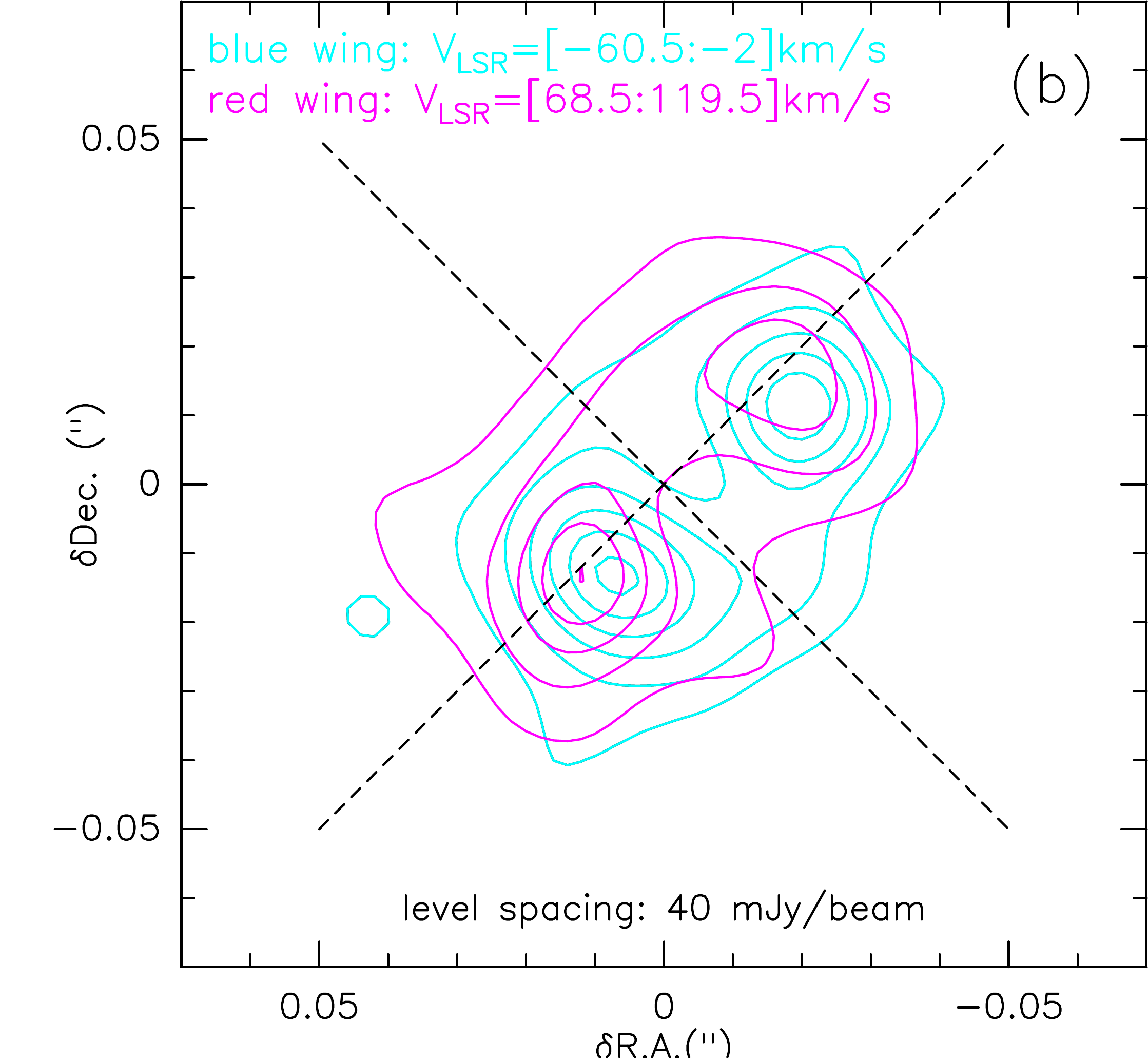}
 \includegraphics*[bb= 95 0 644 594,width=0.222\hsize]{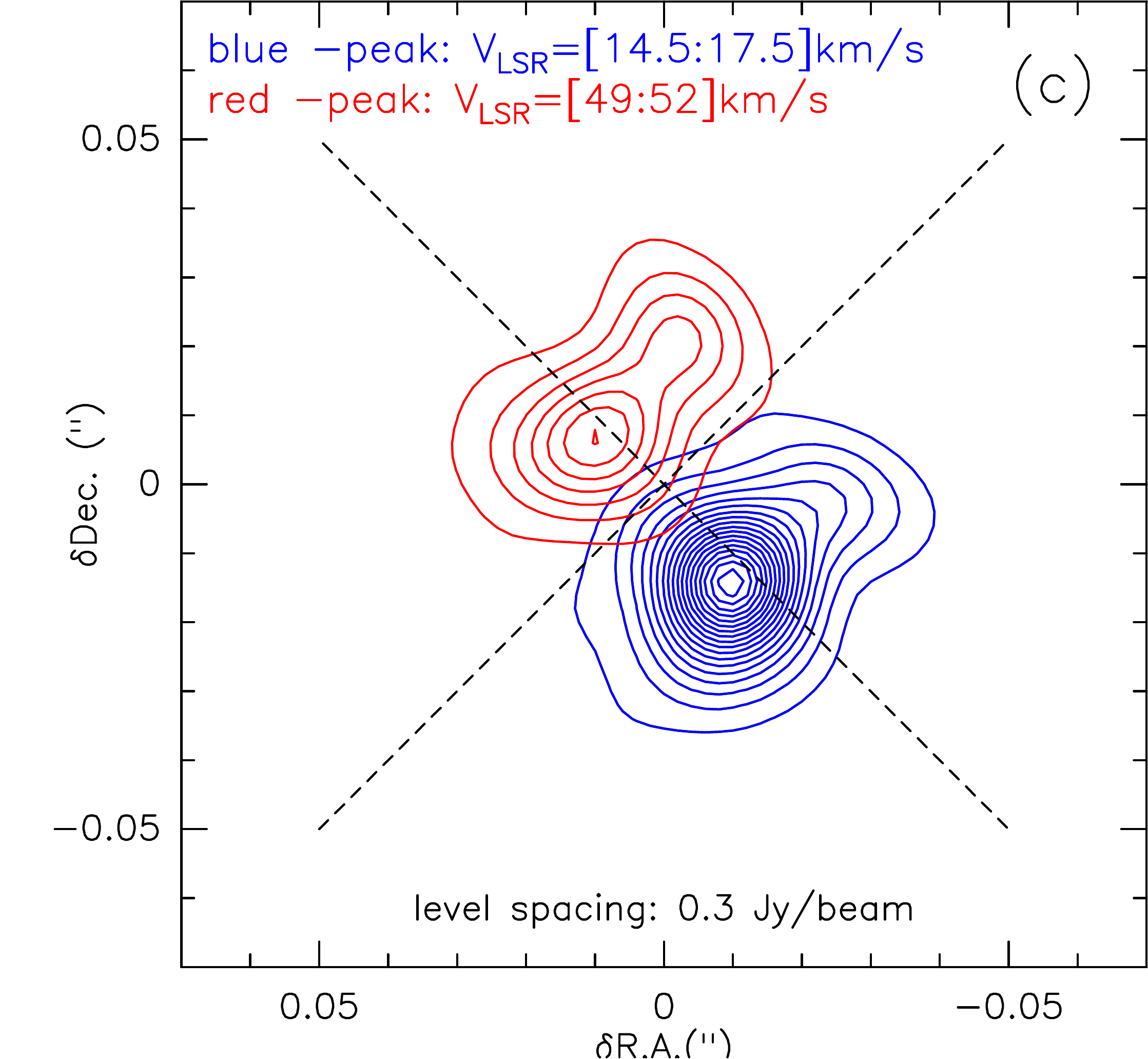}
 \includegraphics*[bb= 95 0 644 594,width=0.222\hsize]{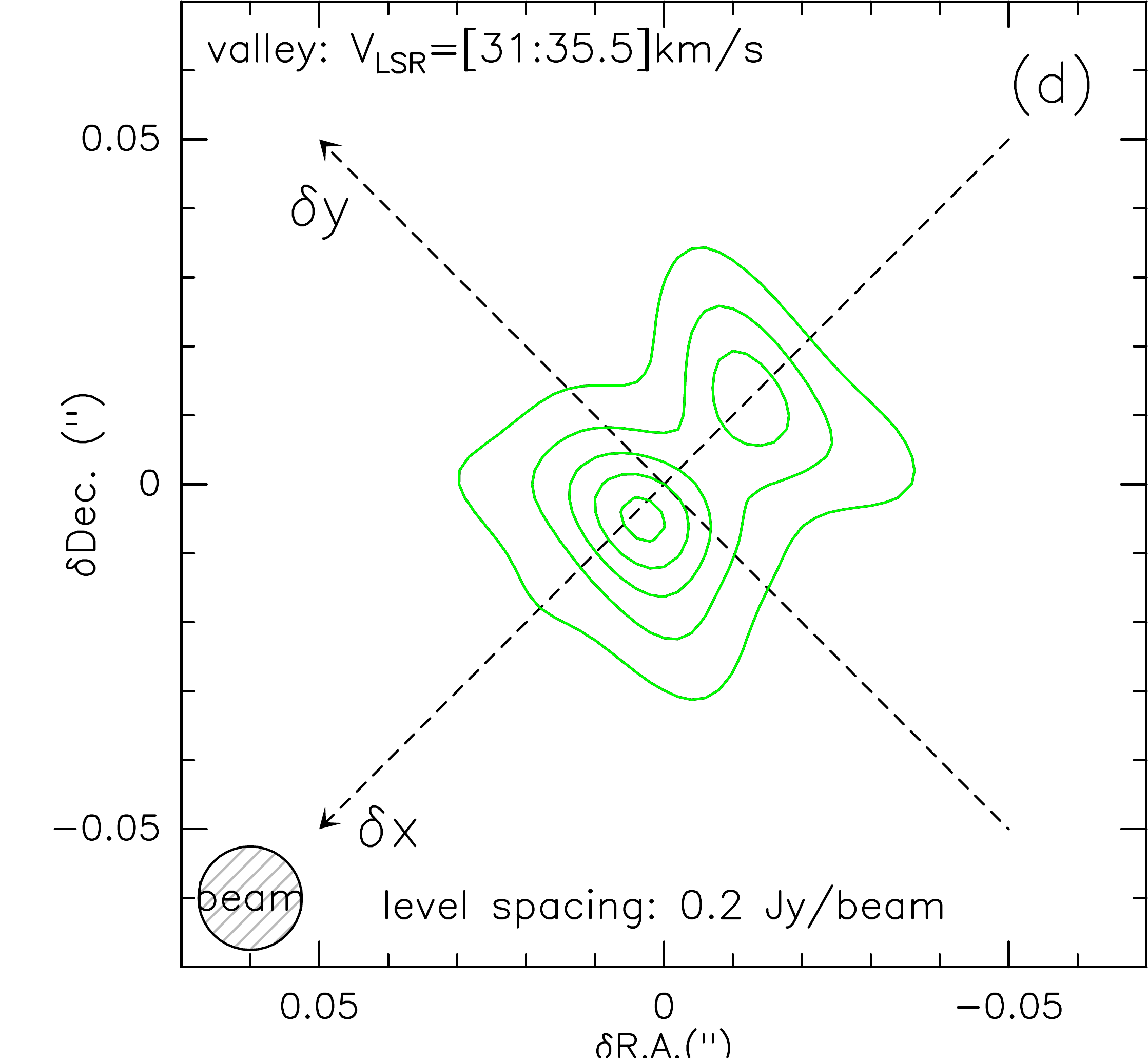} \\ 
    \includegraphics*[bb= 0 0 644 594,width=0.26\hsize]{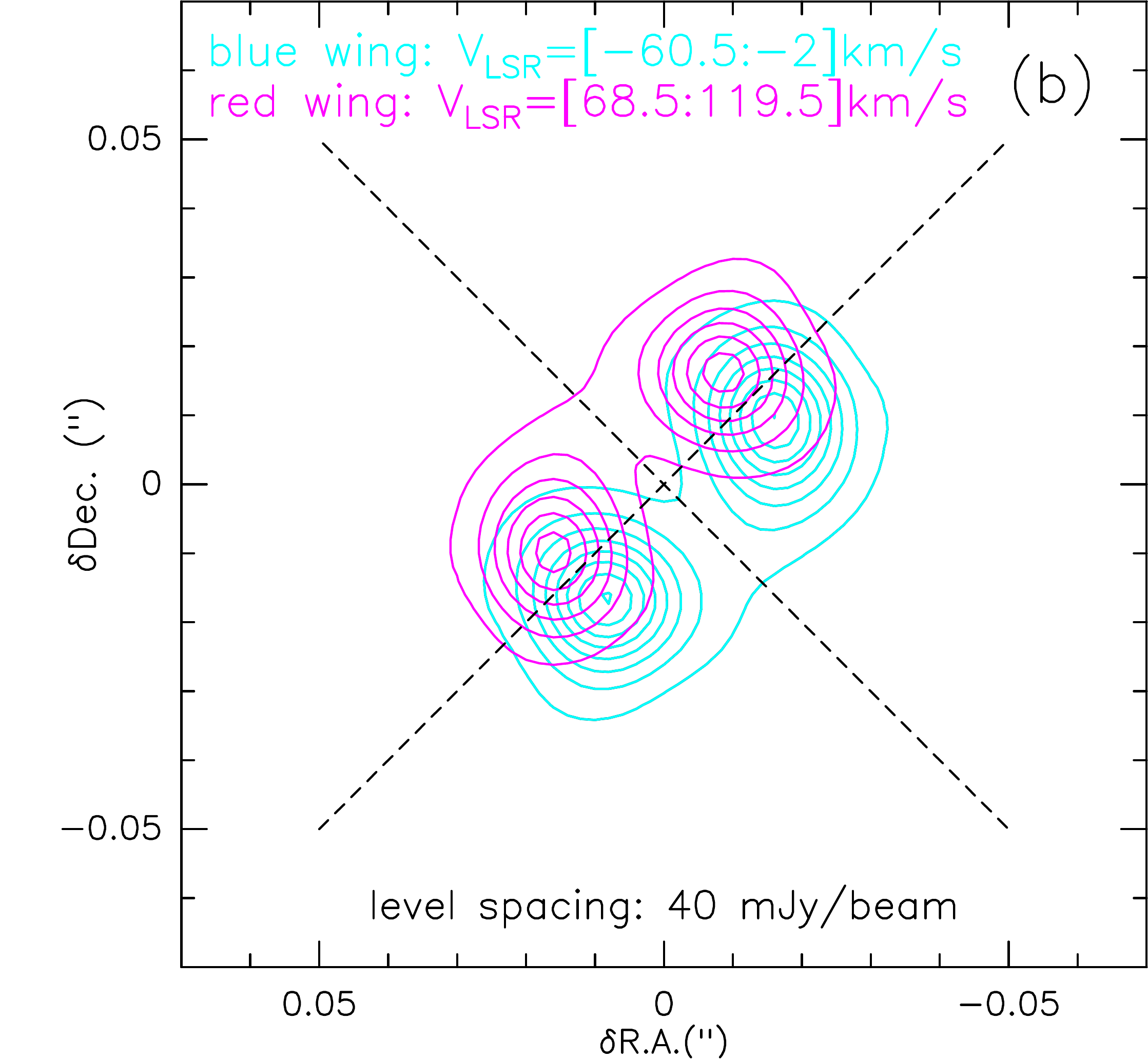}
 \includegraphics*[bb= 95 0 644 594,width=0.222\hsize]{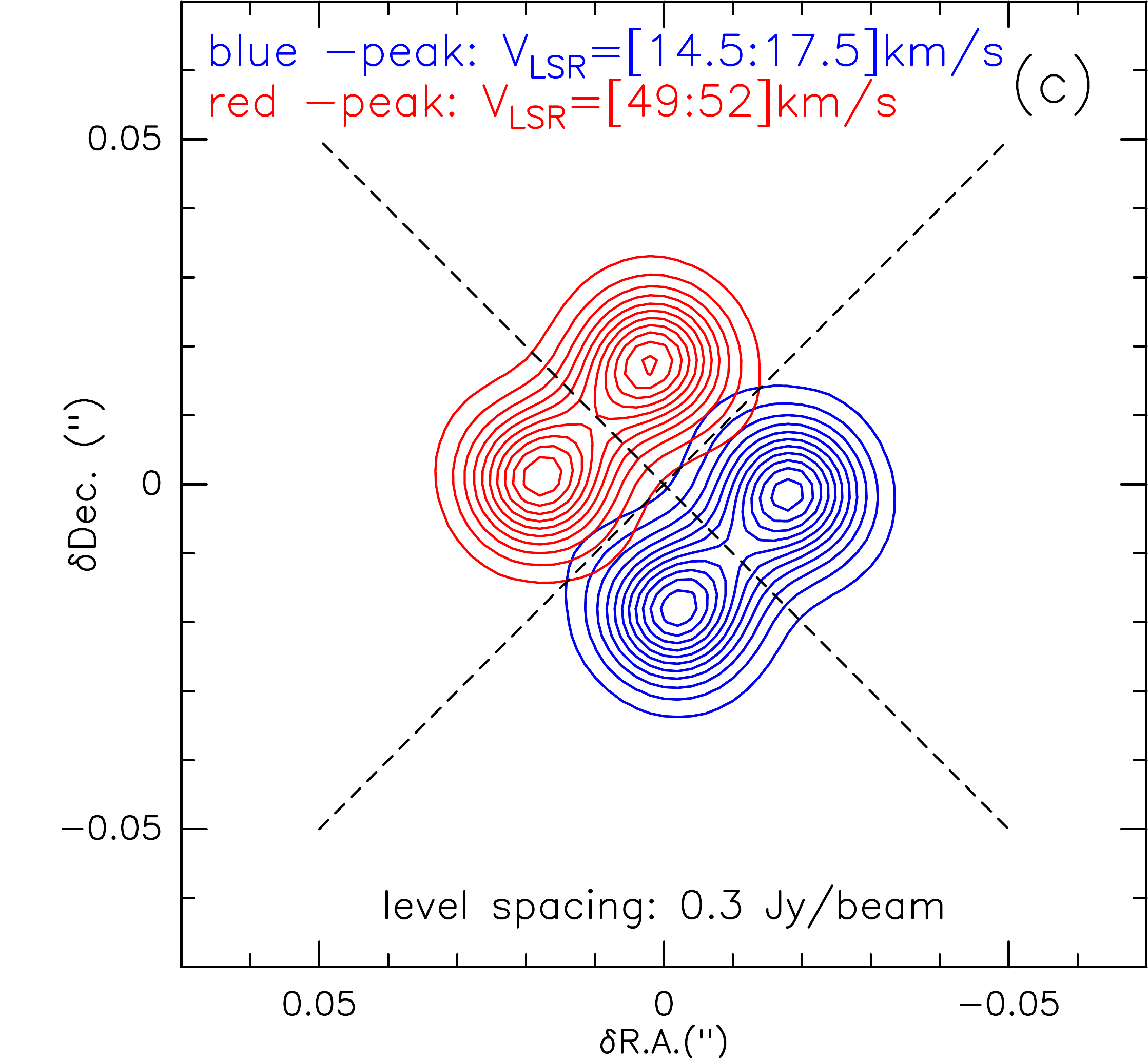}
 \includegraphics*[bb= 95 0 644 594,width=0.222\hsize]{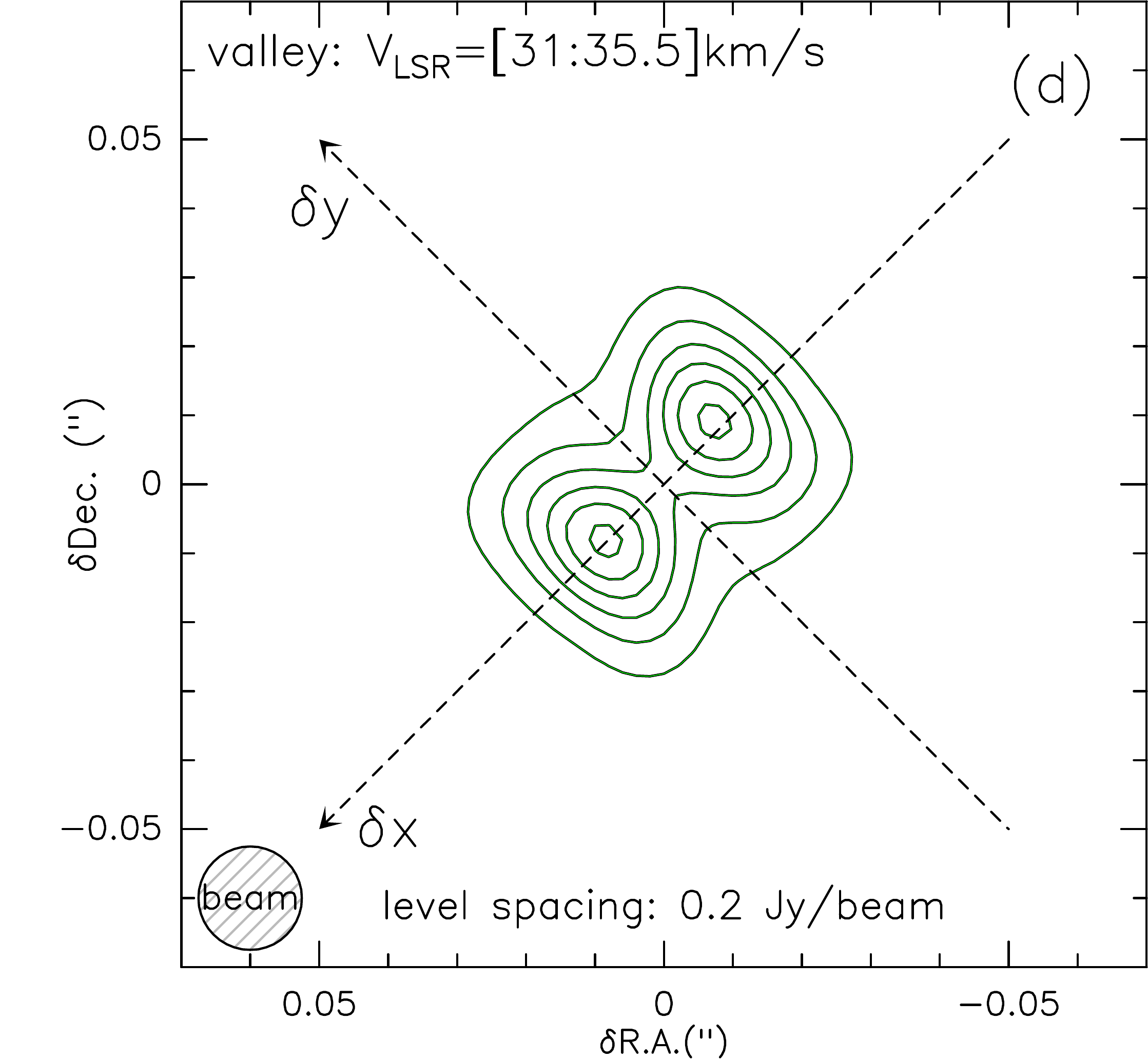} \\
 \includegraphics*[bb= 0 0 644 594,width=0.26\hsize]{h30alpha_wings_HiRes_moreli.pdf}
 \includegraphics*[bb= 95 0 644 594,width=0.222\hsize]{h30alpha_peaks_HiRes_moreli.pdf}
 \includegraphics*[bb= 95 0 644 594,width=0.222\hsize]{h30alpha_valley_HiRes_moreli.pdf} 
  \caption{Same as in Fig.\,\ref{f-h30a-all}(b,c,d) but showing the
   \htal\ super-resolution (15mas) maps (top) and the predictions from
   our model (Table\,\ref{t-moreli}, second row) and a model with no
   rotation in the fast wind (third row). 
 \label{f-h30a-all-moreli-hires}}
   \end{figure*}

\begin{figure*}[htbp!]
     \centering
     \includegraphics*[bb= 0 0 644 594,width=0.26\hsize]{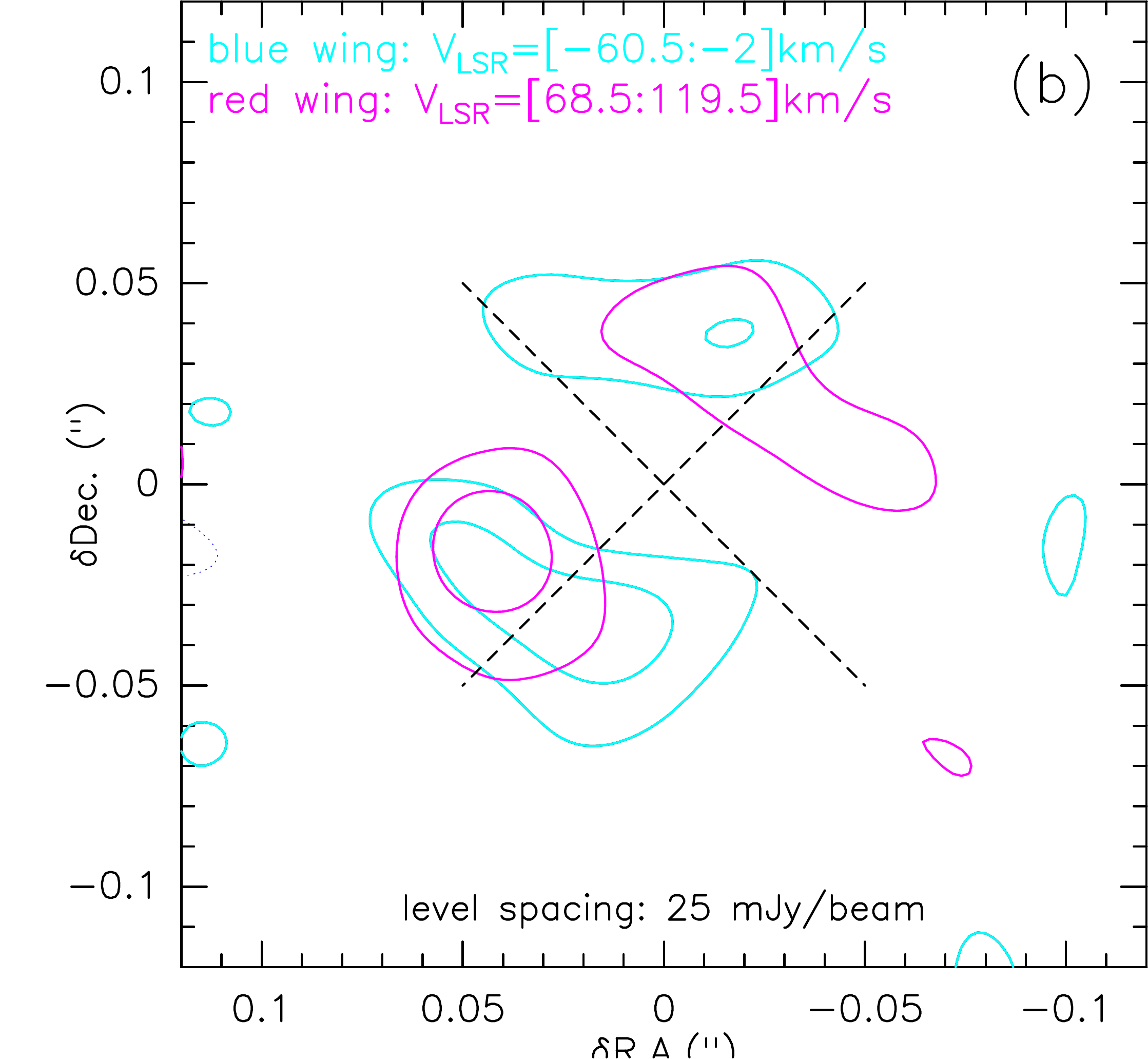}
 \includegraphics*[bb= 95 0 644 594,width=0.222\hsize]{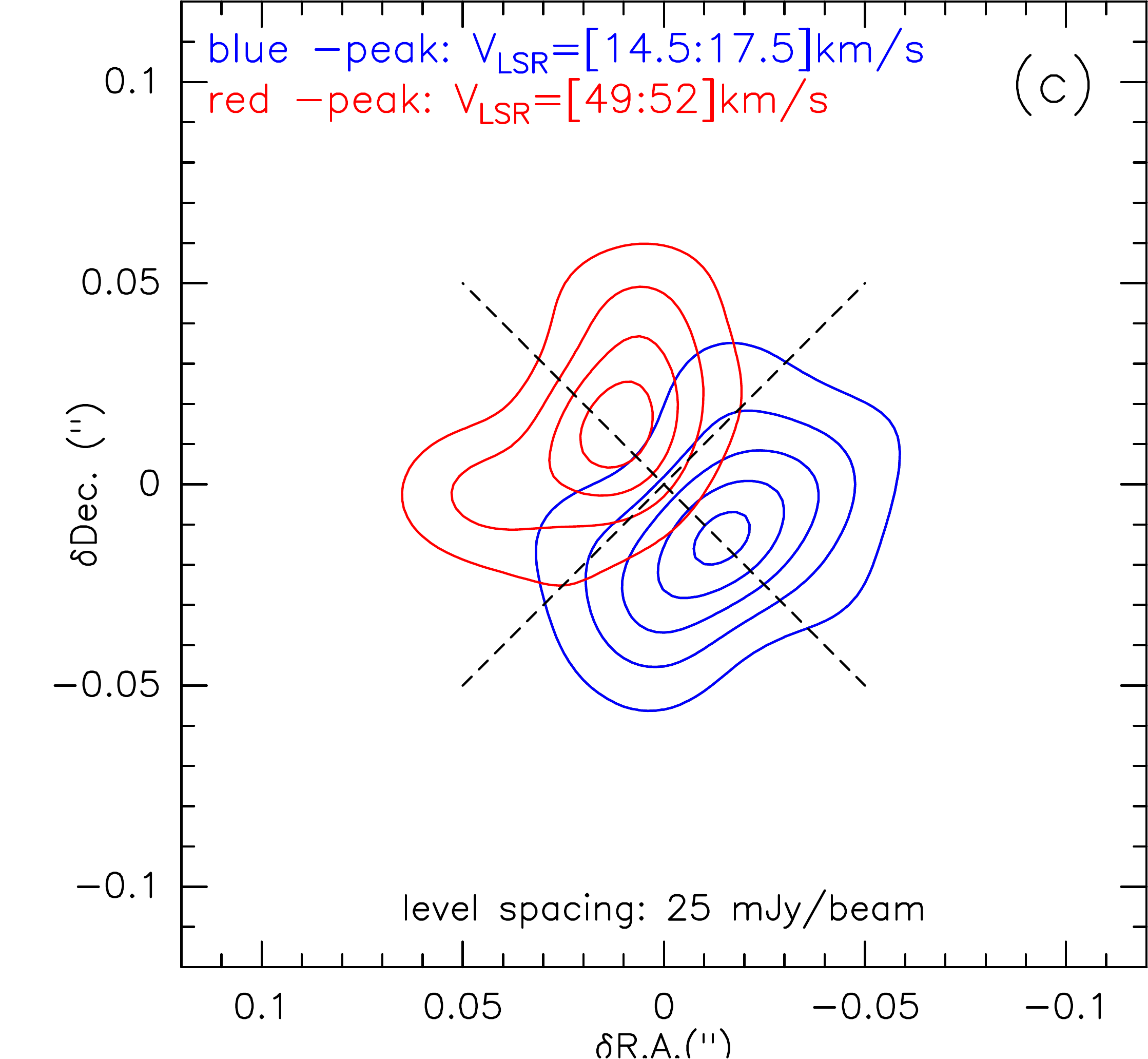}
 \includegraphics*[bb= 95 0 644 594,width=0.222\hsize]{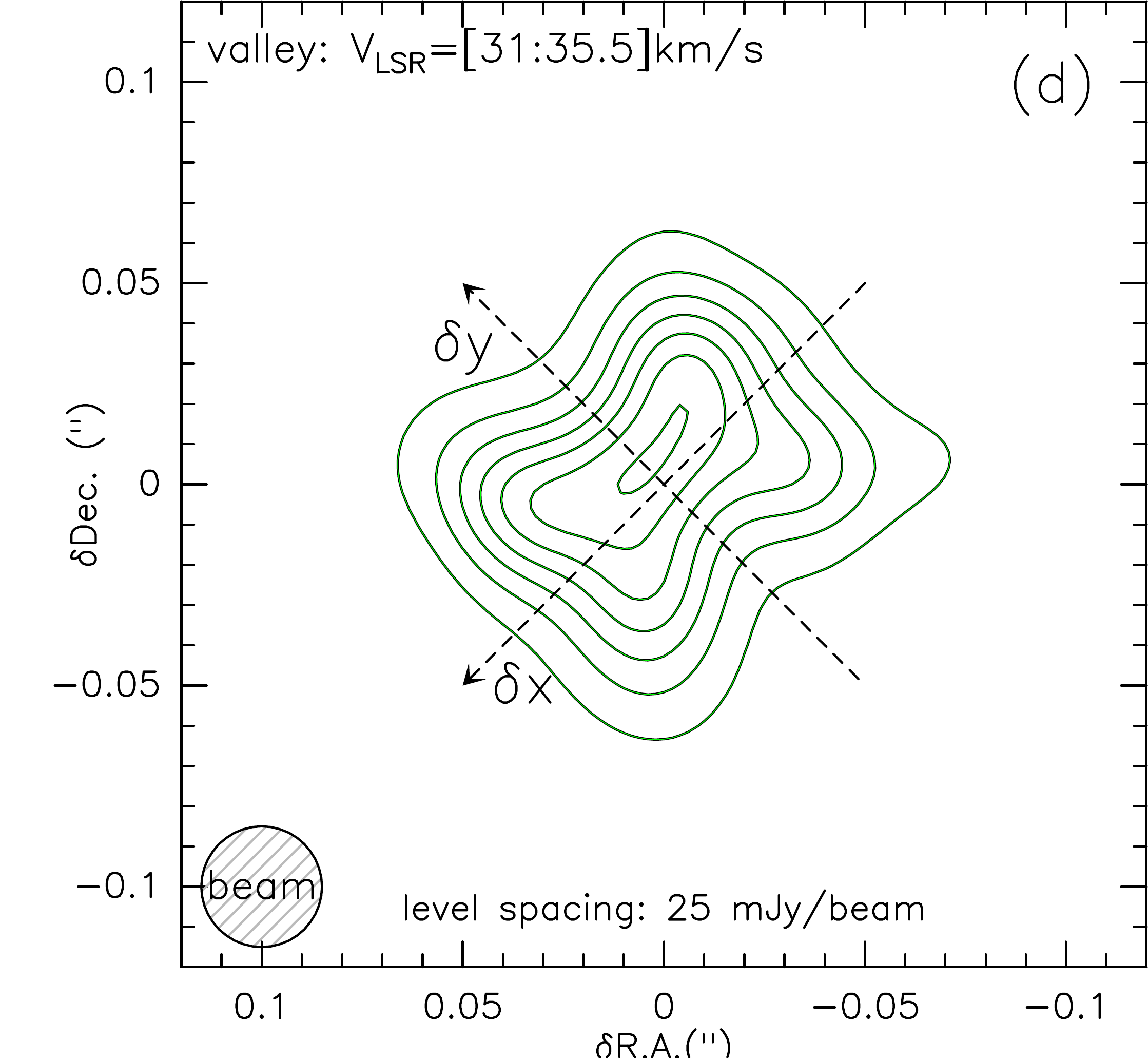} \\ 
 \includegraphics*[bb= 0 0 644 594,width=0.26\hsize]{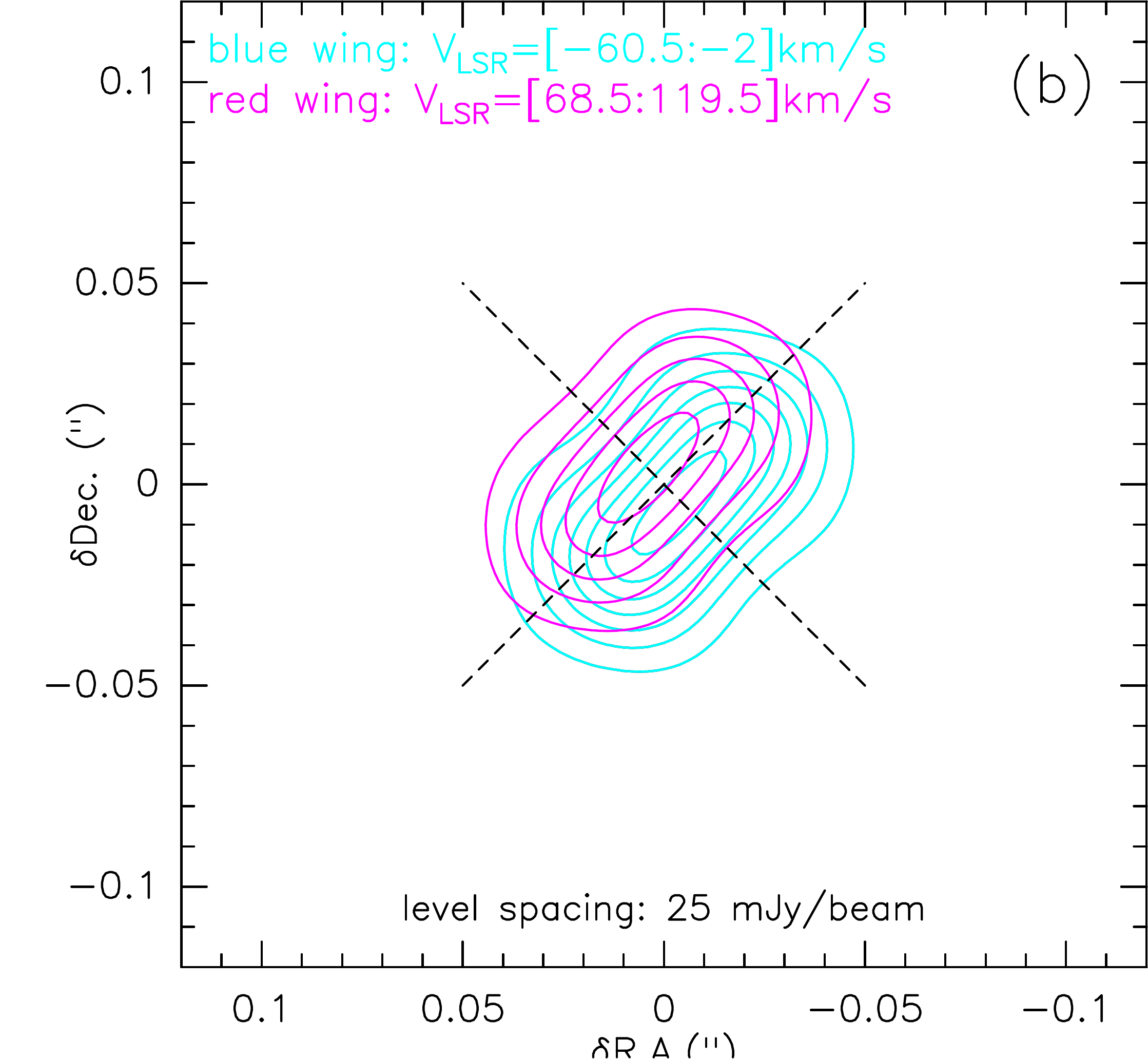}
 \includegraphics*[bb= 95 0 644 594,width=0.222\hsize]{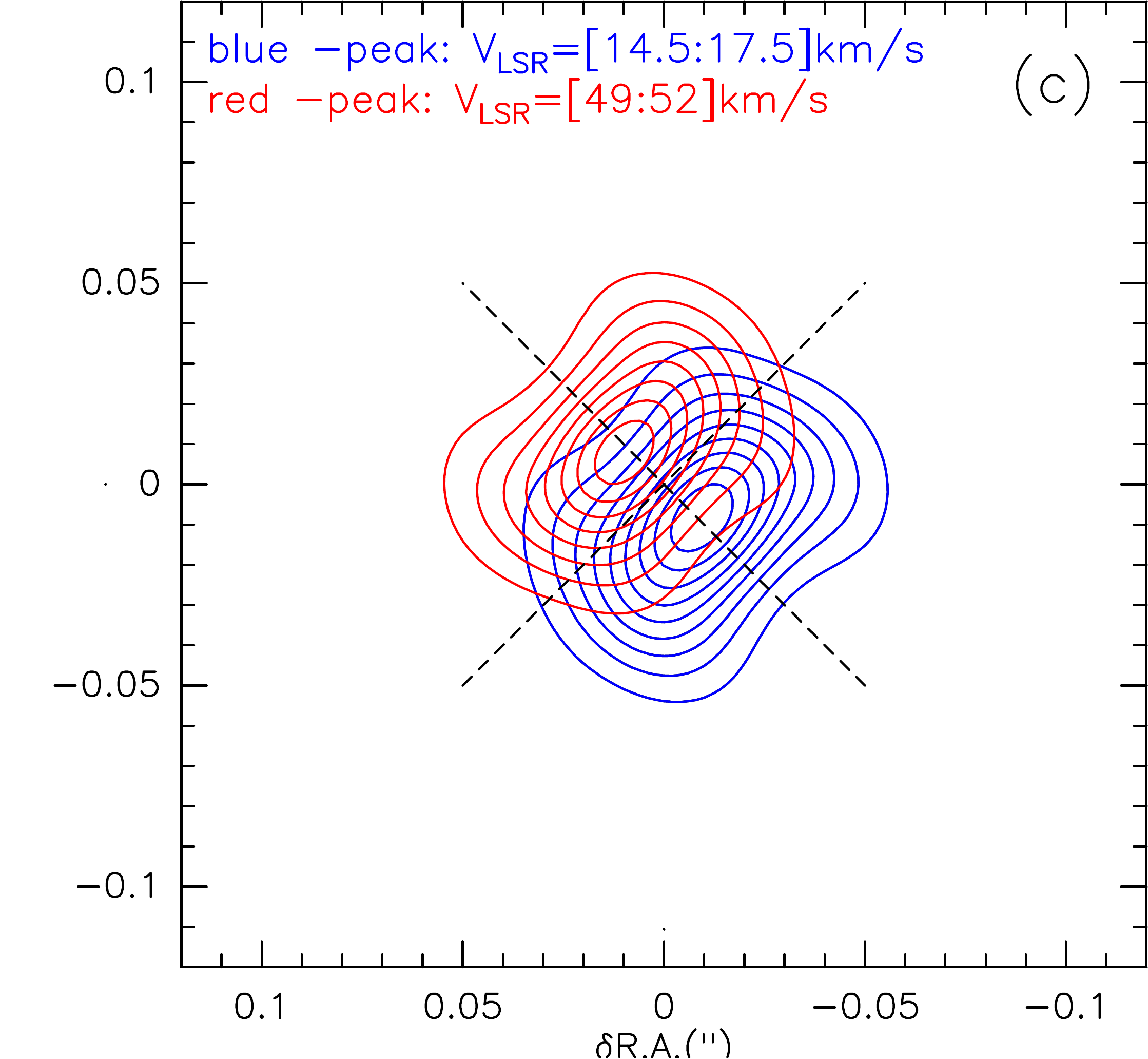}
 \includegraphics*[bb= 95 0 644 594,width=0.222\hsize]{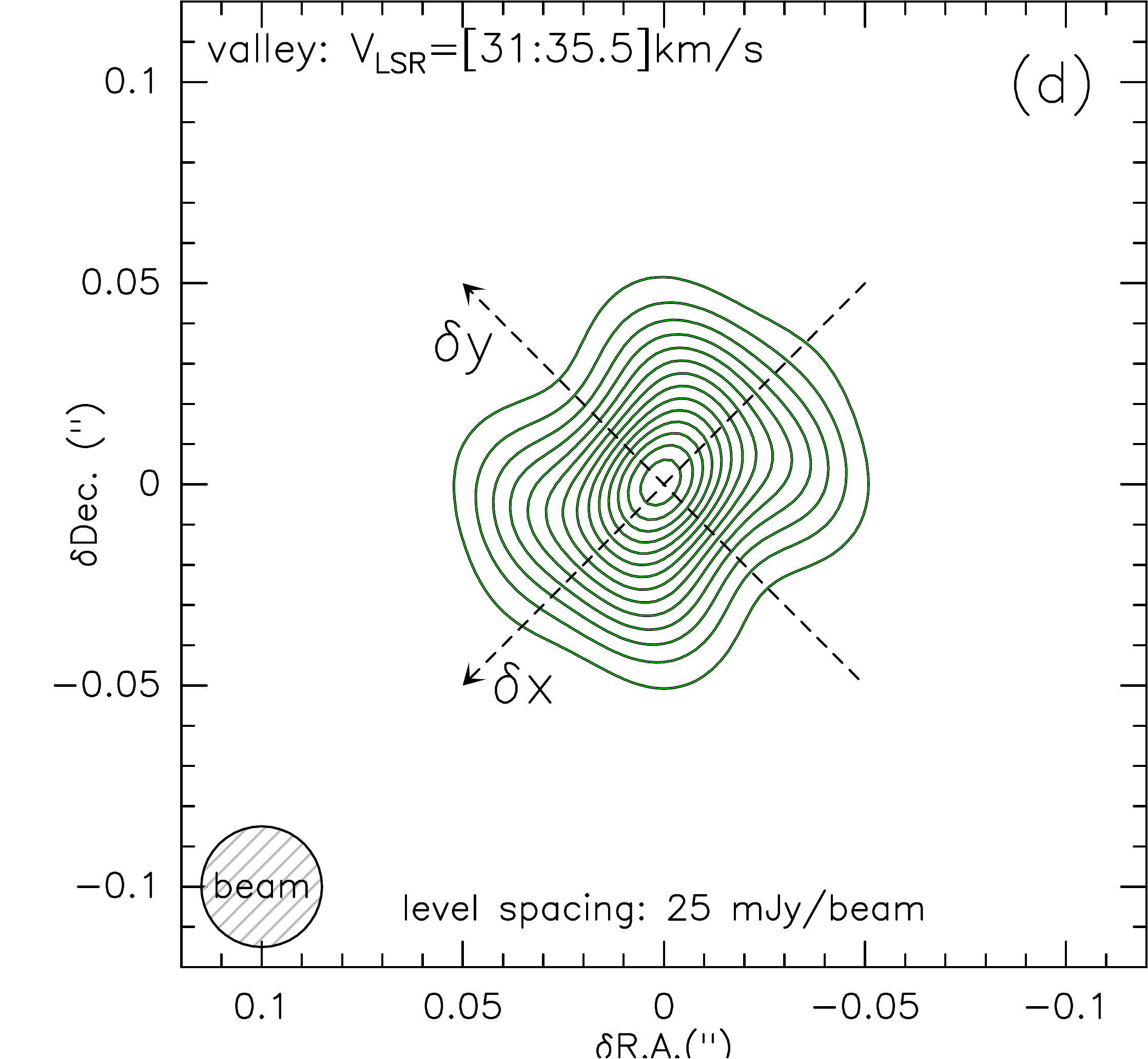}

 \caption{Same as in Fig.\,\ref{f-h39a-all}(b,c,d) but showing the
   \htnal\ super-resolution (30mas) maps (top) and model.
 \label{f-h39a-all-moreli-hires}}
   \end{figure*}

   In Figs.\,\ref{f-h30a-all-moreli-hires} and
   \ref{f-h39a-all-moreli-hires} we summarize the data-model
   comparison for the \htal\ and \htnal\ transitions with
   super-resolution (15 and 30\,mas, respectively). For \htal, we also
   display the predictions by a model where rotation has been
   suppressed in the fast wind to show that, when no
   rotation of the fast wind exist, the peaks of the red and blue-wing
   (peanut-like) emitting regions collapse on the revolution axis
   ($\delta x$), i.e.\,the velocity gradient across the fast wind
   dissapears (Fig.\,\ref{f-h30a-all-moreli-hires}b, bottom row).

   Super-resolution images of the observations and model for the
   \htnal\ line are also shown in
   Fig.\,\ref{f-h39a-all-moreli-hires}. Given the lower S/N ratio,
   rotation of the fast wind is not evident in these maps.

   We note the remarkable asymmetry of the blue-emission peak of the
   \htal\ maser line with respect to the equatorial plane ($y$ axis)
   in the observations (Fig.\,\ref{f-h30a-all-moreli-hires}c, top
   row). This asymmetry does not have a counterpart in the
   \htnal\ emission maps, which are rather symmetric about both the
   $x$- and $y$-axes (Fig.\,\ref{f-h39a-all-moreli-hires}c, top row),
   and is not reproduced either my our model. This difference denotes
   again the non-linear response of the \htal\ maser line to small
   deviations from axial symmetry of the physical conditions in the
   emitting regions.

   As already mentioned (\S\,\ref{moreli}), our model does not
   reproduce the central emission dip of the \htnal\ emission maps at
   high velocities (Fig.\,\ref{f-h39a-all-moreli-hires}b). This
   suggests that the \htnal\ emission arises in regions more distant
   from the center than predicted by our model, which could indicate a
   density radial distribution slightly shallower than adopted in the
   outer layers of the \ion{H}{ii} region best traced by this
   transition.
   


   \subsection{Additional non-mm-RRLs detections}
\begin{figure}
     \centering 
     \includegraphics*[width=0.80\hsize]{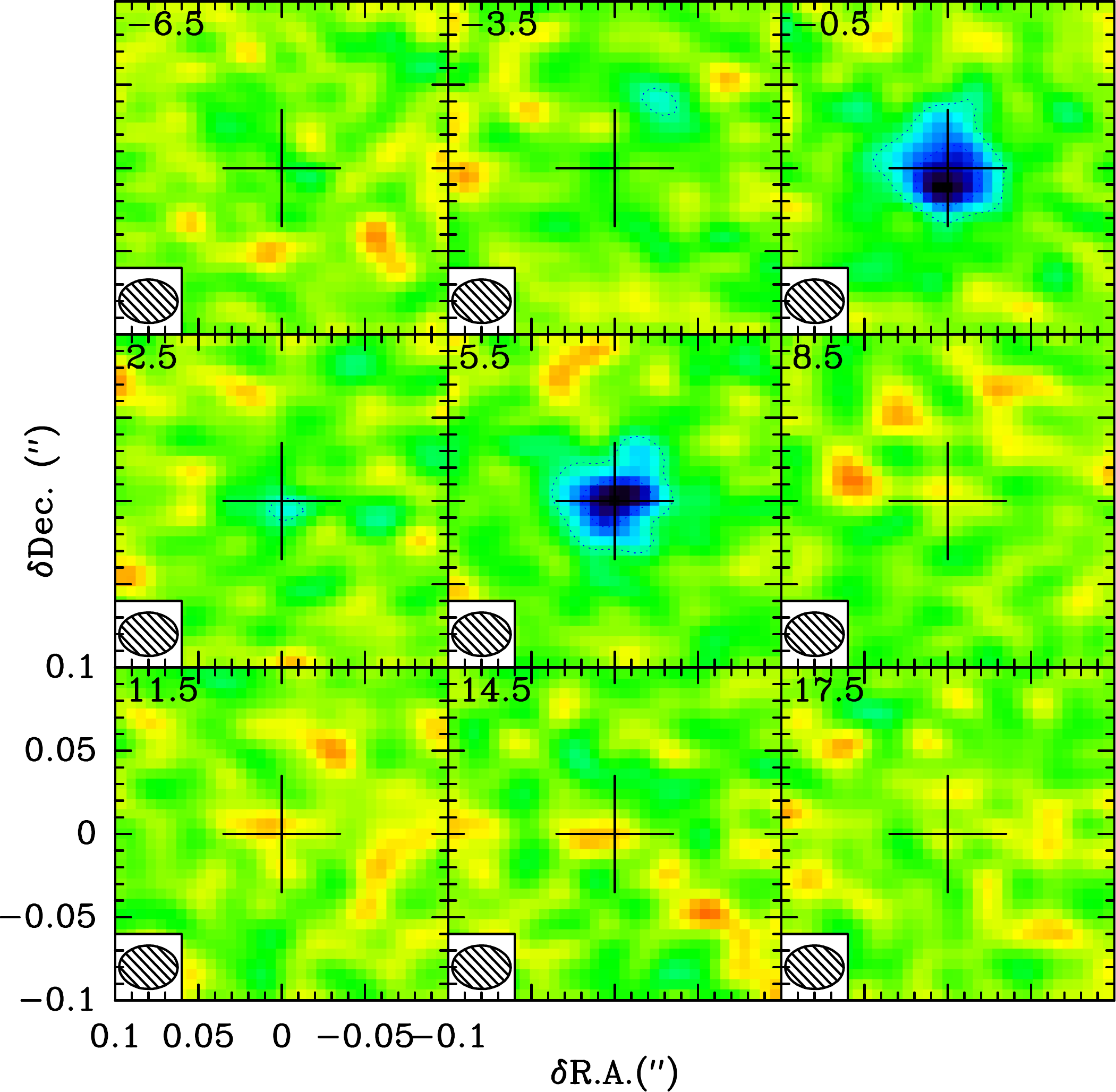} \\
     \vspace{0.25cm} \includegraphics[width=0.80\hsize]{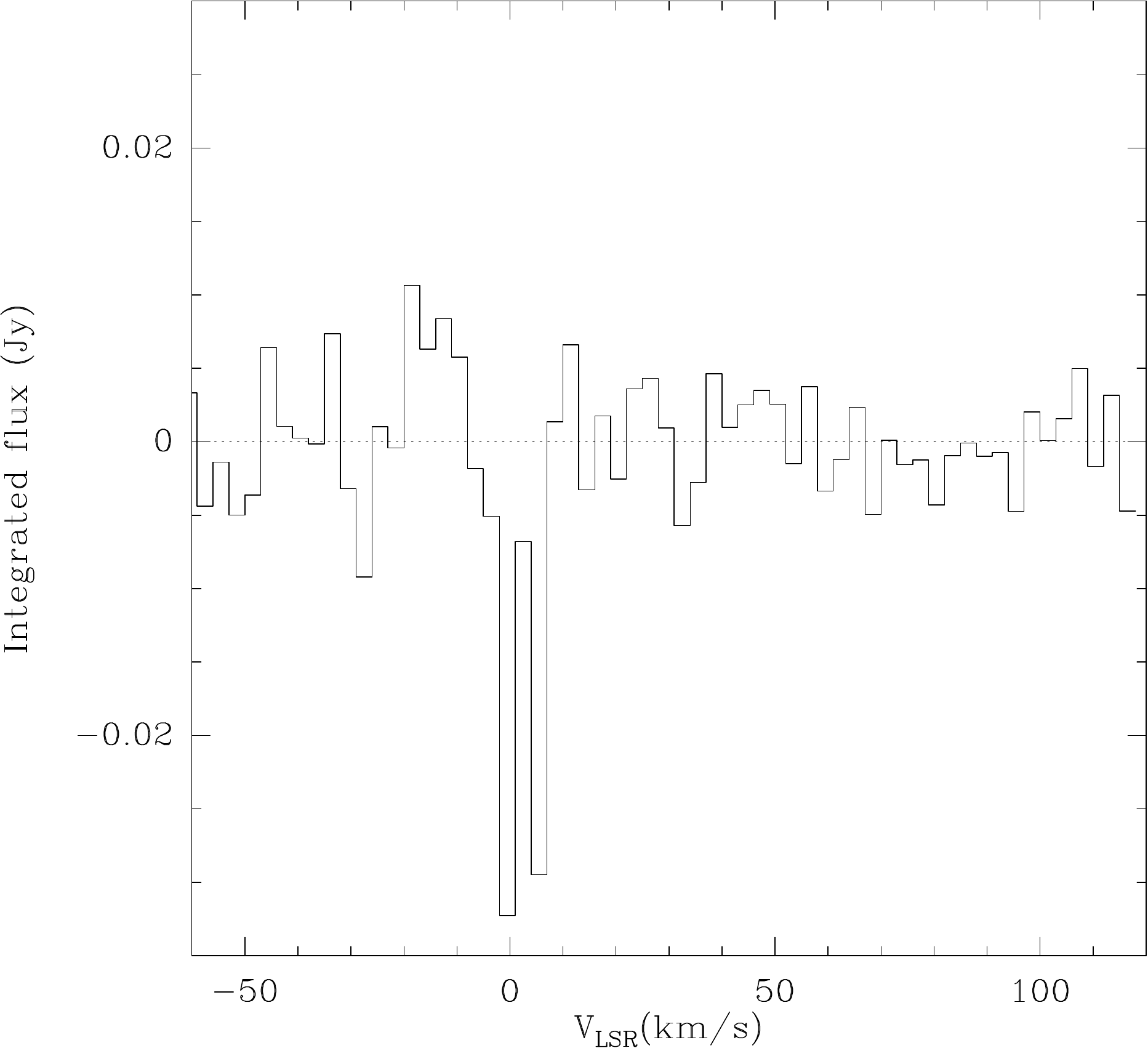}
     \caption{{\bf Top)} \docem\,(2--1) velocity-channel maps around
       the narrow absorption ISM feature detected. Beam is
       0\farc04$\times$0\farc03, PA=89\degr. Level spacing is
       3\,m\jb. {\bf Bottom)} \docem\,(2--1) spectrum towards the
       center of MWC\,922.
       \label{f-co}}
   \end{figure}

   In \mbox{Band 6} we observed, simultaneously with the \htal\ and \hce\ lines,
   the \docem\,(2--1) transition,
   which has been detected in absorption against the compact continuum
   source of MWC\,922 (Fig.\,\ref{f-co}). The two narrow
   ($\sim$2-3\kms-wide) absorption features detected with ALMA are
   centered at \vlsr=$-$0.5 and +5.5\,\kms, consistent with foreground
   ISM absorption. No trace of \docem\,(2--1) emission is found around
   the systemic velocity of MWC\,922
   (\vlsr$\sim$32-33\,\kms). Additional ISM \docem\,(2--1) emission
   features were previously detected towards MWC\,922 with the
   \iram\ single-dish telescope (CSC+17) but these are filtered out in
   our interferometric data. No other molecules (e.g.\,\trecem\,(1--0) included in Band 3)
   are detected neither in absorption or emission in these ALMA
   observations.
   
\end{document}